\RequirePackage{fix-cm}
\documentclass[pdftex,epjc3]{svjour3}
\smartqed  
\usepackage{graphics}
\usepackage{graphicx,epsfig}
\usepackage{subcaption}
\RequirePackage{graphicx}
\usepackage{amsmath}
\usepackage{bm}
\usepackage{amsmath,amssymb}
\usepackage{hyperref}
\usepackage{url}
\usepackage{tabu}
\usepackage{booktabs}
\usepackage{longtable}
\usepackage{nicematrix}
\usepackage[skip=0.5ex]{subcaption}
\usepackage{float}
\usepackage[section]{placeins}
\hypersetup{
    colorlinks=true,       
    linkcolor=red,          
    citecolor=blue,        
    urlcolor=blue           
}
\journalname{Eur. Phys. J. Plus}
\begin{document}

\title{Can we bypass no-go theorem for Ricci-inverse Gravity?
}


\author{Indranil Das\thanksref{e1,addr1}
        \and
        Joseph P Johnson\thanksref{e2,addr2}
        \and
        S. Shankaranarayanan\thanksref{e3,addr2}
}

\thankstext{e1}{e-mail: indranildasdhn@gmail.com}
\thankstext{e2}{e-mail: josephpj@iitb.ac.in}
\thankstext{e3}{e-mail: shanki@phy.iitb.ac.in}


\institute{UM-DAE Centre for Excellence in Basic Sciences, Mumbai 400098, India\label{addr1} \and Department of Physics, Indian Institute of Technology Bombay, Mumbai 400076, India\label{addr2}
}
\institute{UM-DAE Centre for Excellence in Basic Sciences, Mumbai 400098, India\label{addr1} \and Department of Physics, Indian Institute of Technology Bombay, Mumbai 400076, India\label{addr2}}

\date{Received: date / Accepted: date}

\maketitle
\abstract{
Recently, Amendola et al. proposed a geometrical theory of gravity containing higher-order derivative terms~\cite{2020-Amendola.etal-Phys.Lett.B}. The authors introduced anticurvature scalar $(A)$, which is the trace of the inverse of the Ricci tensor ($A^{\mu\nu} = R_{\mu\nu}^{-1}$). In this work, we consider two classes of Ricci-inverse --- Class I and Class II --- models. Class I models are of the form $f(R, A)$ where $f$ is a function of Ricci and anticurvature scalars. Class II models are of the form ${\cal F}(R, A^{\mu\nu}A_{\mu\nu})$ where ${\cal F}$ is a function of Ricci scalar and square of anticurvature tensor. For both these classes of models, we numerically solve the modified Friedmann equations in the redshift range $1500 < z < 0$. We show that the late-time evolution of the Universe, i.e., evolution from matter-dominated epoch to accelerated expansion epoch, \emph{can not} be explained by these two classes of models. Using the reduced action approach, we show that we \emph{can not bypass} the no-go theorem for Ricci-inverse gravity models. Finally, we discuss the implications of our analysis for the early-Universe cosmology.
%
} 
\maketitle
%
\section{Introduction}

General Relativity (GR) has passed many observational tests from weak-gravity regime to strong-gravity regime~\cite{2006-Will-LivingRev.Rel.,2016-Abbott.Others-Phys.Rev.Lett.}. However, GR 
with a Universe containing baryonic matter and radiation can not explain some cosmological observations. The most important among them are the cosmological inflation in the early universe \cite{2000--Cosmologicalinflationlarge}, and the late-time accelerated expansion of the universe \cite{1999-Perlmutter.Others-Astrophys.J.,1998-Riess.Others-Astron.J.}. Even though a solution for an expanding universe exists in the GR framework with baryonic matter/radiation, to explain the accelerated expansion, one needs to make modifications to the theory. One option is to introduce either exotic forms of energy/matter with negative pressure, commonly known as dark energy \cite{2000-Sahni.Starobinsky-Int.J.Mod.Phys.,2006-Copeland.etal-Int.J.Mod.Phys.,2012-Yoo.Watanabe-Int.J.Mod.Phys.D}. 

The most popular dark energy model, the $\Lambda$CDM model, can successfully explain the observational results; however, it is not without problems. The value of $\Lambda$ is many orders smaller than the value of vacuum energy predicted by the standard model of particle physics \cite{1968-Sakharov-Sov.Phys.Dokl.,1989-Weinberg-Rev.Mod.Phys.}. Recently, there also has been disagreement between the value of the cosmological parameters estimated from the high and low redshift observations using the $\Lambda$CDM model \cite{2013-Marra.etal-Phys.Rev.Lett.,2013-Verde.etal-Phys.DarkUniv.}.

Another possible way of explaining the accelerated expansion is by modifying the geometric part of the Einstein Hilbert action, giving rise to the modified gravity theories \cite{2010-DeFelice.Tsujikawa-LivingRev.Rel.,2017-Nojiri.etal-Phys.Rept.,2012-Bamba.etal-Astrophys.SpaceSci.}. It has been shown that modified gravity models with higher-order curvature terms like the $f(R)$ models can lead to the accelerated expansion of the Universe \cite{2010-Sotiriou.Faraoni-Rev.Mod.Phys.,2019-Johnson.Shankaranarayanan-Phys.Rev.D,2011-Nojiri.Odintsov-Phys.Rept.}. However, we still do not have a model which is in perfect agreement with \emph{all} the cosmological observations. For instance, we still don't have an understanding of the late-time acceleration and the associated issues like the $H_0$ tension and $\sigma_8$ tension. \cite{2003-Peebles.Ratra-Rev.Mod.Phys.,2020-DiValentino.etal-Phys.Rev.D,2022-Johnson.etal-JCAP} Hence it is always interesting to consider new possible mechanisms via geometric quantities that can potentially explain the late-time acceleration without invoking exotic physics. 

In Ref. ~\cite{2020-Amendola.etal-Phys.Lett.B} an alternate gravity theory has been proposed in which the Einstein Hilbert action is modified by introducing the anti-curvature tensor ($A^{\mu\nu}$) defined by $A^{\mu\nu} R_{\nu\sigma} = \delta^{\mu}_{\sigma}$ and the anti-curvature scalar $A \equiv g_{\mu\nu}A^{\mu\nu}$. It has to be noted that the anti-curvature scalar $A$ is \emph{not} the inverse of the Ricci scalar $R$ i.e., $A \neq R^{-1}$. Amandola et al. considered modified gravity action containing positive or negative powers of the anti-curvature scalar. The authors proved a general no-go theorem for these models and showed that cosmic trajectories from a decelerated phase could not smoothly join an accelerated phase. 

In this work, we consider generalized anti-curvature models to confirm/infirm the no-go theorem for the Ricci-inverse gravity theories. 
We consider two broad classes of Ricci-inverse --- Class I and Class II --- models. Class I models are of the form $f(R, A)$ where $f$ is a function of  Ricci and anticurvature scalars. Class II models are of the form ${\cal F}(R, A^{\mu\nu}A_{\mu\nu})$ where ${\cal F}$ is a function of Ricci scalar and square of anticurvature tensor. We derive the modified Friedmann equations for the two classes and show that the models have an attractor solution describing the accelerated expansion. Furthermore, we explicitly show the presence of singularities does not allow a smooth transition from a matter-dominated Universe to the phase of accelerated expansion. For a better understanding, we use the reduced action approach for these models, identify these singularities, and show that they are consistent with the results obtained from the study of the evolution equations. To our knowledge, such an analysis has not been carried out for these classes of anti-curvature gravity models. 

In Section (\ref{sec:framework}), we introduce the Ricci and anticurvature tensors and scalars in the flat-space FLRW metric. In Section (\ref{sec:Class I models}), we introduce the $f(R, A)$ model where $f(R, A)$ is an arbitrary function of $R$ and $A$, and then after subsequently analyzing various specific sub-models in this category, we show that it cannot be a viable cosmological model as it fails to describe the transition from the matter-dominated Universe to the late-time accelerated expansion. In Section (\ref{sec:Class II models}) we introduce ${\cal F}(R,A^{\mu\nu}A_{\mu\nu})$ model where ${\cal F}(R,A^{\mu\nu}A_{\mu\nu})$ is an arbitrary function of $R$ and $A^{\mu\nu}$ and derive the general evolution equations. Analyzing the evolution for a few specific examples in this class of models, we show that the problems that plagued the $f(R, A)$ models remains for the ${\cal F}(R, A^{\mu\nu} A_{\mu \nu})$ models as well. In Section (\ref{sec:alternate}), we use the reduced action approach to obtain a fundamental understanding of the reason why the Ricci-inverse models can not explain the evolution of the Universe into the late-time accelerated phase. Section \ref{sec:modgrav} looks at the possibility of mapping the modified gravity models with Ricci-inverse gravity models.
In Section (\ref{sec:conclusion}), we conclude by briefly discussing the result and a possible model that might succeed. Appendices  A and B contain details of the calculations in the main text.

Metric signature $(-,+,+,+)$ and natural units are used throughout the work where $c = \hbar = 8\pi G = 1$. $H$ is the Hubble parameter defined as $H = \dot{a}(t)/a(t)$, where $a(t)$ is the scale factor and is related to the redshift $z$ as $a = 1/(1+z)$. Unless otherwise specified \textit{dot} stands for $d/dt$ and \textit{prime} stands for $d/d \log a$, subscript $``R"$ denotes derivative with
respect to $R$, subscript $``A"$ denotes derivative with respect to $A$, and subscript $``A^2"$ denotes derivative with respect to $A_{\mu\nu} A^{\mu\nu}$.

\section{Ricci-inverse gravity models}\label{sec:framework}

As mentioned above, we aim to confirm/infirm whether we can bypass the no-go theorem for Ricci-inverse gravity. In other words, we want to identify a class of Ricci-inverse models that can describe the evolution of the Universe from the radiation-dominated epoch to the matter-dominated epoch and then to the late-time accelerated expansion epoch. In this section, we list the two classes of models 
that we consider and obtain the modified field equations. 

\subsection{Two classes of Ricci-inverse gravity models}
\label{sec:Twomodels}

In this work, we consider the following two classes of models:
\begin{enumerate}
\item \textbf{Class I models:} In this class we consider the action of the following form: 
\begin{equation}
S = \int d^{4}x \sqrt{-g} \left[ f(R,A)  + \mathcal{L}_m \right]
\label{Class I action}
\end{equation}
where $f(R, A)$ is any general, smooth function of Ricci scalar ($R$) and anticurvature scalar $(A$), and $\mathcal{L}_m$ refers to the matter Lagrangian that is minimally coupled to gravity. One constraint is that, in the appropriate limit, these models should reduce to GR. Varying the above action w.r.t the metric leads to the following equation~\cite{2020-Amendola.etal-Phys.Lett.B}: 

\begin{equation}
\begin{split}
f_{R}R^{\mu\nu} - f_{A}A^{\mu\nu} - \frac{1}{2}fg^{\mu\nu} + g^{\rho\mu}\nabla_{\alpha}\nabla_{\rho}f_{A}A^{\alpha}_{\sigma}A^{\nu\sigma} - \frac{1}{2}\nabla^{2}\left(f_{A}A^{\mu}_{\sigma}A^{\nu\sigma}\right)\\
-\frac{1}{2}g^{\mu\nu}\nabla_{\alpha}\nabla_{\beta}(f_{A}A^{\alpha}_{\sigma}A^{\beta\sigma}) - \nabla^{\mu}\nabla^{\nu}f_{R} + g^{\mu\nu}\nabla^{2}f_{R}= T^{\mu\nu}
\end{split}\label{eomI}
\end{equation}
\item \textbf{Class II models:} In this class we consider the action of the following form: 
\begin{equation}
S = \int d^{4}x \sqrt{-g} \left[  {\cal F}(R,A^{\mu\nu}A_{\mu\nu}) + \mathcal{L}_m \right]
\label{Class II action}
\end{equation}
where ${\cal F}(R,A^{\mu\nu}A_{\mu\nu})$ is any general function containing Ricci scalar and the inverse-Ricci tensor square ($A^{\mu\nu}A_{\mu\nu}$). 
One constraint is that, in the appropriate limit, these models should reduce to GR. 
Varying the above action w.r.t the metric leads to the following equation (\ref{sec:Appendix A} contains the detailed derivation):
\begin{eqnarray}
{\cal F}_{R}R^{\mu\nu} - 2{\cal F}_{A^{2}}A^{\mu\rho}A^{\nu}_{\rho} -\frac{1}{2}{\cal F} g^{\mu\nu} - \nabla^{\mu}\nabla^{\nu}{\cal F}_{R} + g^{\mu\nu}\nabla^{2}{\cal F}_{R} + g^{\rho\nu}\nabla_{\alpha}\nabla_{\rho}({\cal F}_{A^{2}}A_{\sigma\kappa}A^{\sigma\alpha}A^{\mu\kappa}) & & \nonumber \\
-\nabla^{2}({\cal F}_{A^{2}}A_{\sigma\kappa}A^{\sigma\mu}A^{\nu\kappa})
-g^{\mu\nu}\nabla_{\alpha}\nabla_{\rho}({\cal F}_{A^{2}}A_{\sigma\kappa}A^{\sigma\alpha}A^{\rho\kappa})
+2g^{\rho\nu}\nabla_{\rho}\nabla_{\alpha}({\cal F}_{A^{2}}A_{\sigma\kappa}A^{\sigma\mu}A^{\alpha\kappa}) & & \nonumber \\
-g^{\rho\nu}\nabla_{\alpha}\nabla_{\rho}({\cal F}_{A^{2}}A_{\sigma\kappa}A^{\sigma\mu}A^{\alpha\kappa}) = T^{\mu\nu} & &~~~~~
\label{eomII}
\end{eqnarray}
\end{enumerate}
\subsection{Modified Friedmann equations}
\label{sec:MFriedmannEq}

A large amount of cosmological and astrophysical observations provide conclusive evidence that Universe underwent different epochs, starting from the radiation-dominated epoch, followed by the matter-dominated epoch, and finally,  the accelerated epoch with smooth transitions in between. However, to explain the current accelerated expansion, the standard model of cosmology requires modifications either by introducing exotic matter fields or modifying the geometric part. In this work, we consider Ricci-inverse models and aim to describe the evolution of the Universe.

We consider a spatially flat Friedmann-Lema\^itre\\-Robertson-Walker (FLRW) metric: 
\begin{equation}
g_{\mu\nu} = {\rm diag} [-1,a^{2}(t),a^{2}(t),a^{2}(t)]\label{metric in FLRW}
\end{equation}
where $a(t)$ is the scale factor. Various geometric quantities ---  anticurvature tensor ($A^{\mu\nu}$), 
Ricci tensor ($R^{\mu\nu}$), Ricci scalar ($R$) and anticurvature scalar 
($A$)  and anticurvature tensor square ($A^{\mu\nu}A_{\mu\nu}$) --- for the FLRW space-time are given below:
\begin{equation}
\begin{split}
A^{\mu\nu} = {\rm diag} \left[-\frac{a(t)}{3\ddot{a}(t)},\frac{1}{2\dot{a}^{2}(t)+a(t)\ddot{a}(t)},\frac{1}{2\dot{a}^{2}(t)+a(t)\ddot{a}(t)},\frac{1}{2\dot{a}^{2}(t)+a(t)\ddot{a}(t)}\right]\\
= {\rm diag} \left[-\frac{1}{3H^{2}(1+\xi)},\frac{1}{H^{2}(3+\xi)a^{2}(t)},\frac{1}{H^{2}(3+\xi)a^{2}(t)},\frac{1}{H^{2}(3+\xi)a^{2}(t)}\right]\\
\end{split}\label{A+(mu,nu)}
\end{equation}
\begin{equation}
\begin{split}
R_{\mu\nu} = {\rm diag} \left[-\frac{3\ddot{a}(t)}{a(t)},2\dot{a}^{2}(t)+a(t)\ddot{a}(t),2\dot{a}^{2}(t)+a(t)\ddot{a}(t),2\dot{a}^{2}(t)+a(t)\ddot{a}(t)\right]\\
= {\rm diag} \left[-3H^{2}(1+\xi),H^{2}(3+\xi)a^{2}(t),H^{2}(3+\xi)a^{2}(t),H^{2}(3+\xi)a^{2}(t)\right]
\end{split}\label{R-(mu,nu)}
\end{equation} 
\begin{equation}
\begin{split}
R = \frac{6\left(\dot{a}^{2}(t)+a(t)\ddot{a}(t)\right)}{a^{2}(t)}= 6H^{2}(2+\xi)
\end{split}\label{R}
\end{equation}
\begin{equation}
\begin{split}
A = \frac{2a(t)\left(\dot{a}^{2}(t)+5a(t)\ddot{a}(t)\right)}{3\ddot{a}(t)\left(2\dot{a}^{2}(t)+a(t)\ddot{a}(t)\right)} = \frac{2(6+5\xi)}{3H^{2}(1+\xi)(3+\xi)}
\end{split}\label{A}
\end{equation}
\begin{equation}
A^{\mu\nu}A_{\mu\nu} = \frac{4(7\xi^{2}+15\xi+9)}{9H^{4}(1+\xi)^{2}(\xi+3)^{2}}\label{A+(mu,nu)A-(mu,nu)}
\end{equation}
where, $H \equiv \dot{a}/a$ is the Hubble parameter, and 
\begin{equation}
\xi \equiv \frac{H'}{H} = \frac{a\ddot{a}}{\dot{a}^{2}} -1 \, ,
\label{def:xi}
\end{equation}
is the negative of the slow-roll parameter $\epsilon$. Thus, the two modified Friedmann equations are:
\begin{eqnarray}
\nonumber
\left(\frac{\dot{a}}{a}\right)^{2} &=& \dfrac{\rho_{\rm eff}}{3}\hspace{61pt}\\
\left(\frac{\ddot{a}}{a}\right) &=& -\dfrac{1}{6}\left(\rho_{\rm eff}+3p_{\rm eff}\right)
\label{friedmann equations}
\end{eqnarray}
where $\rho_{\rm eff}$ and $p_{\rm eff}$ are the density and pressure of the effective fluid containing multiple components. Using the definition of $\xi$ in Eq \eqref{def:xi}, combining the two equations in (\ref{friedmann equations}) leads to:
\begin{equation}
p_{\rm eff} = -\frac{\rho_{\rm eff}}{3}(3+2\xi) \, . \label{effective equation of state}
\end{equation}
 Thus, one can define an equation of state parameter of the effective fluid as:
\begin{equation}
w_{\rm eff} = -\frac{2\xi}{3} - 1 \, . \label{effective state parameter}
\end{equation}
The value of $\xi$ at any time implies whether the expansion of the Universe is accelerated or decelerated. Any cosmological model which is consistent with observations should describe a smooth transition from a decelerated phase ($\ddot{a}<0$) to the accelerated phase ($\ddot{a}>0$). This corresponds to the transition from $\xi < -1$ to $\xi > -1$.

For the Ricci-inverse models considered here, the Ricci-inverse corrections contribute to the effective fluid density and pressure. Thus, we have:
\begin{eqnarray}
\nonumber
\rho_{\rm eff} &=& \rho_{t}+\rho_{A} \hspace{10pt};\hspace{10pt} \rho_{t} = \rho_{m}+\rho_{r} \hspace{52pt}\\
p_{\rm eff} &=& p_{t}+p_{A} \hspace{10pt};\hspace{10pt} p_{t} = w_{m}\rho_{m} + w_{r}\rho_{r} = p_{r}
\label{def:rhoP}
\end{eqnarray}
where $(\rho_{A}, p_{A})$ refer to the density and pressure of the anticurvature terms, $(\rho_{m}, p_{m})$ refer to the density and pressure of the matter (baryons and dark matter) content in the Universe and $(\rho_{r}, p_{r})$ refer to the density and pressure of the radiation. It has to be noted that the complete set of background evolution equations can be written in terms of variables $\xi$, $\Omega_r$, $\Omega_m$, and the model parameters that remain constant.

Since we assume that the late-time acceleration is due to the Ricci-inverse modifications to GR, we do not include cosmological constant or dark energy components. Substituting the form of $\rho_{\rm eff}$ from Eq. \eqref{def:rhoP} in the first Friedmann equation \eqref{friedmann equations}, we have:
\begin{equation}
\Omega_{m} + \Omega_{r} + \Omega_{A} = 1
\label{eq:energy_const}
\end{equation} 
where $\Omega_{i} = \rho_{i}/3H^{2}$, where the subscript $i$ refers to $r$(radiation) or $m$(matter) or $A$ (anitcurvature). 

In this work, for both classes of models considered, we 
numerically solve the modified Friedmann equations \eqref{friedmann equations} numerically subject to the above constraint Eq.\eqref{eq:energy_const} in the redshift range $1500 < z < 0$ for a range of initial values $-3 \leq \xi_i \leq 1$ that lead to non-diverging solutions. We assume the matter to be pressureless dust. Hence, $w_{m} = 0$ and $w_{r} = 1/3$. We fix the initial values for the remaining parameters to be $\Omega_{r_i} = 0.39$, $\Omega_{m_i} = 0.59$ and $\xi_i' = 0$. The above initial values of $\Omega_r$ and $\Omega_m$ correspond to radiation matter equality at $z \sim 2740$. $\xi'=0$ corresponds to the power-law evolution of the scale factor $a$, which is consistent with the radiation/matter-dominated era.

In the rest of the work, we consider the two classes of Ricci-inverse models introduced in Section \eqref{sec:Twomodels}. We show that the late-time evolution of the Universe, i.e., evolution from matter-dominated epoch to accelerated expansion epoch, can't be explained by this model. 
Using the reduced action approach, we provide a reason why we can not bypass the no-go theorem for Ricci-inverse gravity models. 

The following table contains values of $\xi$ and $w_{\rm eff}$ for different epochs of the Universe. In particular, as we evolve the modified Friedmann equations, $\xi$ needs to evolve from matter-dominated epoch ($\xi = -1.5$) to the current epoch ($\xi \simeq -0.5$).  As we will show, this transition is not possible for these two classes of Ricci-inverse models.
\begin{table}[H]
\begin{center}
\begin{tabular}{cccc}
\hline
\bm{$\xi$} & \bm{$p = -\frac{\rho}{3}(3+2\xi)$} & \bm{$w_{\rm eff}$} & \textbf{Effective matter fluid description}\\
\hline
0 & $p = -\rho$ & -1 & Cosmological constant\\
\hline
-2 & $p = \rho/3$ & 1/3 & Radiation\\
\hline
-1.5 & $p = 0$ & 0 & Matter(pure dust)\\
\hline
-0.495 or -0.5 & $p = -0.67\rho$ & -0.67 & Current Universe\\
\hline
\end{tabular}
\caption{$\xi$ for different epochs of the Universe \label{table:1}}
\end{center}
\end{table}
\section{Class I models}\label{sec:Class I models}

As mentioned earlier, $f(R, A)$ in Eq.~\eqref{Class I action} is an arbitrary, smooth function of Ricci and anticurvature scalars. We further classify this class into three subclasses --- Class Ia, Class Ib, and Class Ic. Class Ia models are polynomials in $R$ and $A$. This model has been discussed in the literature~\cite{2020-Amendola.etal-Phys.Lett.B,2021-Do-Eur.Phys.J.C,2022-Do-Eur.Phys.J.C,2021-Scomparin}. For completeness, we briefly discuss the results of this class of models in \ref{sec:Appendix B}. Class Ib models are of the form $\exp[(R A)^n]$. Class Ic models are non-polynomial functions of $R$ and $A$.

\subsection{Class Ia models}\label{subsec:Class Ia models}

In Ref. \cite{2020-Amendola.etal-Phys.Lett.B}, the authors proved a no-go theorem for the case when $f(R, A)$ is a polynomial function of $R$ and $A$. Specifically, the authors showed that for polynomial function of $R$ and $A$, cosmic  trajectories around $\xi = -1.5$ will never evolve to $\xi = -0.5$. \ref{sec:Appendix B} contains the plots for these classes of models. In Ref.~\cite{2021-Do-Eur.Phys.J.C} the author showed that one gets a no-go theorem in the Ricci-inverse model for both the isotropic and anisotropic inflation. This was further extended by including the second-order anticurvature scalar term in Ref.~\cite{2022-Do-Eur.Phys.J.C} where the author showed that it is impossible to have stable inflation within FLRW cosmology.  

This naturally leads to the following question: Can we bypass the no-go theorem for an arbitrary power of $R$ and $A$? To address this, we consider class Ib models.

\subsection{Class Ib models}\label{subsec:Class Ib models}

In this class, we consider the following form of $f(R, A)$:
\begin{equation}
\label{action:class1b}
f(R,A) = Re^{\alpha (RA)^{n}} \, ,
\end{equation}
where $n$ is an integer, and $\alpha$ (a dimensionless constant) decides the deviation from GR. We have made the above choice for the following reasons: First, it is equivalent to an infinite polynomial involving a combination of powers of the Ricci and anticurvature scalars. Second, in the limit of $\alpha \to 0$, it reduces to Einstein-Hilbert action. The modified Friedman equations for this class of models are:
\begin{equation}
\begin{split}
\rho_{t} = 3H^{2}e^{4^{n}\alpha\left(\frac{(2+\xi)(6+5\xi)}{(1+\xi)(3+\xi)}\right)^{n}}\hspace{2pt}\left[\hspace{2pt} 1+\frac{2\times(3+2\xi)4^{n}\alpha n\xi}{(1+\xi)(3+\xi)(6+5\xi)}\left(\frac{(2+\xi)(6+5\xi)}{(1+\xi)(3+\xi)}\right)^{n}\right.\\
\left.+\frac{2\times4^{n}\alpha n \left(\frac{(2+\xi)(6+5\xi)}{(1+\xi)(3+\xi)}\right)^{n}\xi'}{(1+\xi)^{2}(2+\xi)(3+\xi)^{2}(6+5\xi)^{2}}\hspace{2pt}\Bigg(\hspace{2pt}3(36 + 84\xi + 75\xi^{2} + 32\xi^{3})\right.\\
\left.+6n\xi^{2}(3+4\xi)\left(1+4^{n}\alpha\left(\frac{(2+\xi)(6+5\xi)}{(1+\xi)(3+\xi)}\right)^{n}\right)\right.\\
\left.+\xi^{4}\left(17+8n+8n\times4^{n}\alpha\left(\frac{(2+\xi)(6+5\xi)}{(1+\xi)(3+\xi)}\right)^{n}\right)\Bigg)\right]\\
\end{split}\label{density eom}
\end{equation}

\begin{equation}
\begin{split}
p_{t} = \Omega_{r}H^{2} = H^{2}e^{4^{n}\alpha\left(\frac{(2+\xi)(6+5\xi)}{(1+\xi)(3+\xi)}\right)^{n}}\hspace{2pt}\Bigg[\hspace{2pt}-(3+2\xi)-\frac{2\times4^{n}n\alpha\xi(3+2\xi)^{2}}{(1+\xi)(6+5\xi)}\left(\frac{(2+\xi)(6+5\xi)}{(1+\xi)(3+\xi)}\right)^{n}\\
-2\times4^{n}n\alpha\left(\frac{(2+\xi)(6+5\xi)}{(1+\xi)(3+\xi)}\right)^{n}\hspace{2pt}\Bigg(\hspace{2pt}\frac{3\xi'}{(1+\xi)^{2}(6+5\xi)^{2}(3+\xi)^{2}}\hspace{2pt}\Bigg(\hspace{2pt}108+252\xi+225\xi^{2}\\
+96\xi^{3}+17\xi^{4}+2n(3+2\xi)^{2}+2n(3+2\xi)^{2}\xi^{2}4^{n}\alpha\left(\frac{(2+\xi)(6+5\xi)}{(1+\xi)(3+\xi)}\right)^{n}\Bigg)\\
+\frac{(\xi')^{2}}{(1 + \xi)^{3}(2 + \xi)^{2}(3 + \xi)^{3}(6 + 5\xi)^{3}}\Bigg(-9720-388556\xi-65070\xi^{2}-61371\xi^{3}-35532\xi^{4}-12834\xi^{5}\\
-2782\xi^{6}-255\xi^{7}-12n\xi(2+\xi)(3+2\xi)(-27-36\xi+15\xi^{3}+5\xi^{4})\left(1+4^{n}\alpha\left(\frac{(2+\xi)(6+5\xi)}{(1+\xi)(3+\xi)}\right)^{n}\right)\\
+4n^{2}\xi^{3}(3+2\xi)^{3}\left(1+2^{n}\alpha\left(3\times2^{n}\left(\frac{(2+\xi)(6+5\xi)}{(1+\xi)(3+\xi)}\right)^{n}+8^{n}\alpha\left(\frac{(2+\xi)(6+5\xi)}{(1+\xi)(3+\xi)}\right)^{2n}\right)\right)\Bigg)\\
+\frac{\xi''(108+252\xi+225\xi^{2}+96\xi^{3}+17\xi^{4}+2n\xi(3+\xi)^{2}\left(1+4^{n}\alpha\left(\frac{(2+\xi)(6+5\xi)}{(1+\xi)(3+\xi)}\right)^{n}\right)}{(1 + \xi)^{2}(2 + \xi)(3 + \xi)^{2}(6 + 5\xi)^{2}}\Bigg)\Bigg]
\end{split}\label{pressure eom}
\end{equation}
Since the matter and radiation are non-interacting, they individually 
satisfy the following energy conservation equations:
\begin{equation}
\dot{\Omega}_{r} = -H(t) \Omega_{r} \left[4 +\frac{2\dot{H}}{H^{2}}\right]
, \quad
\dot{\Omega}_{m} = -H(t) \Omega_{m}\left[3 +\frac{2\dot{H}}{H^{2}}\right]
\end{equation}
In terms of $\ln(a)$, the above equations lead to:
\begin{equation}
\Omega_{r}'+ \Omega_{r}(4+ 2 \xi) = 0, \quad \Omega_{m}'+\Omega_{m}(3+2\xi) = 0 \label{energy density equation}
\end{equation}
Setting $\alpha = 0$ in the above equations leads to standard GR equations. 
Since $n$ is an arbitrary parameter, different values of $n$ lead to varied
evolution. To keep calculations tractable, we consider the following four values of $n$: $n = 1,2,-1$ and $-2$. For each of these values of $n$, we choose different values of $\alpha$ and a large range of initial conditions for $\xi$. 

We evolve the above equations (\ref{density eom}, \ref{pressure eom}, \ref{energy density equation}) numerically in the redshift range $1500 < z < 0$ for the initial values of the three parameters to be $\Omega_{r_i} = 0.39$, $\Omega_{m_i} = 0.59$, $-3 \leq \xi_i \leq 1$  and $\xi_i' = 0$. 

\begin{figure}[H]
\begin{subfigure}{.55\textwidth}
  \centering
  \includegraphics[width=.8\linewidth]{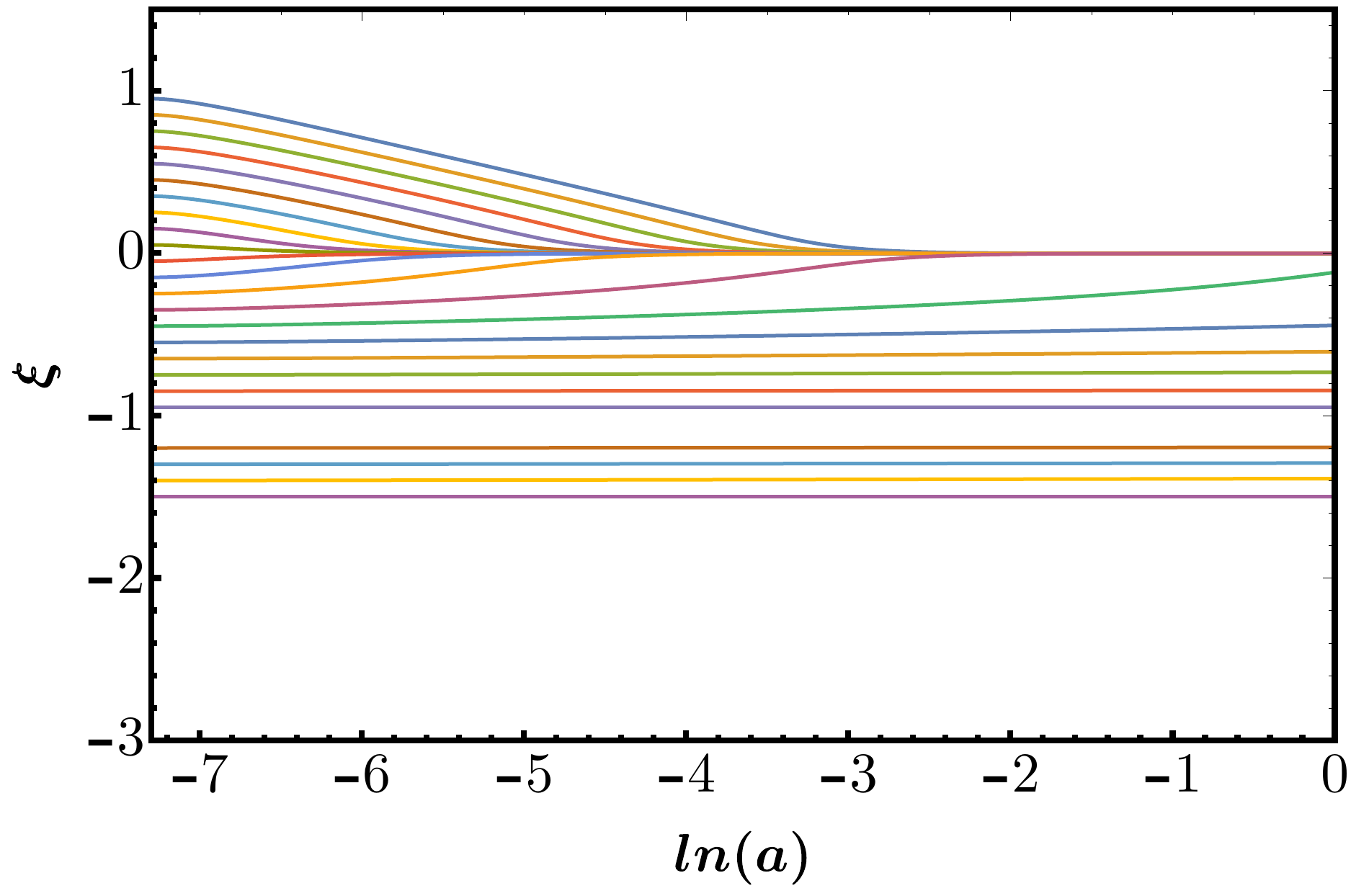}  
  \caption{$\xi(a)$ vs ln(a),$\alpha = 23$}
\end{subfigure}
\begin{subfigure}{.55\textwidth}
  \centering
  \includegraphics[width=.8\linewidth]{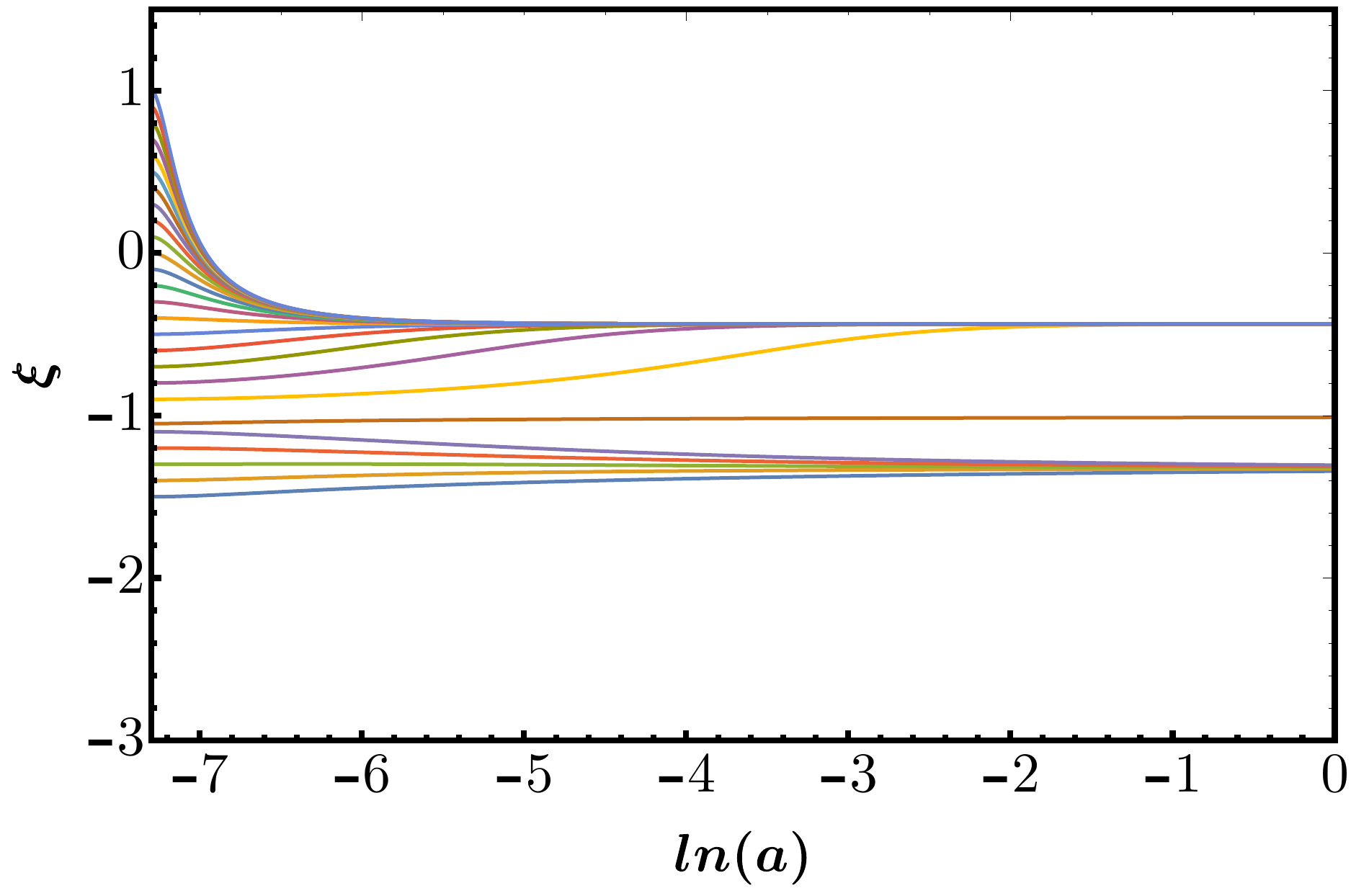}  
  \caption{$\xi(a)$ vs ln(a),$\alpha = 0.07$}
\end{subfigure}
\newline
\begin{subfigure}{.55\textwidth}
  \centering
  \includegraphics[width=.8\linewidth]{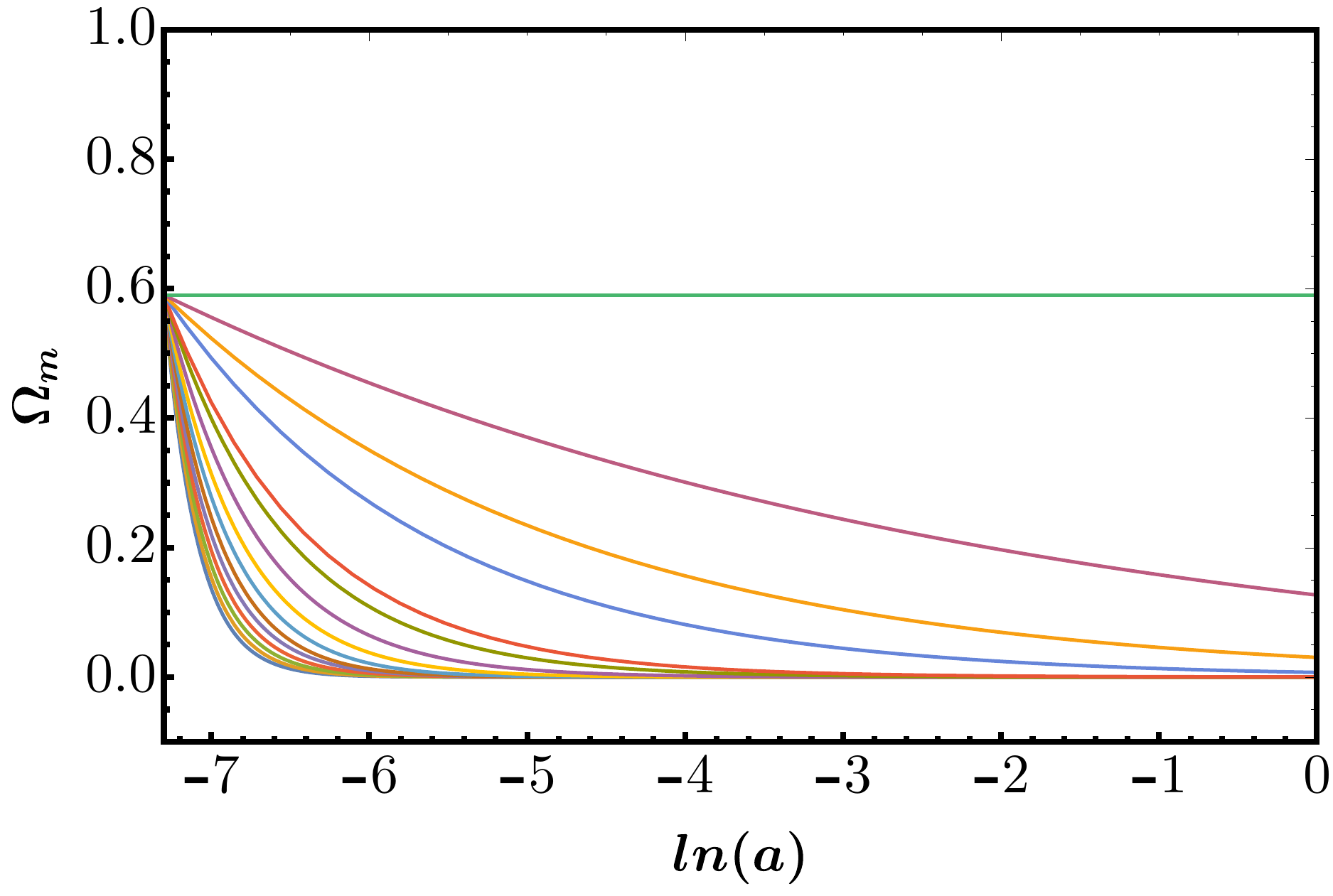}  
  \caption{$\Omega_{m}(a)$ vs ln(a),$\alpha = 23$}
\end{subfigure}
\begin{subfigure}{.55\textwidth}
  \centering
  \includegraphics[width=.8\linewidth]{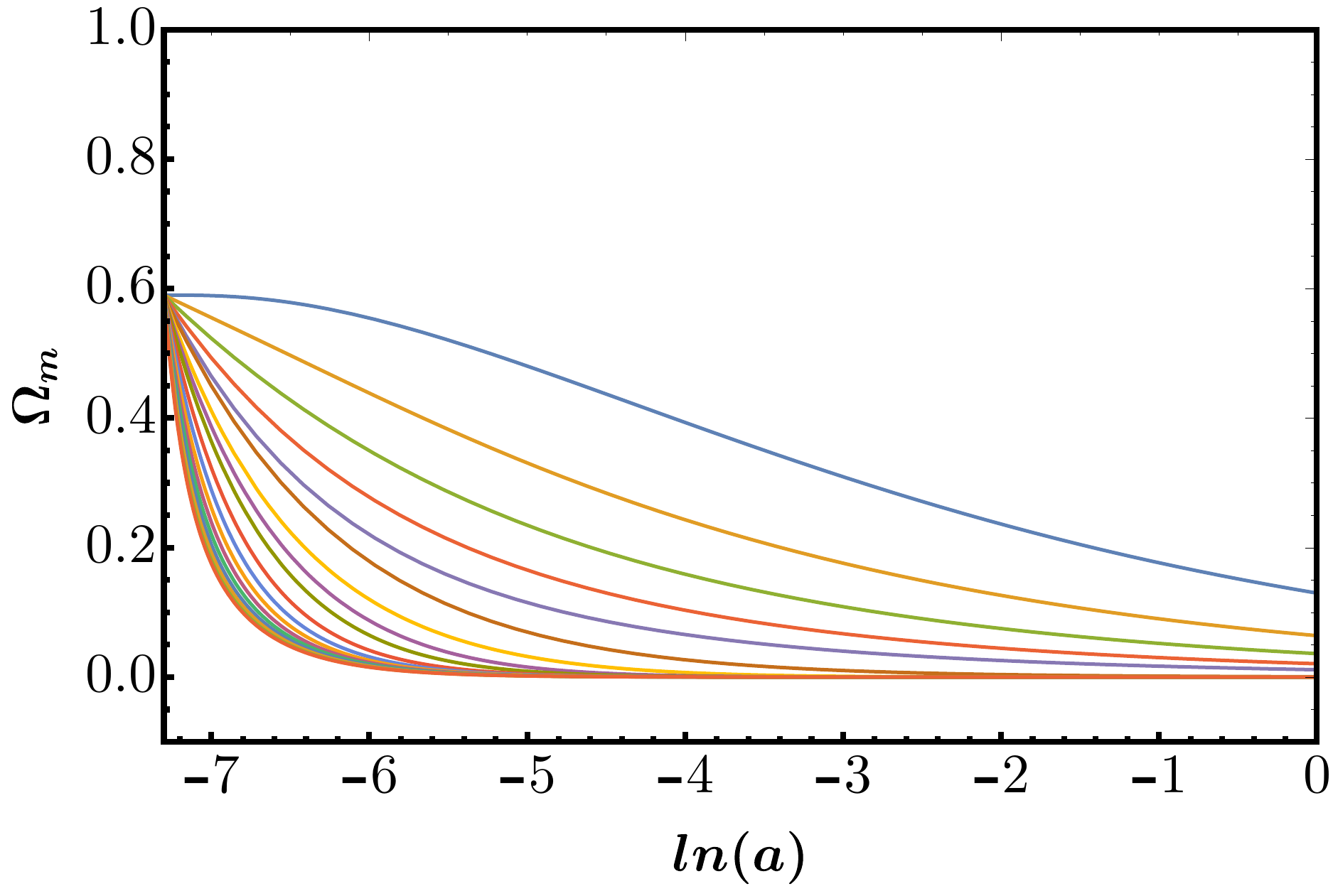}  
  \caption{$\Omega_{m}(a)$ vs ln(a),$\alpha = 0.07$}
\end{subfigure}
\newline
\begin{subfigure}{.55\textwidth}
  \centering
  \includegraphics[width=.8\linewidth]{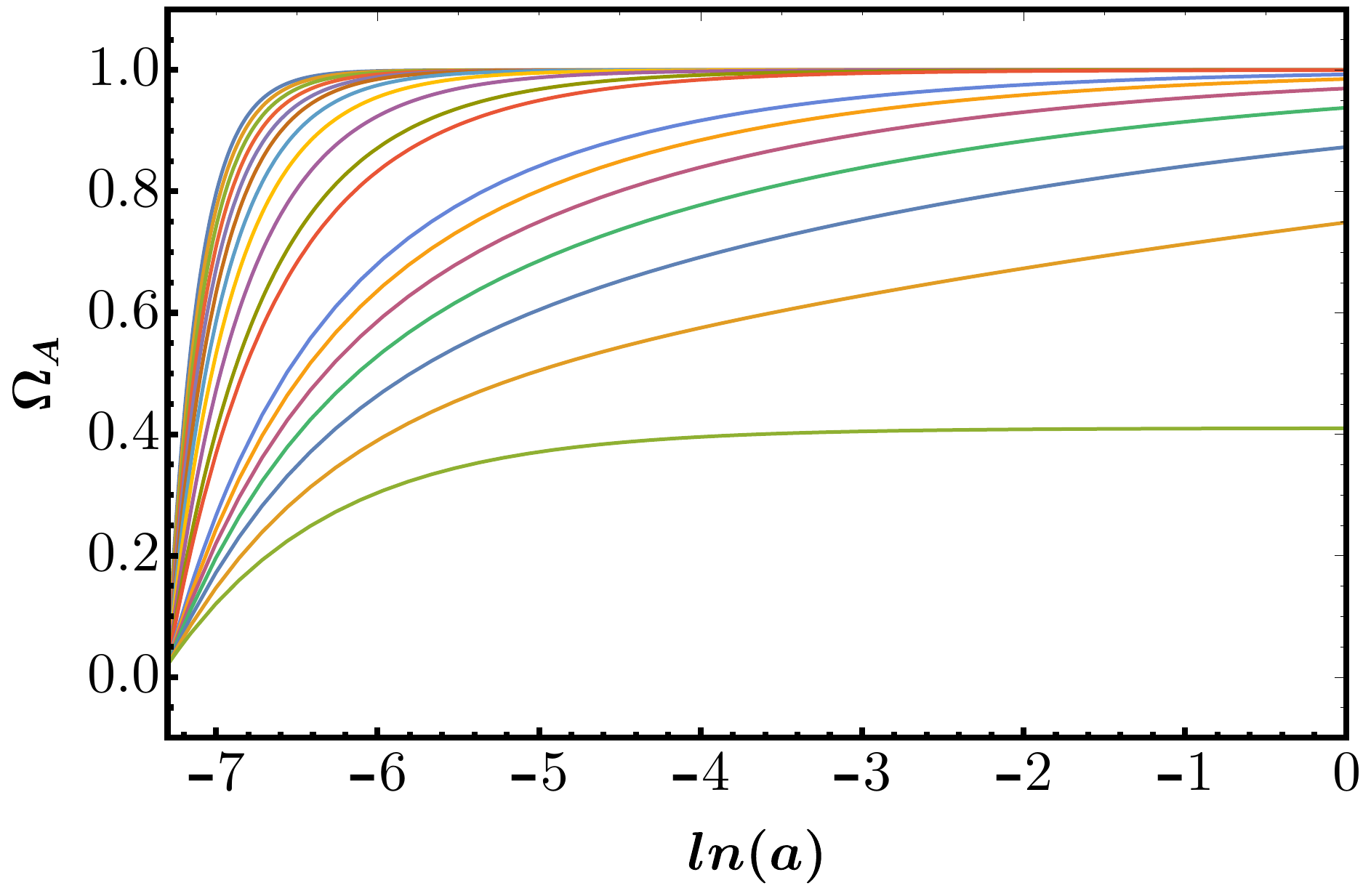}  
  \caption{$\Omega_{A}(a)$ vs ln(a),$\alpha = 23$}
\end{subfigure}
\begin{subfigure}{.55\textwidth}
  \centering
  \includegraphics[width=.8\linewidth]{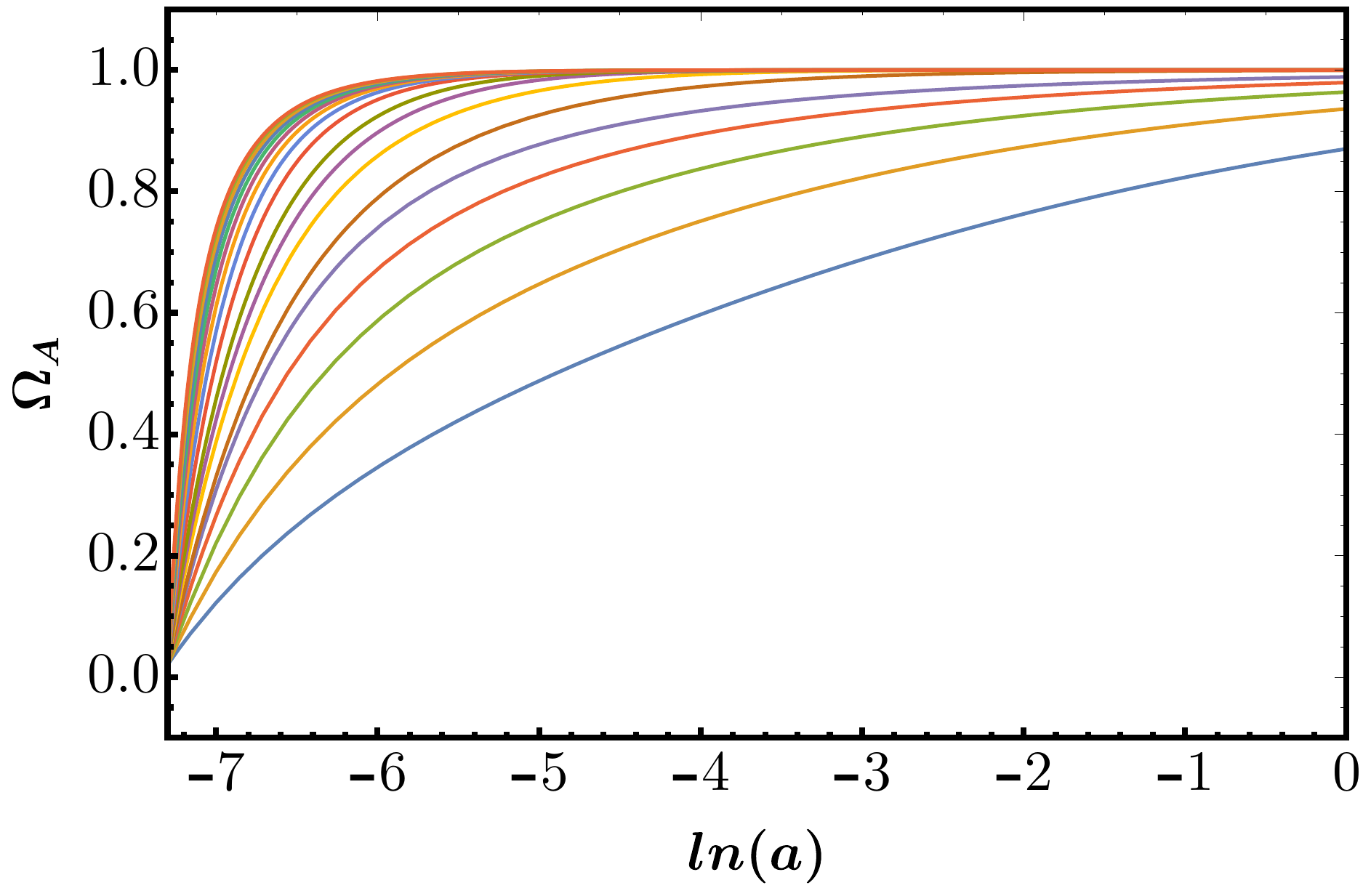}  
  \caption{$\Omega_{A}(a)$ vs ln(a),$\alpha = 0.07$}
\end{subfigure}
\caption{Plot of $\xi(a), \Omega_m(a), \Omega_A(a)$ as a function of $\ln(a)$ for $f(R, A) = Re^{\alpha (RA)}$. Plots on the left are for $\alpha =  23$ and the plots are the right are for $\alpha = 0.07$.}
\label{fig:(ra)}
\end{figure}

In Figs. \eqref{fig:(ra)} - \eqref{fig:(ra)-2,2}, we have plotted $\xi(a), \Omega_r(a), \Omega_m(a),$ and $\Omega_A(a)$ as functions of $\ln(a)$ (corresponding to the redshift range $1500 < z < 0$) for different values of $n$. In each set of figures, we have plotted for at least two distinct $\alpha$ values. For each value of $\alpha$ we have plotted for different initial values of $\xi$. At $z = 1500$ (or $\ln(a) = -7.31$),  $\xi$ is in the range $[-5, 0.6]$. This is because we want to obtain a Universe that evolves from a radiation-dominated epoch to a matter-dominated epoch and then to a late-time accelerated expansion epoch (cf. Table \ref{table:1}).

 
In the rest of this section, we discuss the features of each of these scenarios and identify whether, for a range of initial values of $\xi$, we can obtain a Universe that evolves from a radiation-dominated epoch to a matter-dominated epoch and then to a late-time accelerated expansion epoch.

Fig. \ref{fig:(ra)} contains the plots for $n = 1$ (in Eq. \eqref{action:class1b}) for two values of $\alpha$. Plots on the left are for $\alpha =  23$ and the plots are the right are for $\alpha = 0.07$.  Different colors in the plots refer to different initial values 
of $\xi$ in the range $[-5, 0.6]$. From these plots, we infer the following:
\begin{longtable}{|l|l|l|l|l|l|l|l|}
    \hline
    \multicolumn{4}{|c|}{$\alpha=23$} &\multicolumn{4}{|c|}{$\alpha=0.07$}\\
    \hline
    $\xi_i$& $\xi_{f}$ & $\Omega_{{A}_{f}}$ & $\Omega_{{m}_{f}}$  & $\xi_i$ & $\xi_{f}$ & $\Omega_{{A}_{f}}$ & $\Omega_{{m}_{f}}$ \\
    \hline
    $\xi_i >-1$& $\xi_{f}=0$ & $\Omega_{{A}_{f}}=1$ & $\Omega_{{m}_{f}}=0$ &  $\xi_i >-0.9$ & $\xi_{f} \approx -0.5 $ & $\Omega_{{A}_{f}}=1$ & $\Omega_{{m}_{f}}=0$ \\
    \hline
    $\xi_i \in [-1.2,-1]$& $\xi_f \rightarrow \pm \infty$  & $\Omega_{{A}_{f}}=1$ & $\Omega_{{m}_{f}}=0$  & $\xi_i \in [-1.1,-1.0]$ & $\xi_{f} \rightarrow \pm \infty $ & $\Omega_{{A}_{f}}\rightarrow \pm \infty$ & $\Omega_{{m}_{f}}\rightarrow \pm \infty$ \\
    \hline
    $\xi_i <-1.2$& $ \approx \xi_i$ & $\Omega_{{A}_{f}} < 1 $ & $\Omega_{{m}_{f}} \neq 0$  & $\xi_i \in [-1.5,-1.1]$ & $\xi_{f} \approx -1.4 $ & $\Omega_{{A}_{f}}\rightarrow 1$ & $\Omega_{{m}_{f}}\rightarrow 0$ \\
    \hline
\end{longtable}
\begin{enumerate}
\item For $\alpha = 23$ and initial values of $\xi > -1$, we get a de Sitter attractor ($\xi = 0$). Corresponding to these initial values, $\Omega_{r}$ and $\Omega_{m}$ starts at $0.39$ and $0.59$ respectively, and quickly converge to $0$; while $\Omega_{A}$ starts at 0.02 and quickly converges to $1$.

\item For $\alpha = 23$ and for the initial values of $\xi$ in the range $[-1.2, -1]$, 
the values ($\xi, \Omega_m, \Omega_r)$ diverge. 

\item For $\alpha = 23$ and initial values of $\xi <  -1.2$, $\xi$ is nearly constant and $\Omega_{r}$ and $\Omega_{m}$ does not quickly converge. Importantly, for these initial values, $\Omega_{m}$ always approaches a non-zero value. For these initial values of $\xi$, $\Omega_{A}$ starts at $0.02$ and does not quickly converge and always approaches a value less than $1$.

\item For $\alpha = 0.07$ and initial values of $\xi >  -0.9$, we get $\xi \approx -0.5$ as an attractor ($w_{\rm eff} \approx -0.67$). Corresponding to these initial values $\Omega_{r}$ and $\Omega_{m}$ start at $0.39$ and $0.59$ respectively and quickly converge to $0$. $\Omega_{A}$ starts at 0.02 and quickly converges to 1.

\item For $\alpha = 0.07$ and for the initial values of $\xi$ in the range $[-1.1, -1]$, 
the values ($\xi, \Omega_m, \Omega_r)$ diverge. 

\item For $\alpha = 0.07$ and initial values of $\xi$ in the range $[-1.5,-1.1]$ lead to an attractor at around $\xi = -1.4$ (very close to matter dominated Universe).$\Omega_{r}$ and $\Omega_{m}$ starts at $0.39$ and $0.59$ respectively and do not quickly converge. Like in the case of $\alpha = 23$, for these initial values, $\Omega_{m}$ slowly approaches zero. For these initial values of $\xi$, $\Omega_{A}$ starts at $0.02$ and slowly converges to $1$. 
\end{enumerate}

Fig. \ref{fig:(ra)2,1} contains the plots of $n = 2$ (in Eq. \eqref{action:class1b}) for 
$\alpha = 0.0016$. Plots on the left are for the initial values of $\xi$ in the range 
$[-2, 0.6]$ and the plots are the right are for the initial values of $\xi$ in the range 
$[-2.9, -2.1]$. Different colors in the plots correspond to different initial values 
of $\xi$ in these ranges. From these plots, we infer the following:
\begin{enumerate}
\item For initial values of $\xi > -0.9$, $\xi \approx -0.5$ is an attractor ($w_{\rm eff} \approx -0.67$ corresponding to the current Universe). Corresponding to these initial values, $\Omega_{r}$ and $\Omega_{m}$ starts at $0.39$ and $0.59$, respectively, and quickly converge to $0$; while $\Omega_{A}$ starts at $0.02$ and quickly converges to 1.

\item For the initial values of $\xi$ in the range $[-2, -1.1]$, 
the values ($\xi, \Omega_m, \Omega_r)$ diverge. 

\item For initial values of $\xi < -2$,  $\xi = -2.8$ is an attractor. For these initial values, $\Omega_{r}$ and $\Omega_{m}$ start at $0.39$ and $0.59$, respectively, and evolve to unphysical values, which contradicts the observed trend. Hence,   these initial values are excluded.
\end{enumerate}
%
%
\begin{figure}[H]
\begin{subfigure}{.55\textwidth}
 %
  \includegraphics[width=.8\linewidth]{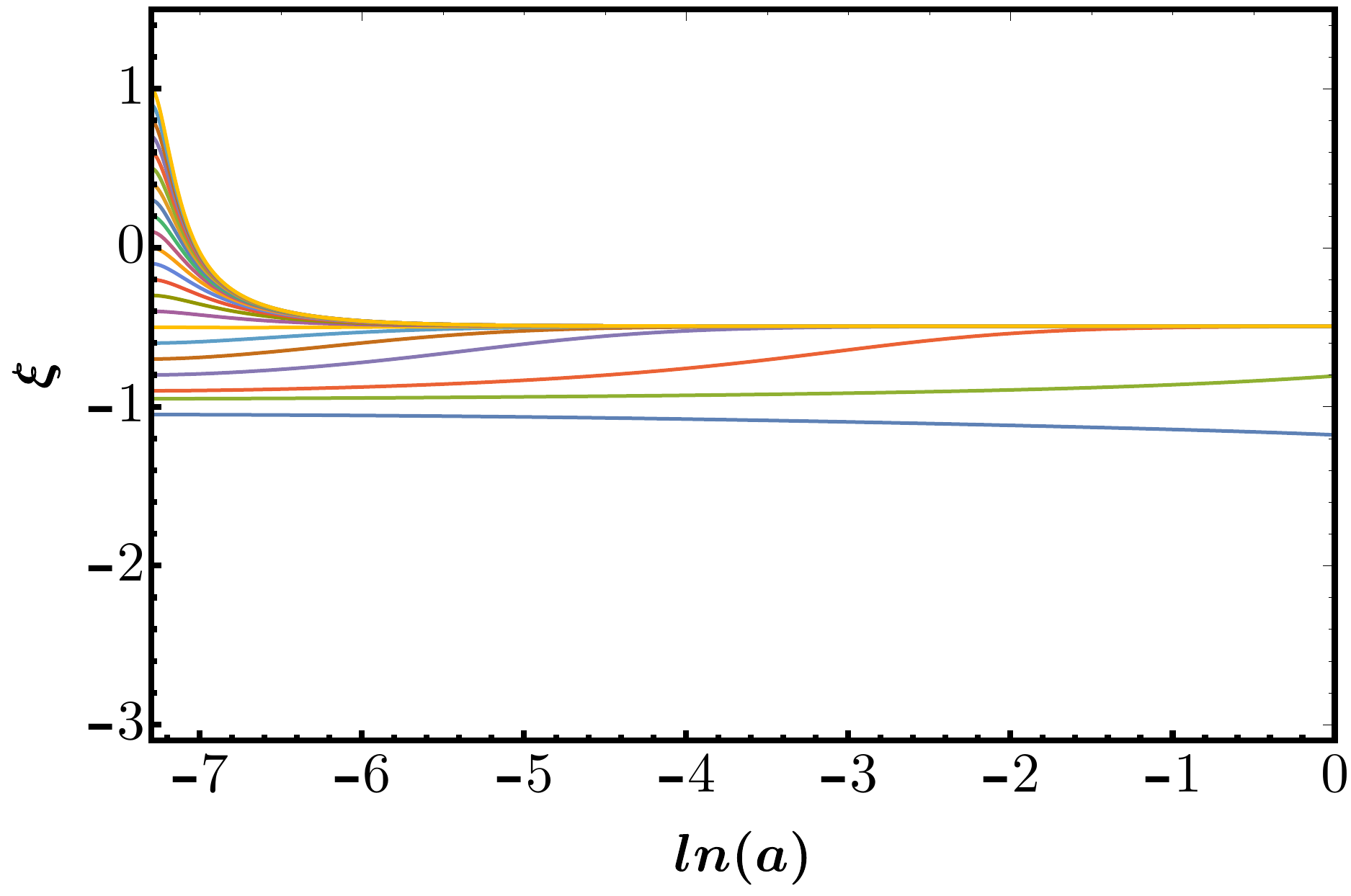}  
  \caption{$\xi(a)$ vs ln(a)}
\end{subfigure}
\begin{subfigure}{.55\textwidth}
 %
  \includegraphics[width=.8\linewidth]{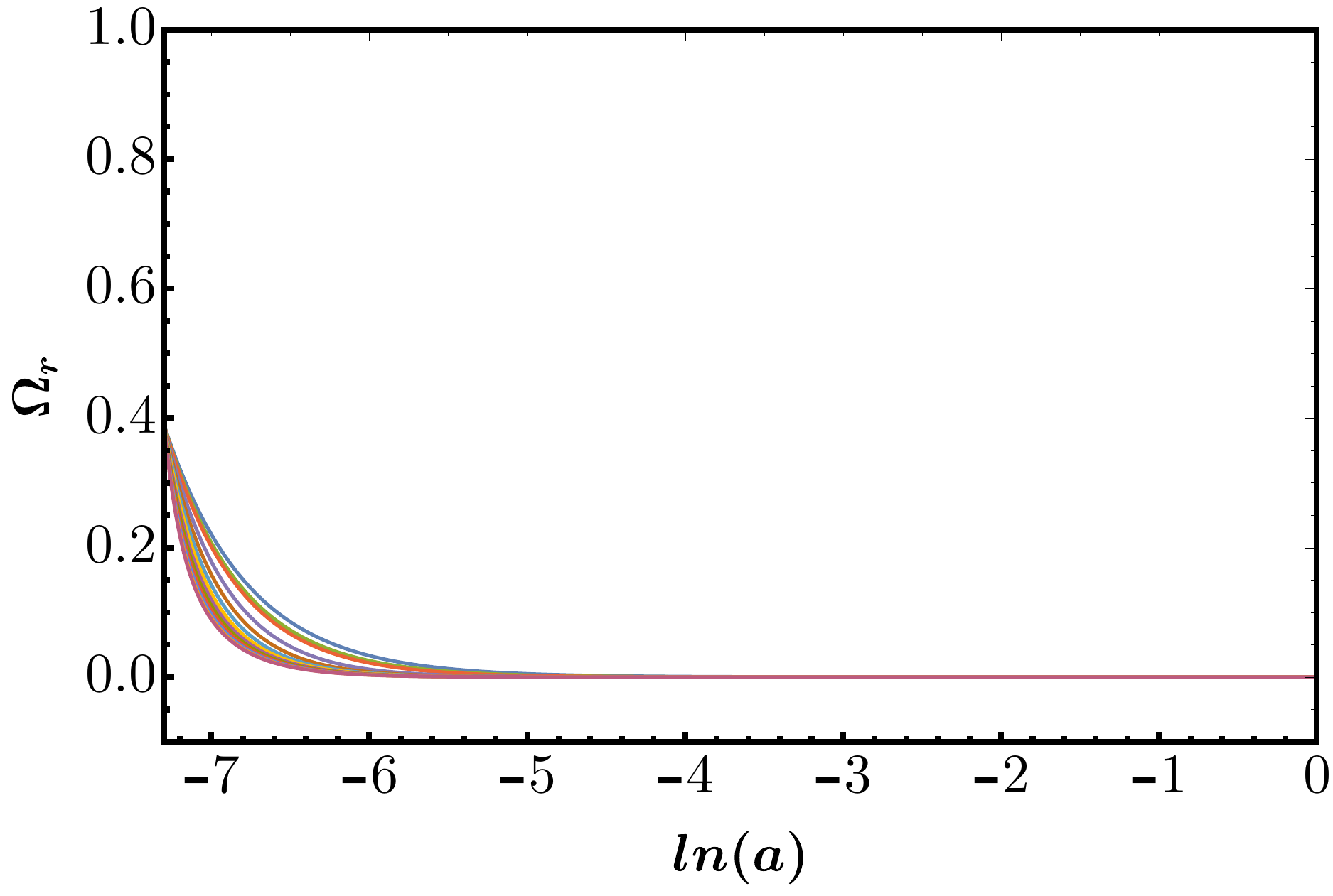}  
  \caption{$\Omega_r(a)$ vs ln(a)}
\end{subfigure}
\newline
\begin{subfigure}{.55\textwidth}
  %
  \includegraphics[width=.82\linewidth]{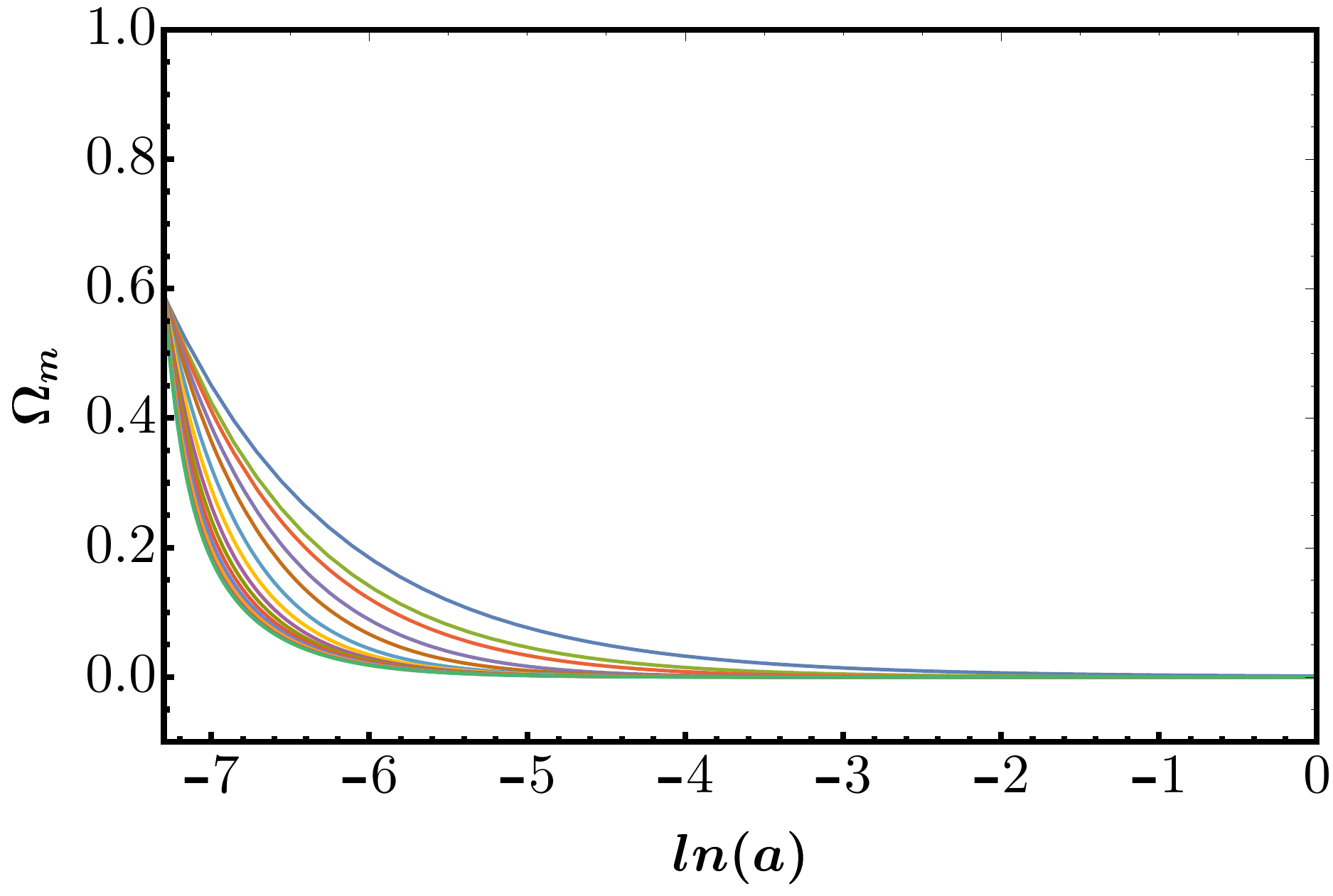}  
  \caption{$\Omega_{m}(a)$ vs ln(a)}
\end{subfigure}
\begin{subfigure}{.55\textwidth}
  %
  \includegraphics[width=.82\linewidth]{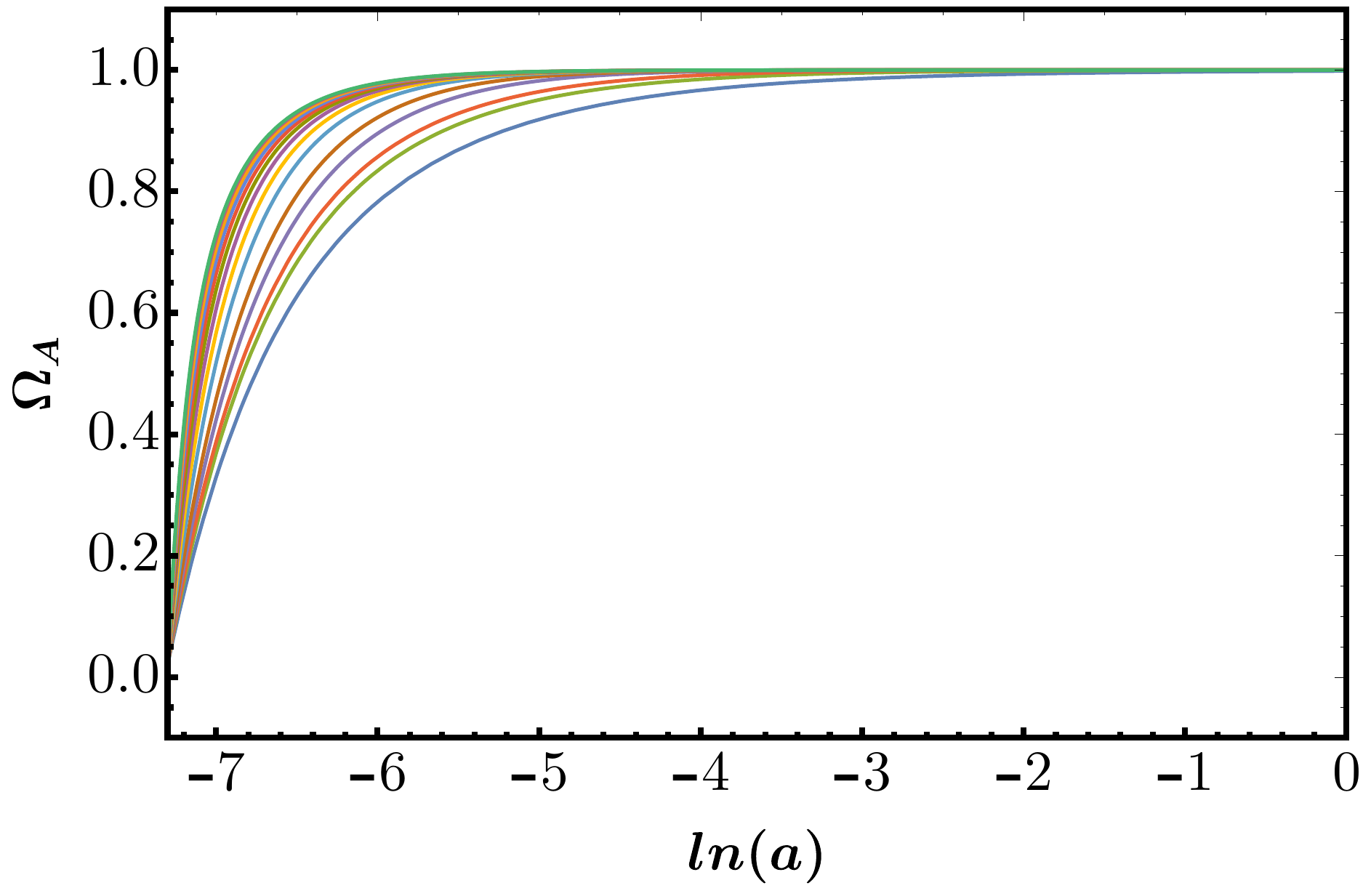}  
  \caption{$\Omega_{A}(a)$ vs ln(a)}
\end{subfigure}
\caption{Plot of $\xi(a), \Omega_r(a), \Omega_m(a), \Omega_A(a)$ as a function of $\ln(a)$ for $f(R, A) = Re^{\alpha (RA)^2}$ and $\alpha = 0.0016$ . 
}
\label{fig:(ra)2,1}
\end{figure}

Fig. \ref{fig:(ra)2,2} contains the plots of $n = 2$ (in Eq. \eqref{action:class1b}) for two values of $\alpha$. Plots on the left are for $\alpha =  -0.015$ and the plots are the right are for $\alpha = 0.1$.  Different colors in the plots correspond to different initial values of $\xi$ in the range $[-5, 0.6]$. From these plots, we infer the following:
\begin{enumerate}
\item For $\alpha = 0.1$ and initial values of $\xi > -0.9$, we get a de Sitter attractor ($\xi = 0$). Corresponding to these initial values, $\Omega_{r}$ and $\Omega_{m}$ starts at $0.39$ and $0.59$ respectively, and quickly converge to $0$; while $\Omega_{A}$ starts at 0.02 and quickly converges to $1$.
\item For $\alpha = 0.1$ and initial values of $\xi$ in the range $[-2, -1.5]$, 
$\xi$ diverges. 
\item For $\alpha = 0.1$ and other initial values of $\xi$ in the range $[-1.5, -1.1]$, $\xi$ is almost a constant, and $\Omega_{r}$ and $\Omega_{m}$ does not quickly converge. Importantly, for these initial values, $\Omega_{m}$ always approaches a non-zero value. For these initial values of $\xi$, $\Omega_{A}$ starts at $0.02$ and does not quickly converge and always approaches a value less than $1$.
\item For $\alpha = 0.1$ and initial values of $\xi  <  -2$, we observe something akin to convergence, but $\xi$ doesn't converge within the physical range of 
$\ln(a)$.  For these initial values, $\Omega_{r}$ and $\Omega_{m}$ start at $0.39$ and $0.59$, respectively, and evolve to very large values. Hence, these initial values are not physically relevant. 
\item For $\alpha = - 0.015$ and initial values of $\xi > -0.6$, we get a nearly de Sitter attractor ($\xi \simeq 0.2$). Corresponding to these initial values, $\Omega_{r}$ and $\Omega_{m}$ starts at $0.39$ and $0.59$ respectively, and quickly converge to $0$; while $\Omega_{A}$ starts at $0.02$ and quickly converges to $1$.
\item For $\alpha = -0.015$ and initial values of $\xi$ in the range $[-2, -1.6]$,  ($\xi, \Omega_m, \Omega_r)$ diverge. 
\end{enumerate}
\begin{figure}[H]
\begin{subfigure}{.55\textwidth}
  %
  \includegraphics[width=.8\linewidth]{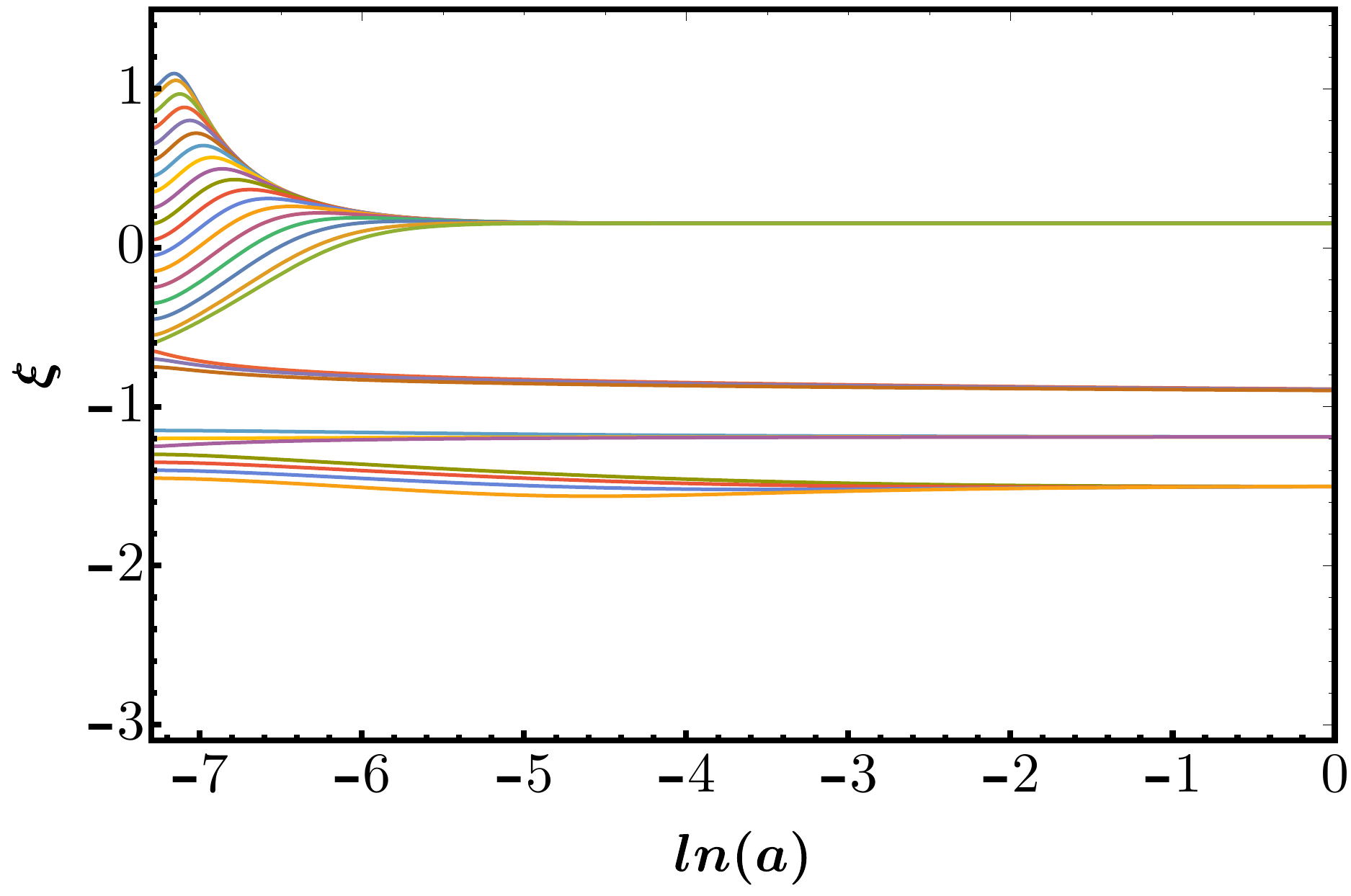}  
  \caption{$\xi(a)$ vs ln(a),$\alpha = -0.015$}
\end{subfigure}
\begin{subfigure}{.55\textwidth}
  %
  \includegraphics[width=.8\linewidth]{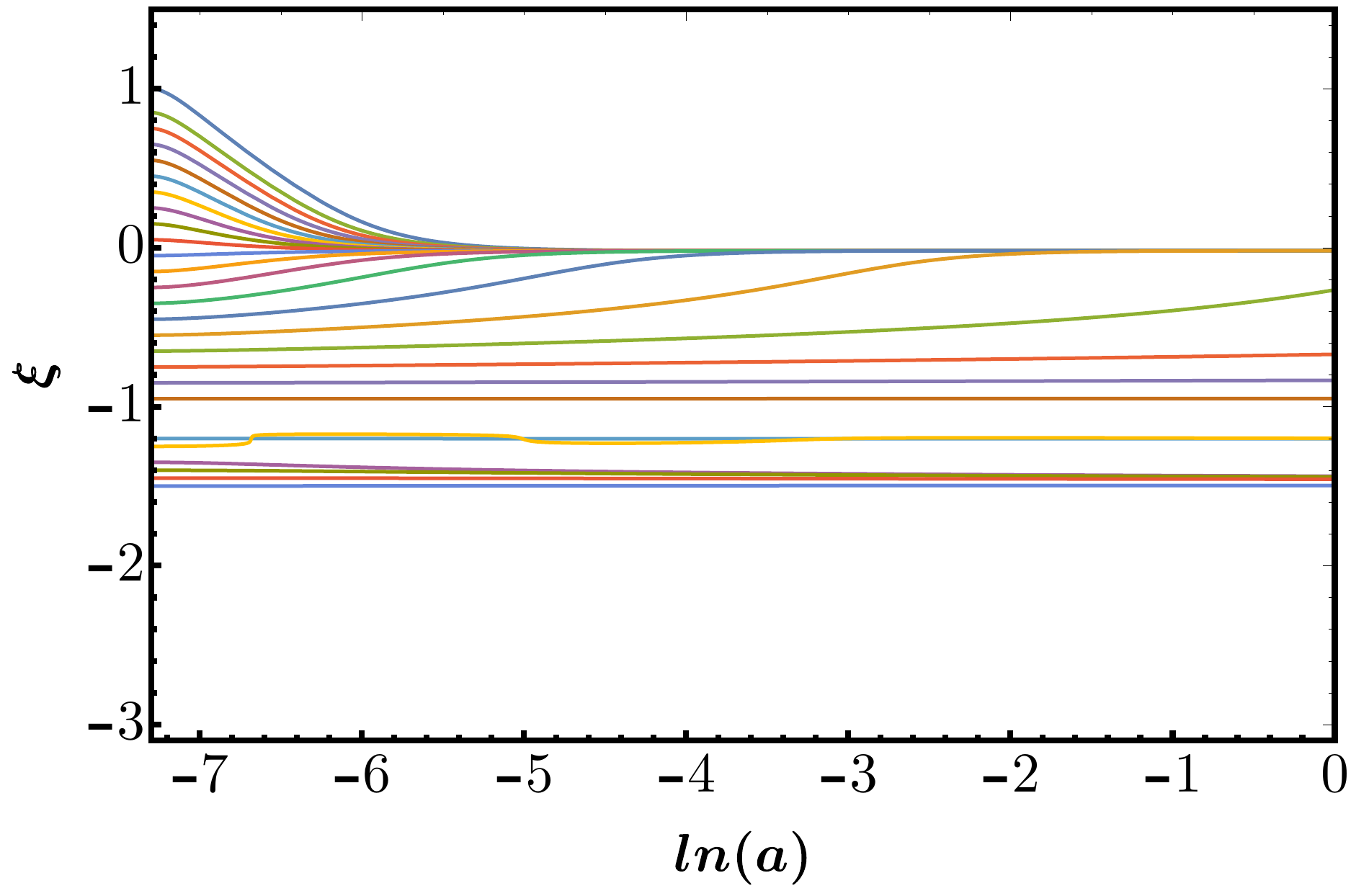}  
  \caption{$\xi(a)$ vs ln(a),$\alpha = 0.1$}
\end{subfigure}
\newline
\begin{subfigure}{.55\textwidth}
  %
  \includegraphics[width=.82\linewidth]{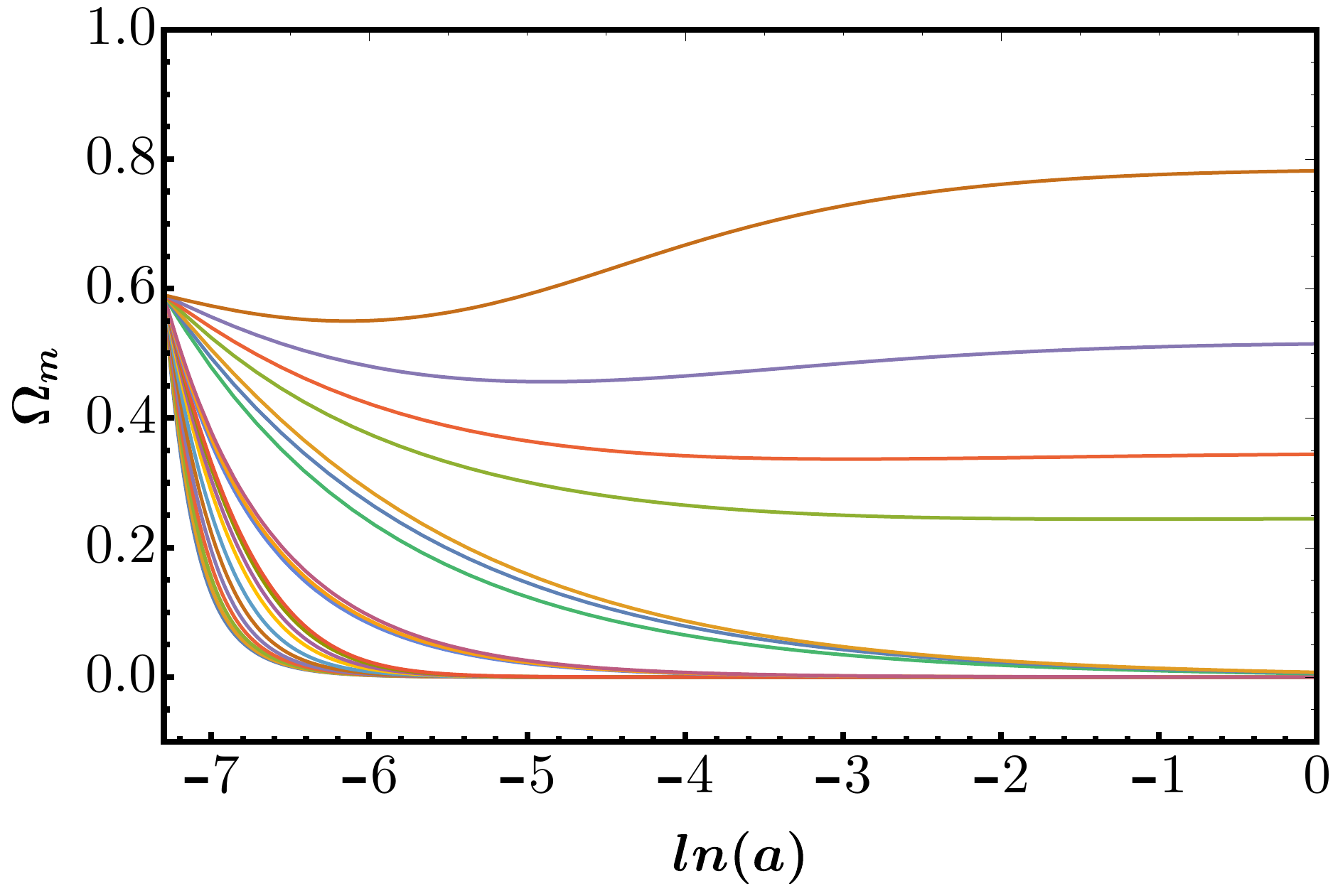}  
  \caption{$\Omega_{m}(a)$ vs ln(a),$\alpha = -0.015$}
\end{subfigure}
\begin{subfigure}{.55\textwidth}
  %
  \includegraphics[width=.82\linewidth]{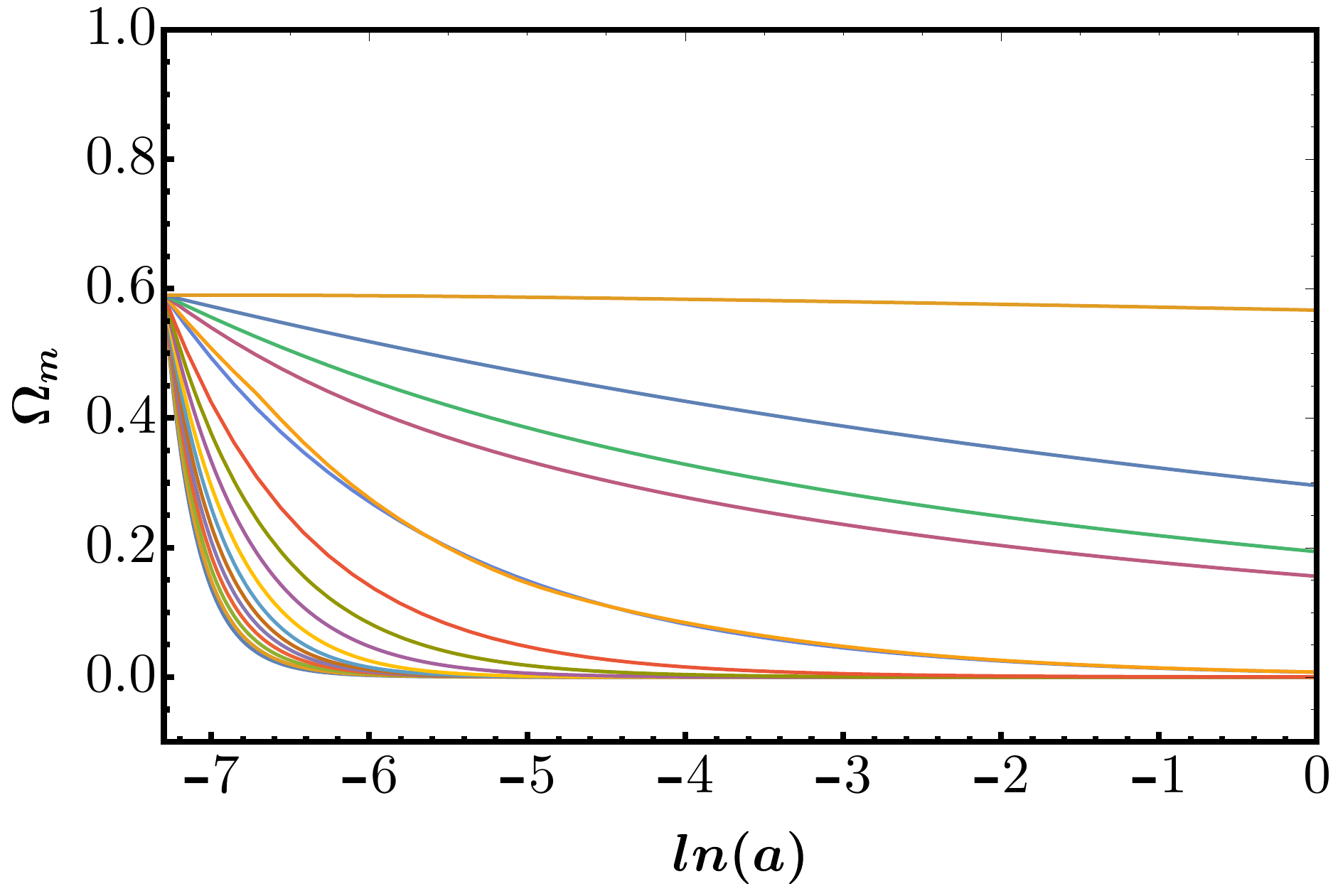}  
  \caption{$\Omega_{m}(a)$ vs ln(a),$\alpha = 0.1$}
\end{subfigure}
\newline
\begin{subfigure}{.55\textwidth}
  %
  \includegraphics[width=.82\linewidth]{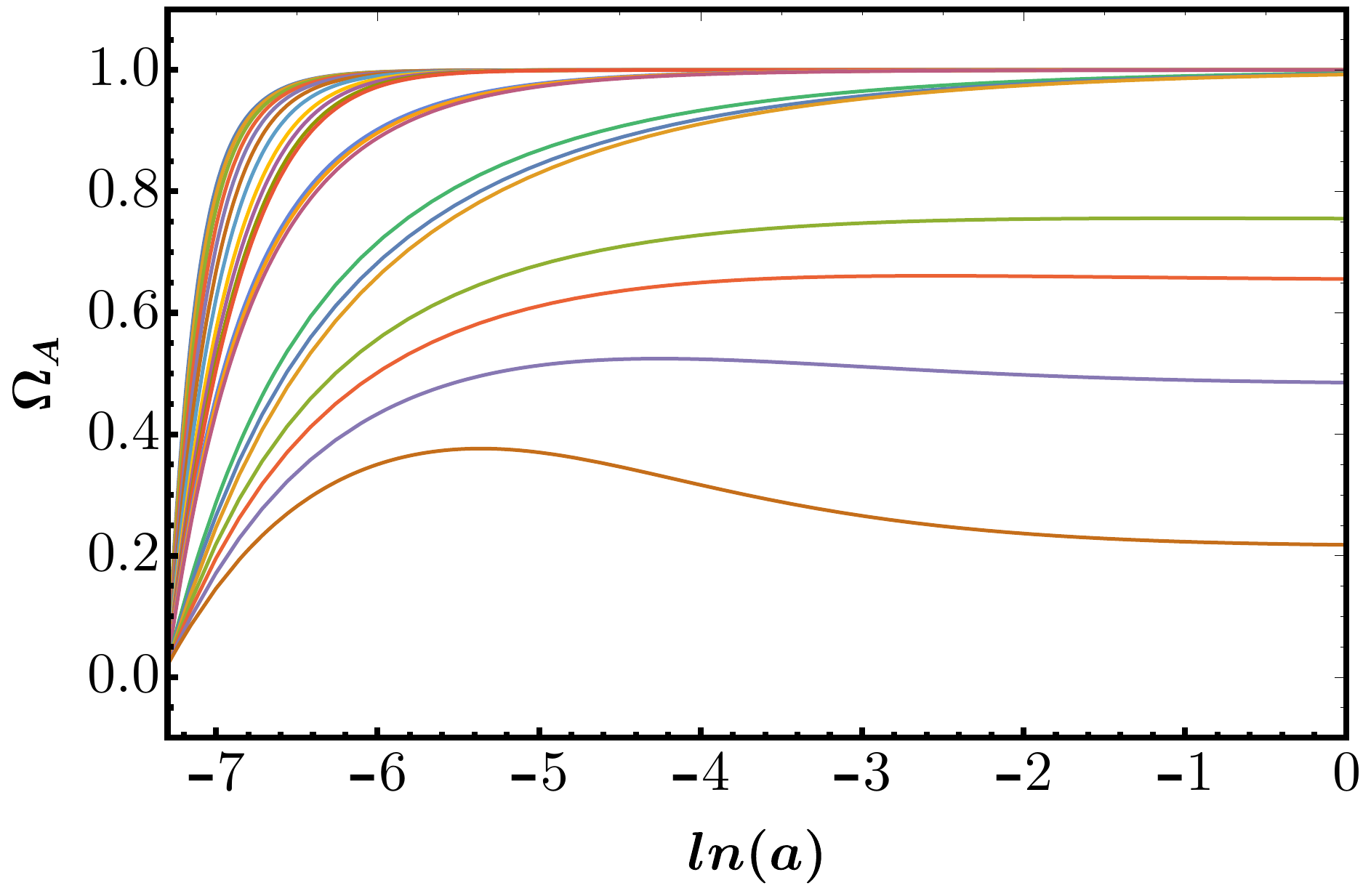}  
  \caption{$\Omega_{A}(a)$ vs ln(a),$\alpha = -0.015$}
\end{subfigure}
\begin{subfigure}{.55\textwidth}
  %
  \includegraphics[width=.82\linewidth]{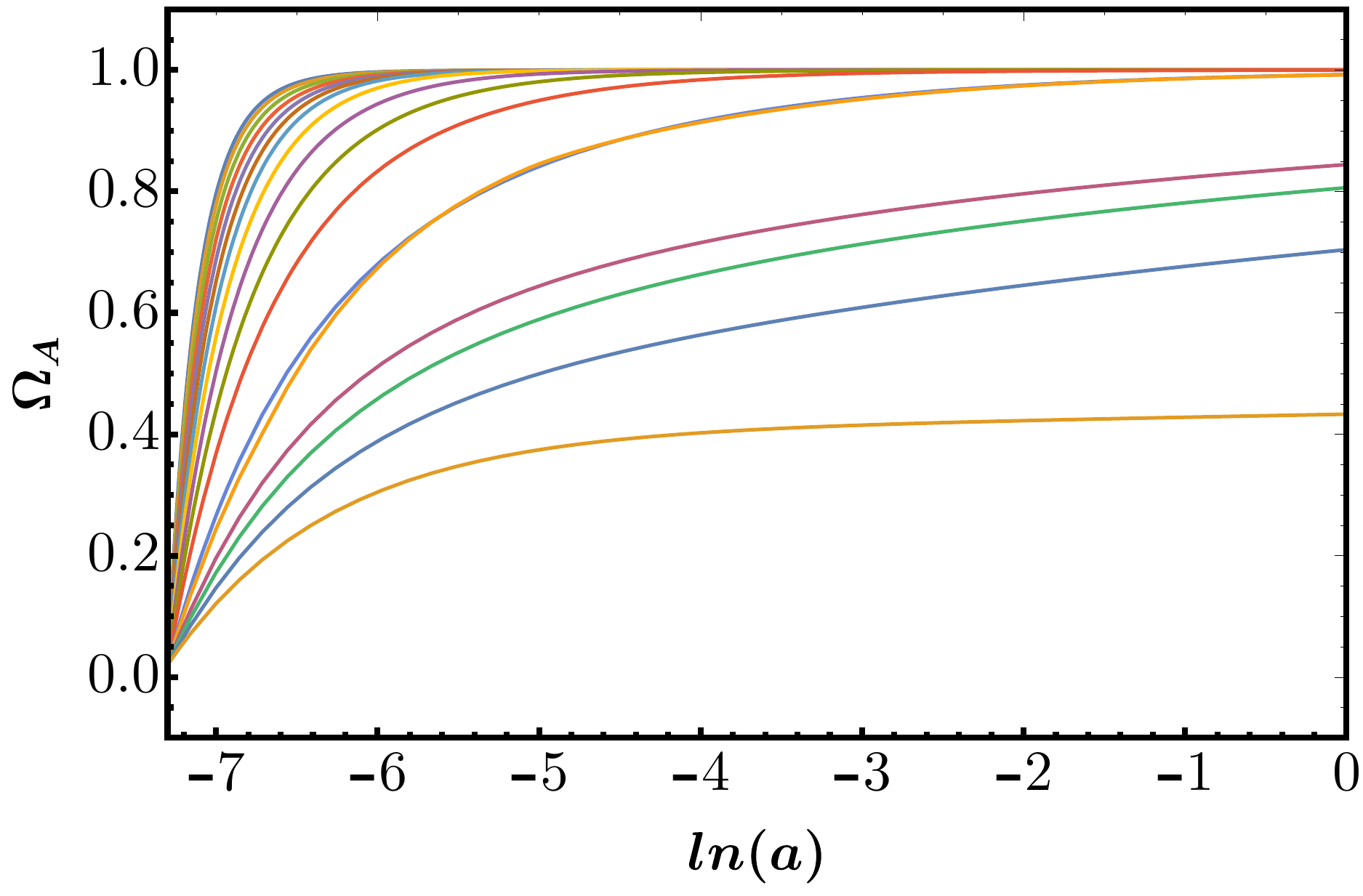}  
  \caption{$\Omega_{A}(a)$ vs ln(a),$\alpha = 0.1$}
\end{subfigure}
\caption{Plot of $\xi(a), \Omega_m(a), \Omega_A(a)$ as a function of $\ln(a)$ for $f(R, A) = Re^{\alpha (RA)^2}$. Plots on the left are for $\alpha =  -0.015$ and the plots are the right are for $\alpha = 0.1$.}
\label{fig:(ra)2,2}
\end{figure}
Fig. \ref{fig:(ra)-1} contains the plots of $n = -1$ (in Eq. \eqref{action:class1b}) for two values of $\alpha$. Plots on the left are for $\alpha =  -15$ and the plots are the right are for $\alpha = 700$.  Different colors in the plots refer to different initial values of $\xi$ in the range $[-5, 0.6]$. From these plots, we infer the following:

\begin{enumerate}
\item For $\alpha = -15$ and initial values of $\xi > -1.2$, we get $\xi = -0.5$ as an attractor ($w_{\rm eff} \simeq -0.67$ corresponding to the current Universe). For these initial values, $\Omega_{r}$ and $\Omega_{m}$ start at $0.39$ and $0.59$ respectively, and quickly converge to $0$; while $\Omega_{A}$ starts at 0.02 and quickly converges to $1$.

\item For $\alpha = -15$ and initial values of $\xi$ in the range $[-2, -1.2]$, 
($\xi, \Omega_m, \Omega_r)$ diverge. 

\item For $\alpha = -15$ and initial values of $\xi < - 2$, we get another attractor at $\xi = -2$ (radiation dominated epoch).  For these initial values, $\Omega_{r}$ and $\Omega_{m}$ starts at $0.39$ and $0.59$, respectively and diverge. Hence, these initial values are not physically relevant.

\item For $\alpha = 700$ and initial values of $\xi$ in the range $[-0.4, 0.4]$, we get a de Sitter attractor ($\xi = 0$). Corresponding to these initial values, $\Omega_{r}$ and $\Omega_{m}$ starts at $0.39$ and $0.59$ respectively, and quickly converge to $0$; while $\Omega_{A}$ starts at 0.02 and quickly converges to $1$.

\item For $\alpha = 700$ and initial values of $\xi$ in the range 
$[-1.2, -1]$, ($\xi, \Omega_m, \Omega_r)$ diverge. 

\item For $\alpha = 700$ and initial value of $\xi = -1.2$, $\xi$ is almost a constant. For the initial values in the range $[-1,-0.5]$, $\Omega_{r}$ and $\Omega_{m}$ starts at 0.39 and 0.59 respectively and converges to 0; while, $\Omega_{A}$ starts at 0.02 and converges to 1. 
\end{enumerate} 

\begin{figure}[H]
\begin{subfigure}{.55\textwidth}
  %
  \includegraphics[width=.82\linewidth]{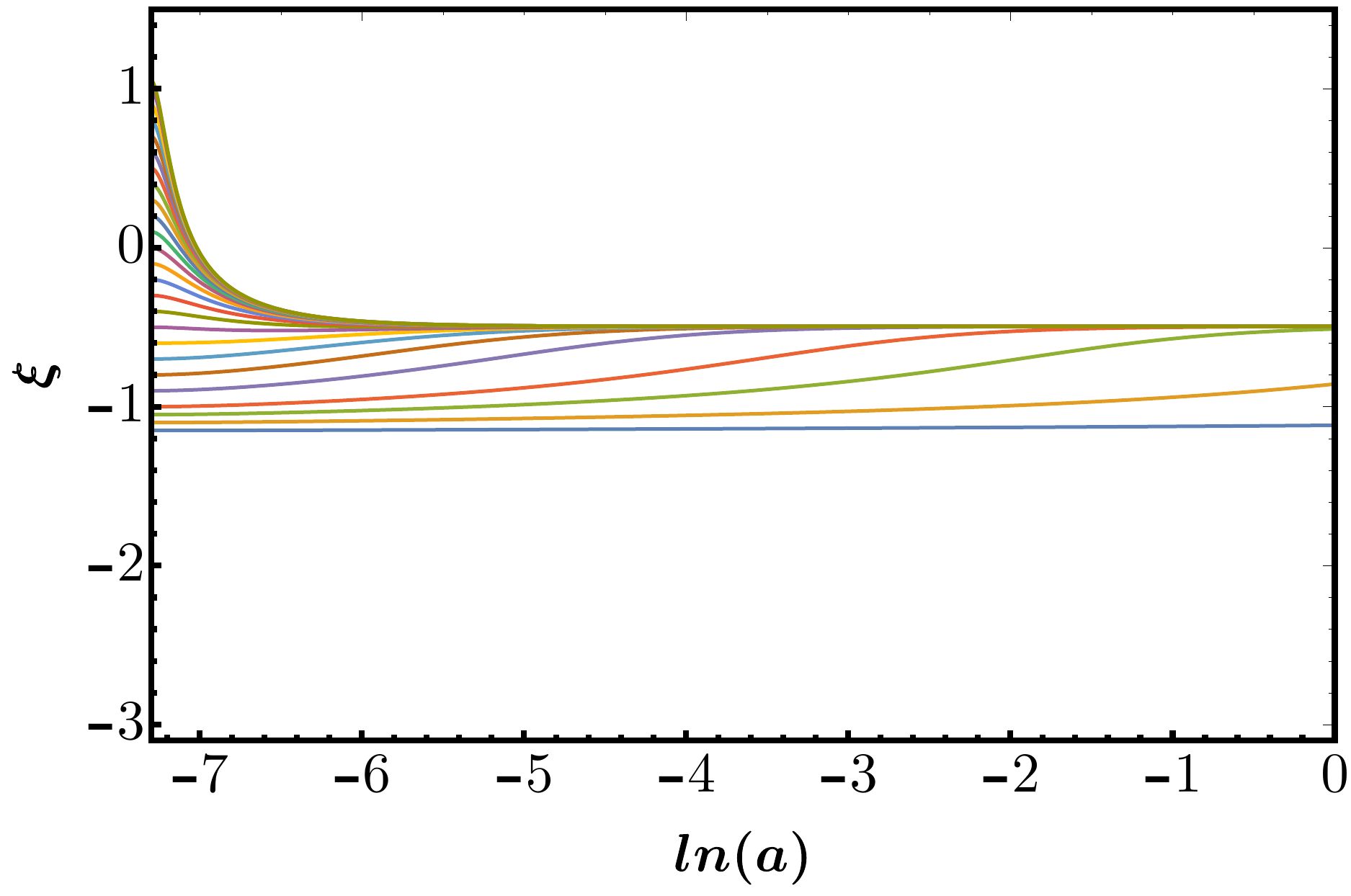}  
  \caption{$\xi(a)$ vs ln(a),$\alpha = -15$}
\end{subfigure}
\begin{subfigure}{.55\textwidth}
  %
  \includegraphics[width=.82\linewidth]{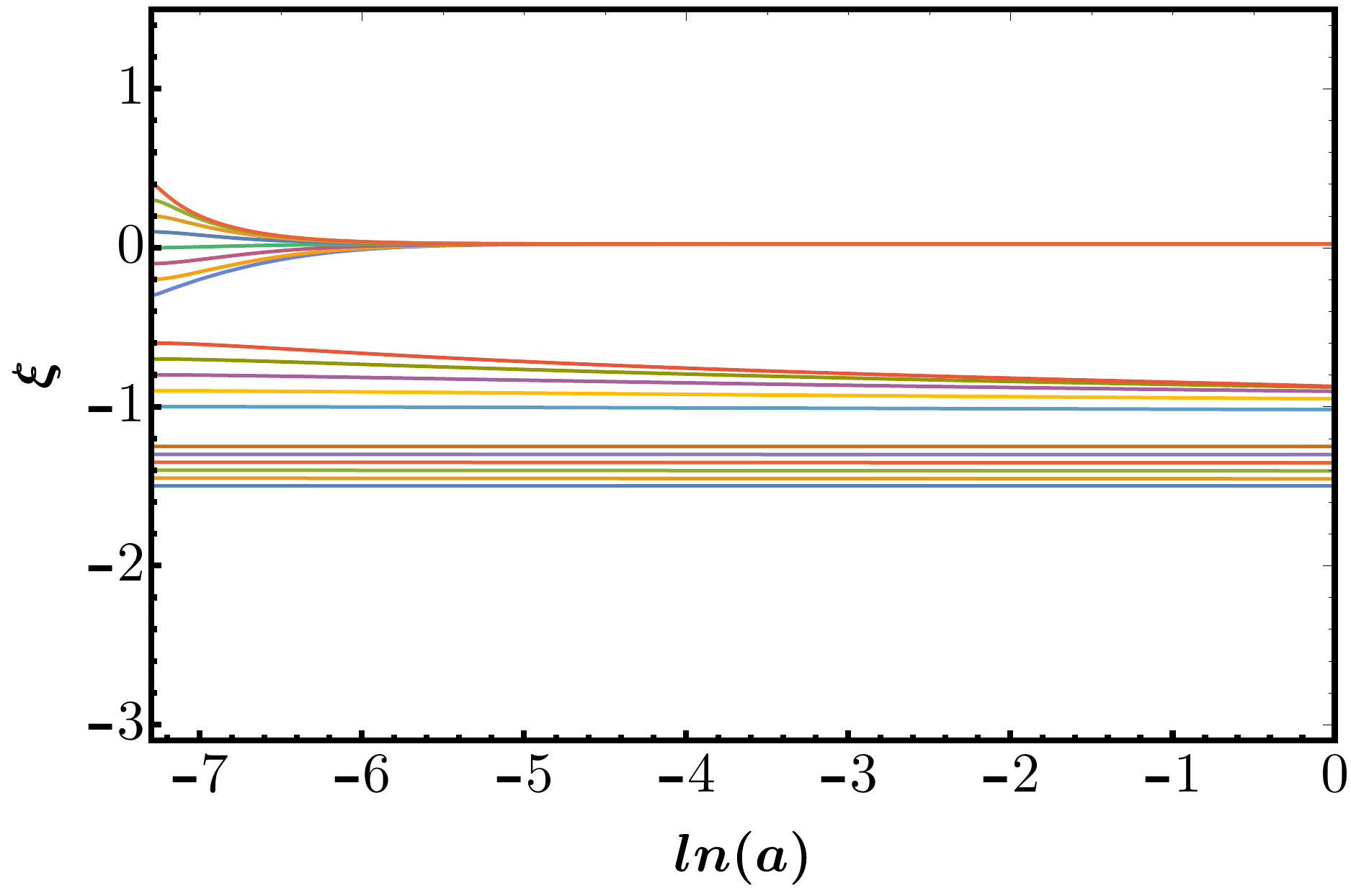}  
  \caption{$\xi(a)$ vs ln(a),$\alpha = 700$}
\end{subfigure}
\begin{subfigure}{.55\textwidth}
  %
  \includegraphics[width=.82\linewidth]{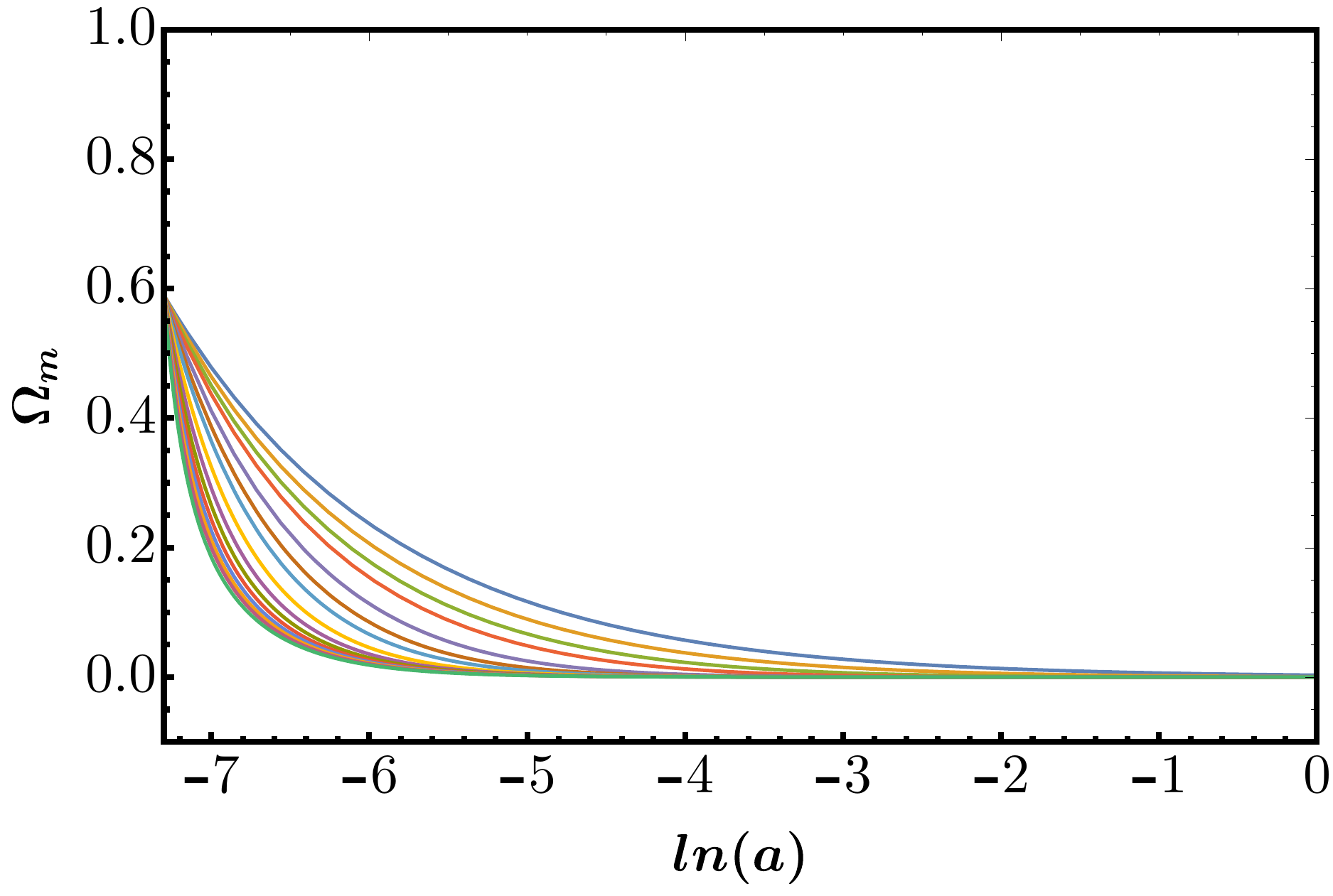}  
  \caption{$\Omega_{m}(a)$ vs ln(a),$\alpha = -15$}
\end{subfigure}
\begin{subfigure}{.55\textwidth}
  %
  \includegraphics[width=.82\linewidth]{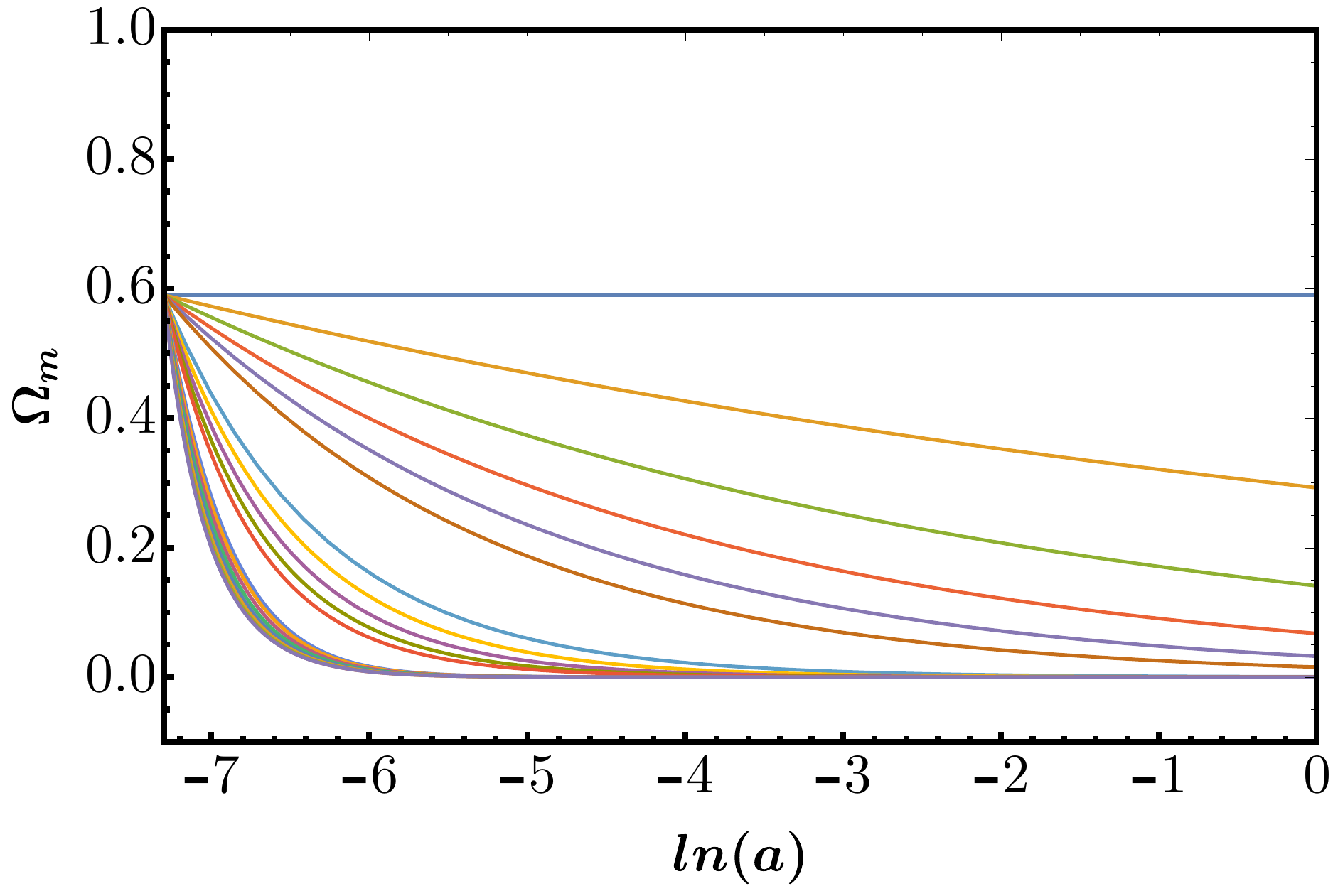}  
  \caption{$\Omega_{m}(a)$ vs ln(a),$\alpha = 700$}
\end{subfigure}
\end{figure}
\begin{figure}[H]\ContinuedFloat
\begin{subfigure}{.55\textwidth}
  %
  \includegraphics[width=.82\linewidth]{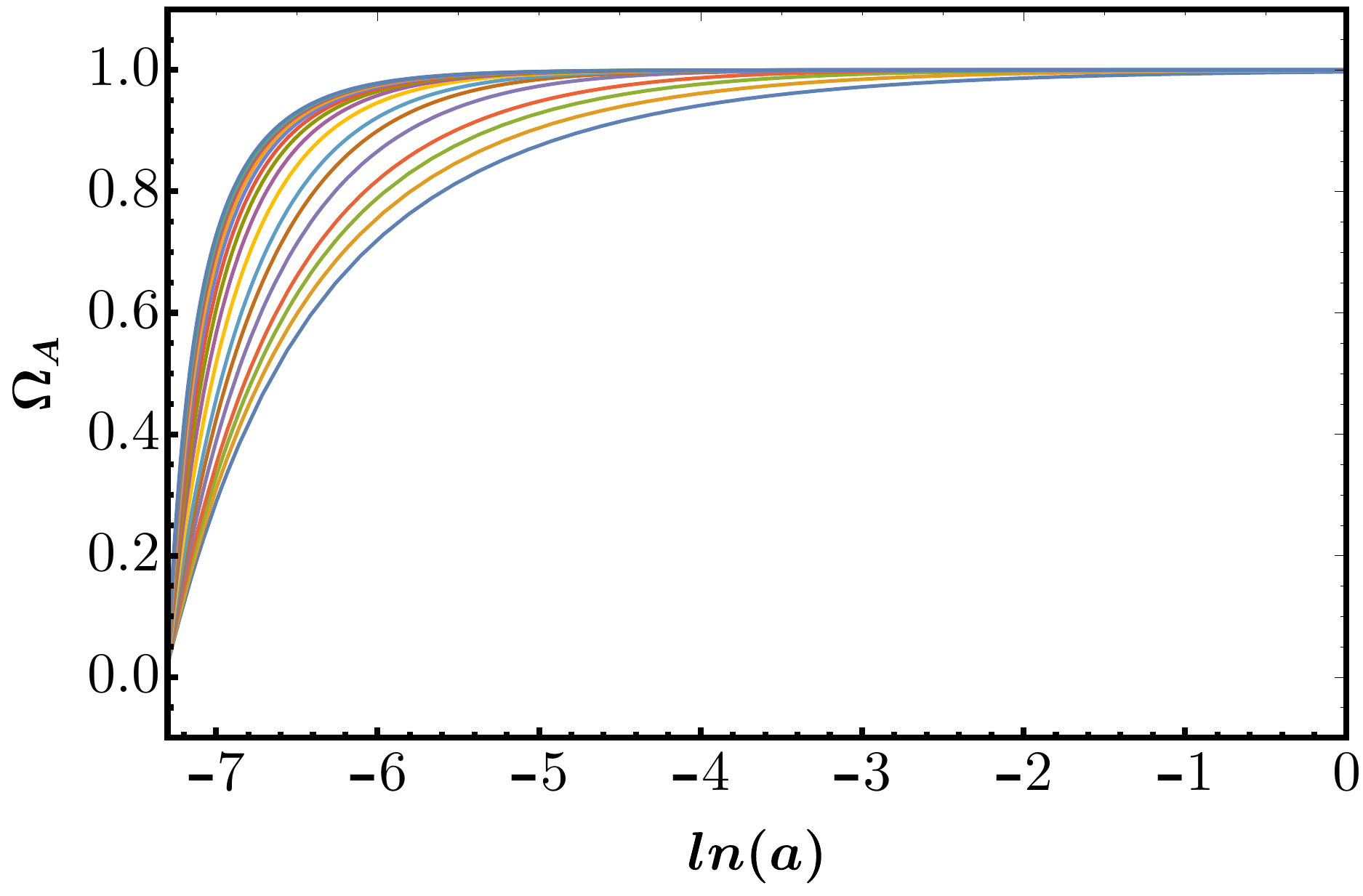}  
  \caption{$\Omega_{A}(a)$ vs ln(a),$\alpha = -15$}
\end{subfigure}
\begin{subfigure}{.55\textwidth}
  %
  \includegraphics[width=.82\linewidth]{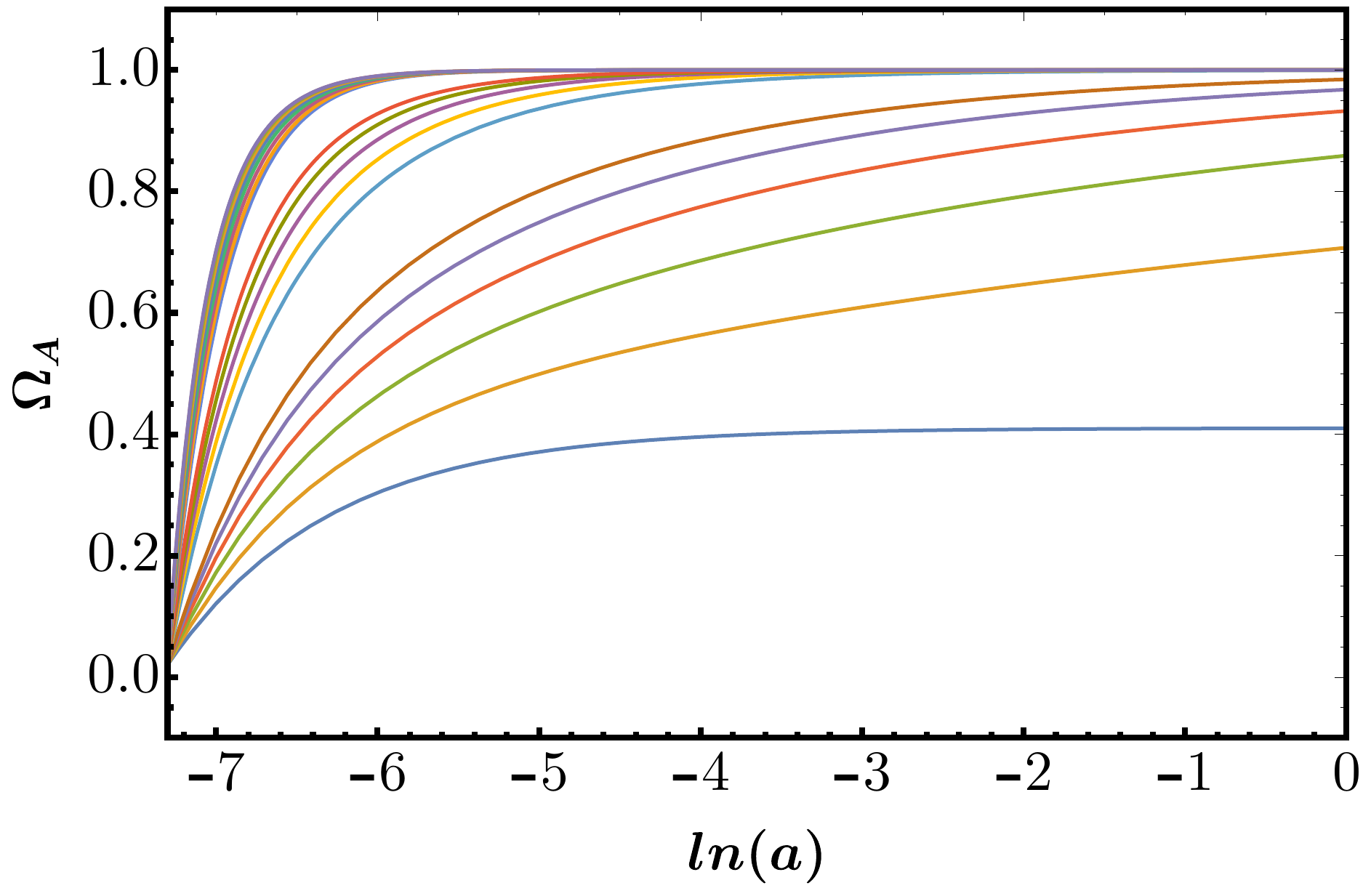}  
  \caption{$\Omega_{A}(a)$ vs ln(a),$\alpha = 700$}
\end{subfigure}
\caption{Plot of $\xi(a), \Omega_m(a), \Omega_A(a)$ as a function of $\ln(a)$ for $f(R, A) = Re^{\alpha (RA)^{-1}}$. Plots on the left are for $\alpha =  -15$ and the plots are the right are for $\alpha = 700$.}
\label{fig:(ra)-1}
\end{figure}

Fig. \ref{fig:(ra)-2,1} contains the plots of $n = -2$ (in Eq. \eqref{action:class1b}) for two values of $\alpha$. Plots on the left are for $\alpha =  -1100$ and the plots are the right are for $\alpha = -120$.  Different colors in the plots refer to different initial values of $\xi$ in the range $[-5, 0.6]$. From these plots, we infer the following:
\begin{enumerate}
\item For $\alpha = -1100$ and initial value of $\xi > -0.9$, we get  $\xi = -0.02$ (pure dust) as a attractor. For initial value $\xi = -1$, $\xi$ is a 
constant. For any other initial value of $\xi < -1$ (like $\xi = -1.1$), 
$\xi$ diverges.

\item For $\alpha = -120$ and initial value of $\xi > -0.9$, we get  $\xi = -0.5$ (corresponding to the current Universe) as a attractor. For initial values of 
$\xi = -1$ and $-1.1$, $\xi$ is  a constant.  For any other initial value of 
$\xi < -1.2$ (like $\xi = -1.2$), $\xi$  diverges.

\item For both values of $\alpha$ and initial values of $\xi > -1$, $\Omega_{r}$ and $\Omega_{m}$ starts at $0.39$ and $0.59$ respectively, and converge to 0; while, $\Omega_{A}$ starts at $0.02$ and converges to $1$.
\end{enumerate}


\begin{figure}[H]
\begin{subfigure}{.55\textwidth}
  %
  \includegraphics[width=.82\linewidth]{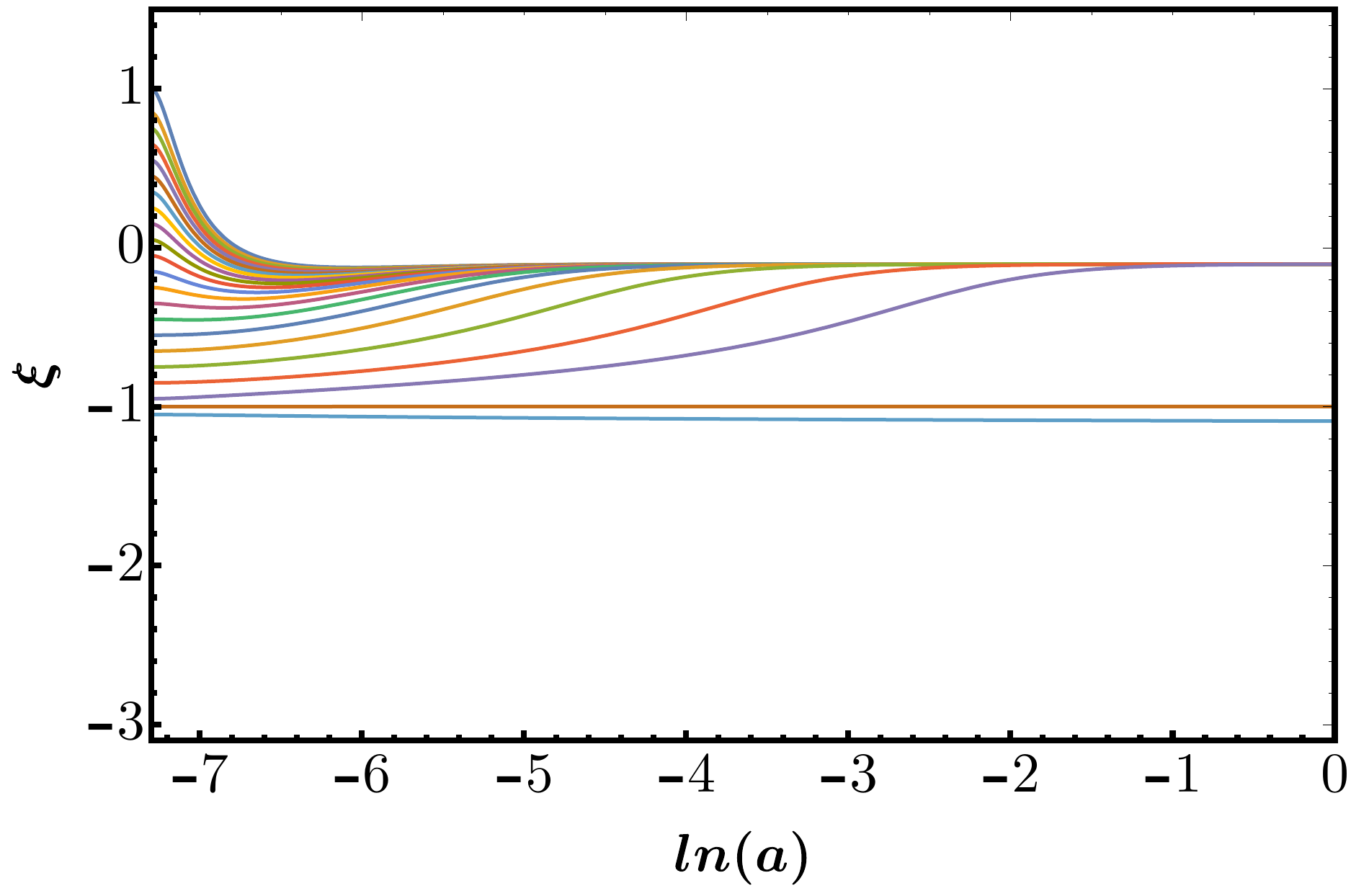}  
  \caption{$\xi(a)$ vs ln(a),$\alpha = -1100$}
\end{subfigure}
\begin{subfigure}{.55\textwidth}
  %
  \includegraphics[width=.82\linewidth]{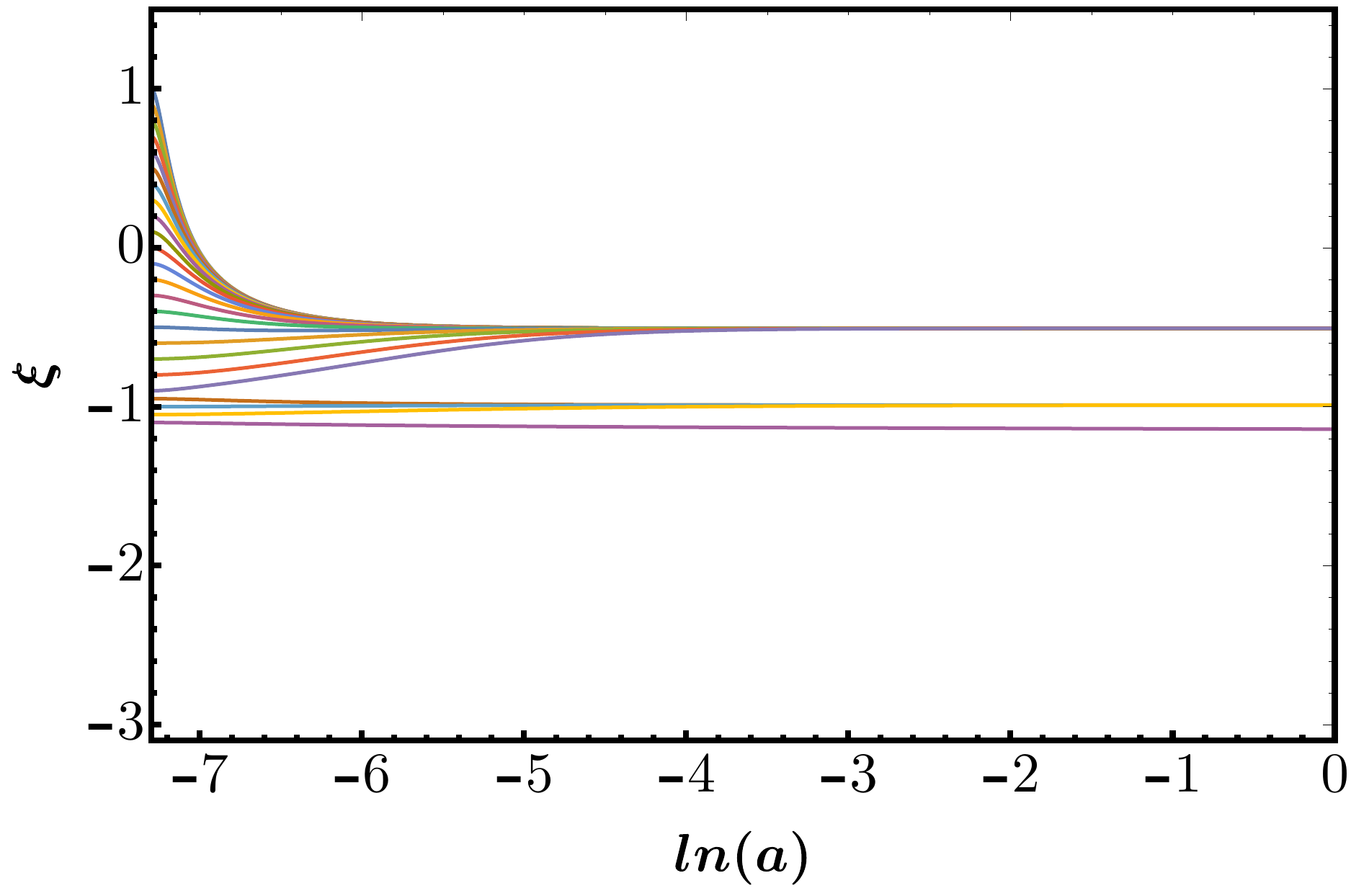}  
  \caption{$\xi(a)$ vs ln(a),$\alpha = -120$}
\end{subfigure}
\end{figure}
\begin{figure}[H]\ContinuedFloat
\begin{subfigure}{.55\textwidth}
  %
  \includegraphics[width=.82\linewidth]{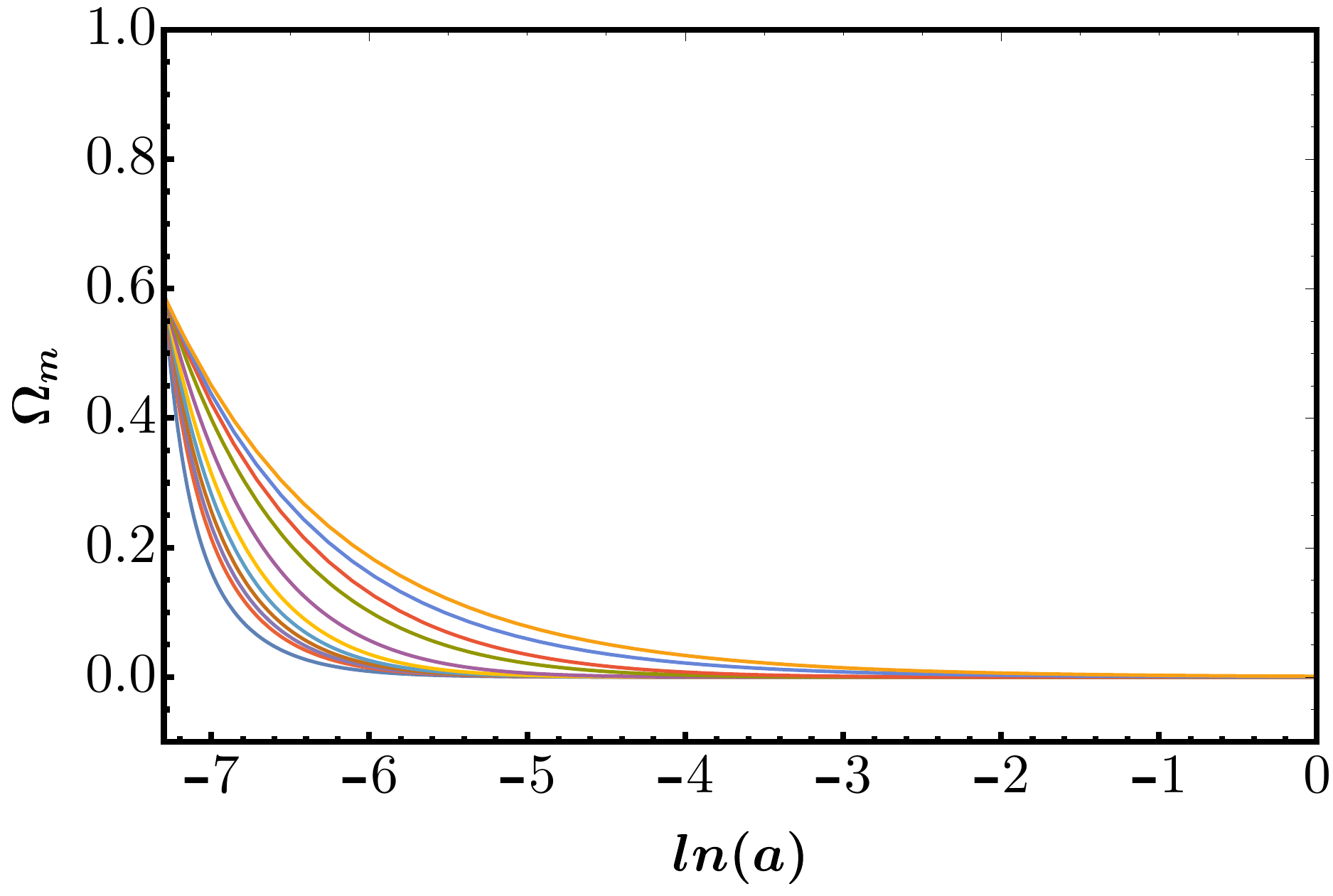}  
  \caption{$\Omega_{m}(a)$ vs ln(a),$\alpha = -1100$}
\end{subfigure}
\begin{subfigure}{.55\textwidth}
  %
  \includegraphics[width=.82\linewidth]{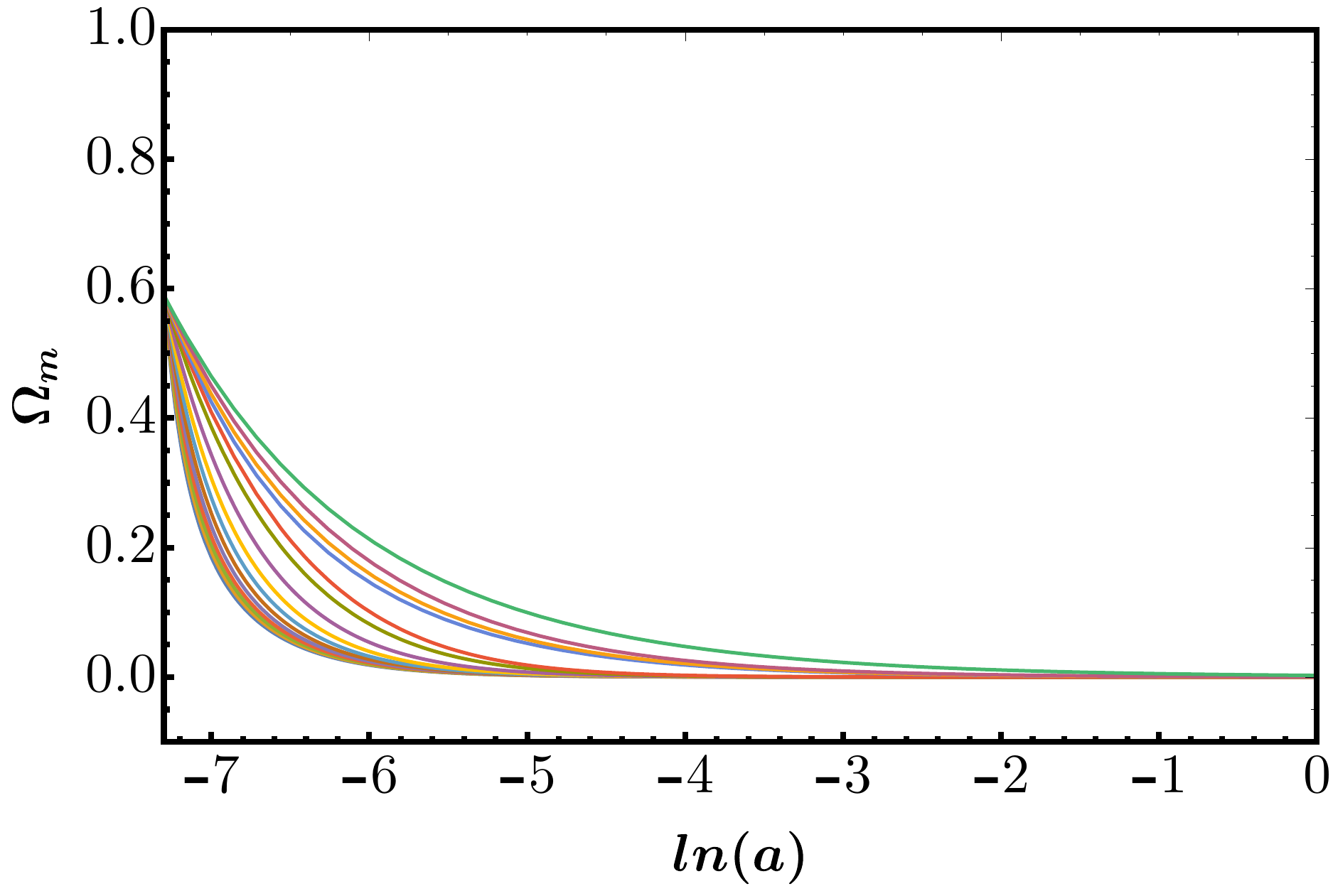}  
  \caption{$\Omega_{m}(a)$ vs ln(a),$\alpha = -120$}
\end{subfigure}
\newline
\begin{subfigure}{.55\textwidth}
  %
  \includegraphics[width=.82\linewidth]{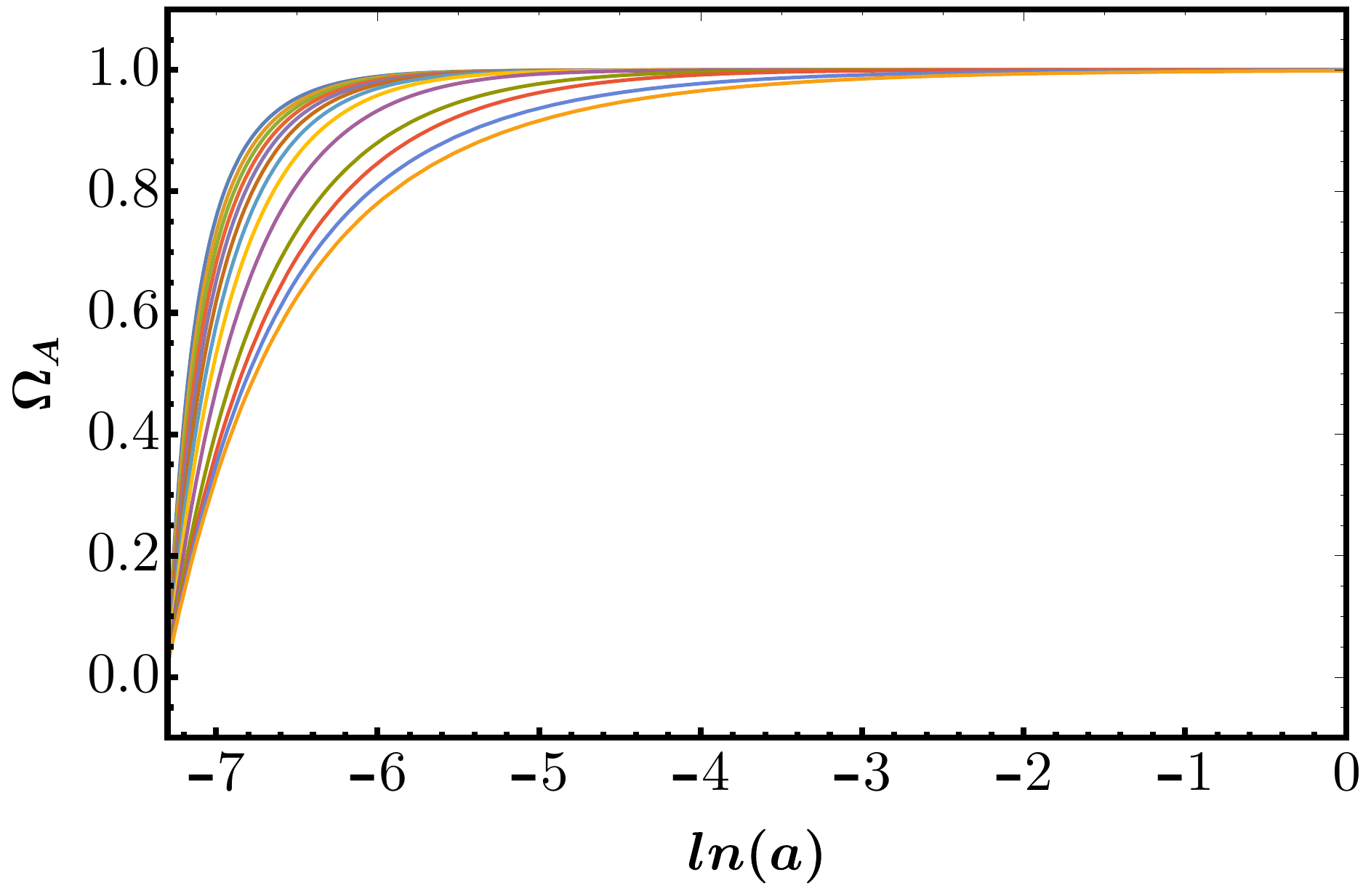}  
  \caption{$\Omega_{A}(a)$ vs ln(a),$\alpha = -1100$}
\end{subfigure}
\begin{subfigure}{.55\textwidth}
  %
  \includegraphics[width=.82\linewidth]{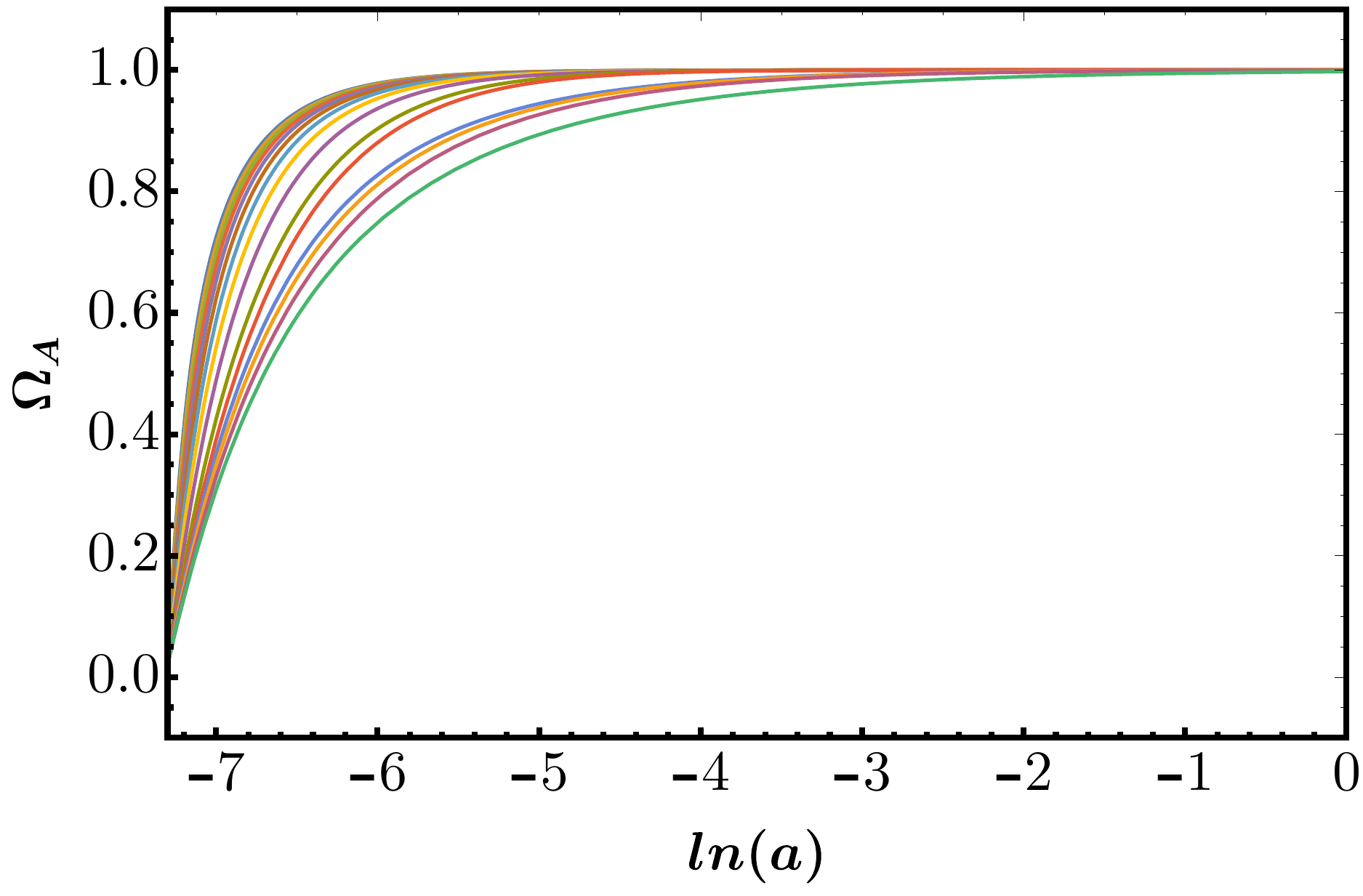}  
  \caption{$\Omega_{A}(a)$ vs ln(a),$\alpha = -120$}
\end{subfigure}
\caption{Plot of $\xi(a), \Omega_m(a), \Omega_A(a)$ as a function of $\ln(a)$ for $f(R, A) = Re^{\alpha (RA)^{-2}}$. Plots on the left are for $\alpha =  -1100$ and the plots are the right are for $\alpha = -120$.}
\label{fig:(ra)-2,1}
\end{figure}
%
\begin{figure}[H]
\begin{subfigure}{.55\textwidth}
  %
  \includegraphics[width=.82\linewidth]{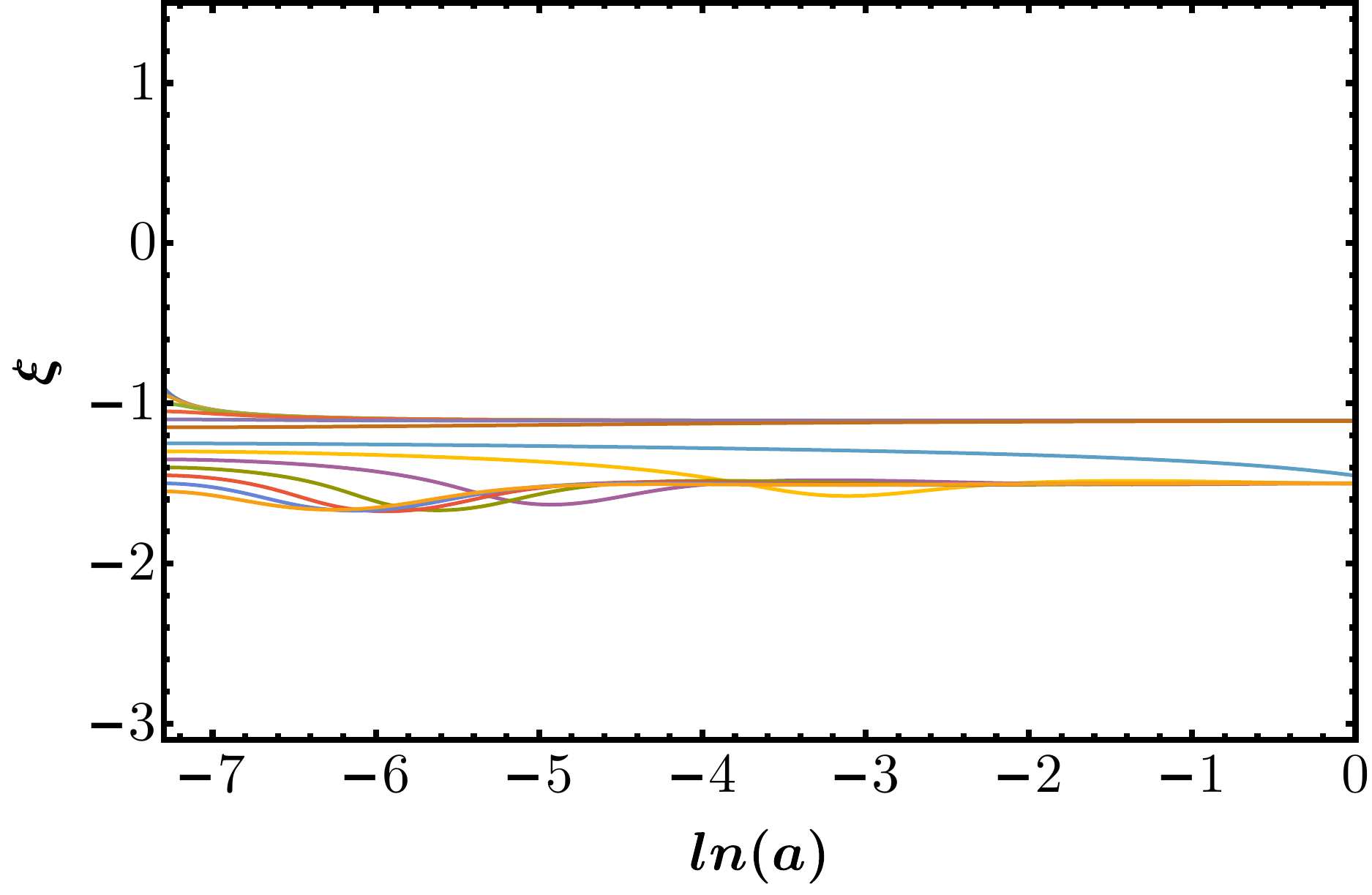}  
  \caption{$\xi(a)$ vs ln(a),$\alpha = 0.9$}
\end{subfigure}
\begin{subfigure}{.55\textwidth}
  %
  \includegraphics[width=.82\linewidth]{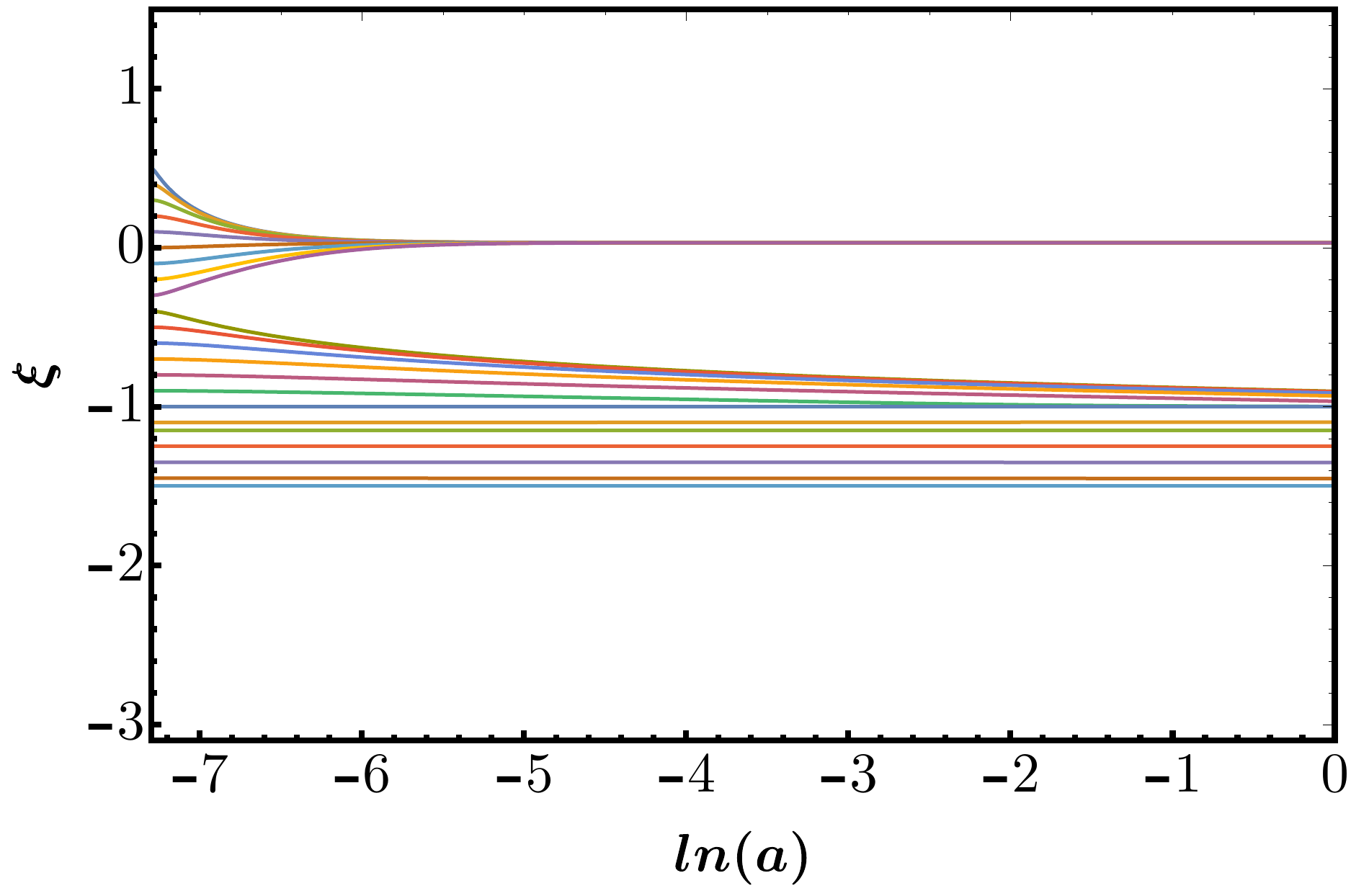}  
  \caption{$\xi(a)$ vs ln(a),$\alpha = 4250$}
\end{subfigure}
\end{figure}
\begin{figure}[H]\ContinuedFloat
\begin{subfigure}{.55\textwidth}
  %
  \includegraphics[width=.82\linewidth]{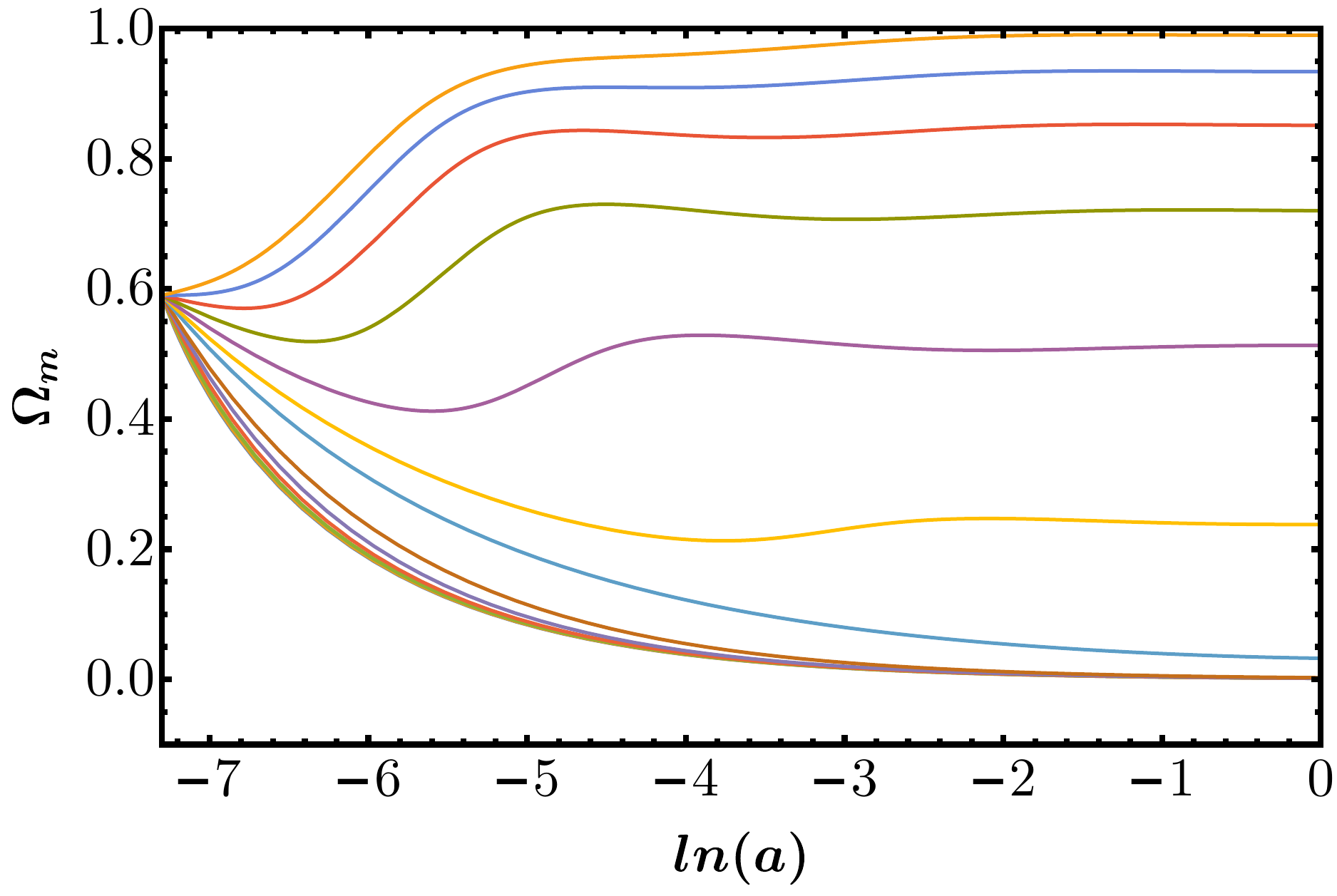}  
  \caption{$\Omega_{m}(a)$ vs ln(a),$\alpha = 0.9$}
\end{subfigure}
\begin{subfigure}{.55\textwidth}
  %
  \includegraphics[width=.82\linewidth]{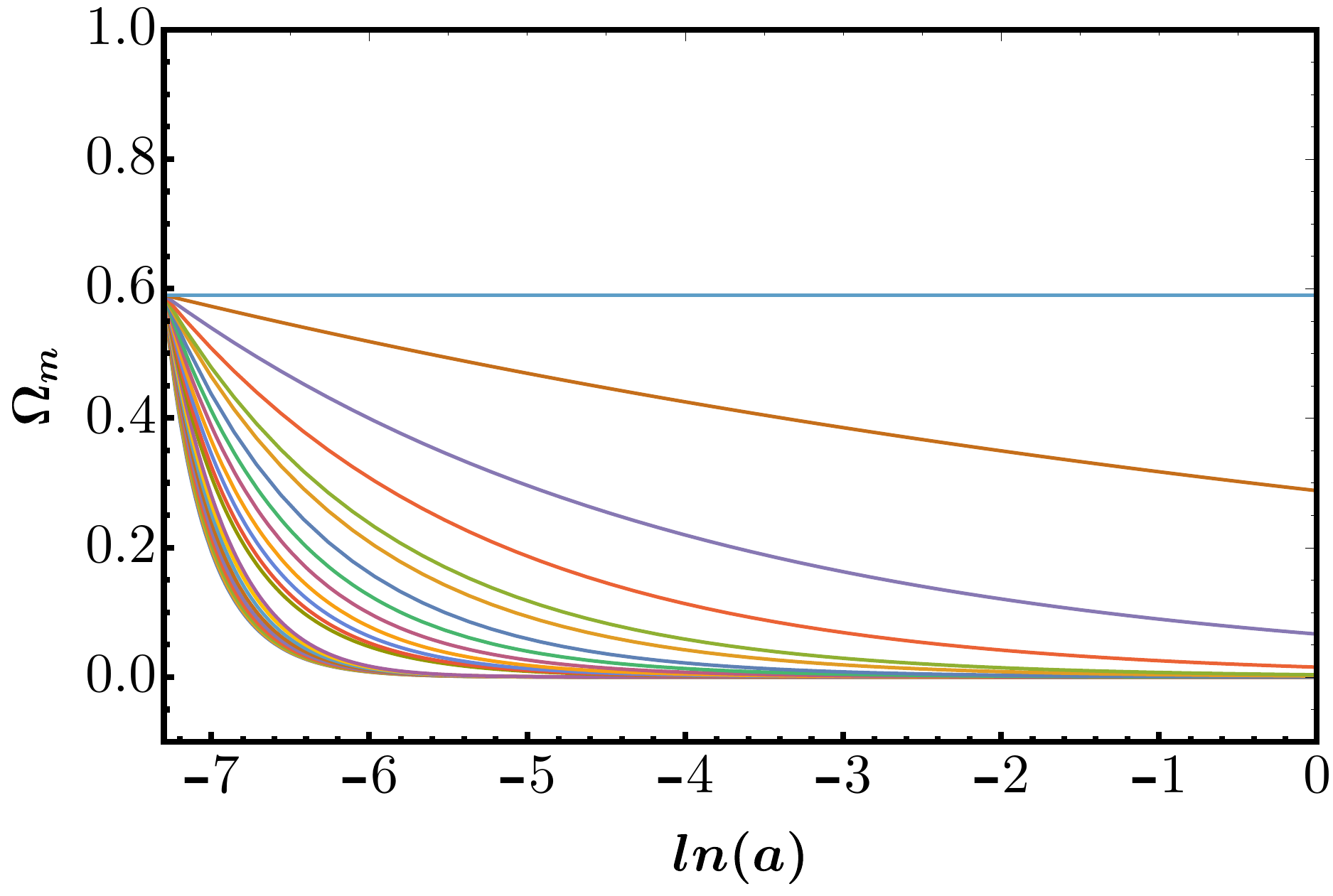}  
  \caption{$\Omega_{m}(a)$ vs ln(a),$\alpha = 4250$}
\end{subfigure}
\newline
\begin{subfigure}{.55\textwidth}
  %
  \includegraphics[width=.82\linewidth]{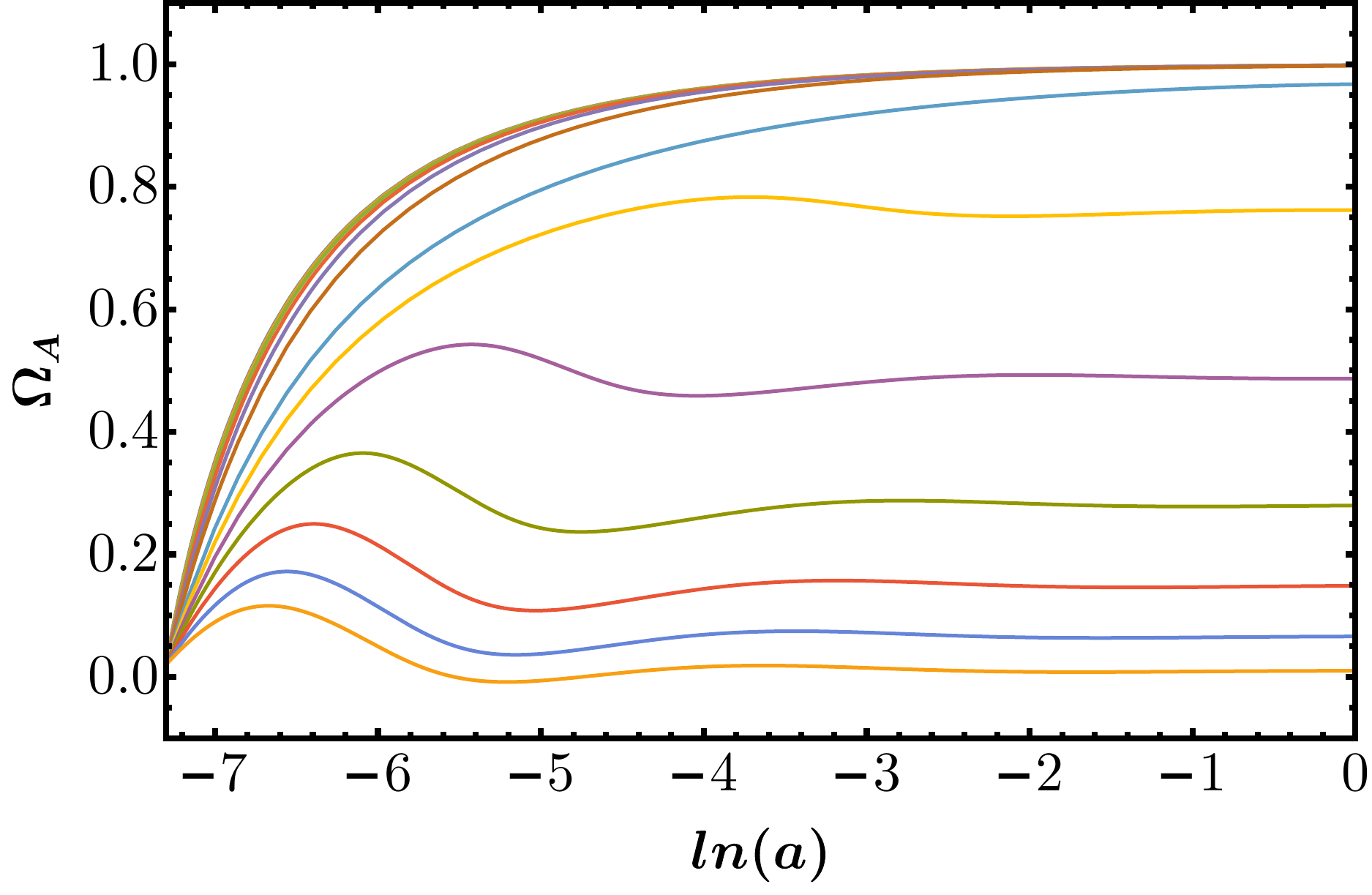}  
  \caption{$\Omega_{A}(a)$ vs ln(a),$\alpha = 0.9$}
\end{subfigure}
\begin{subfigure}{.55\textwidth}
  %
  \includegraphics[width=.82\linewidth]{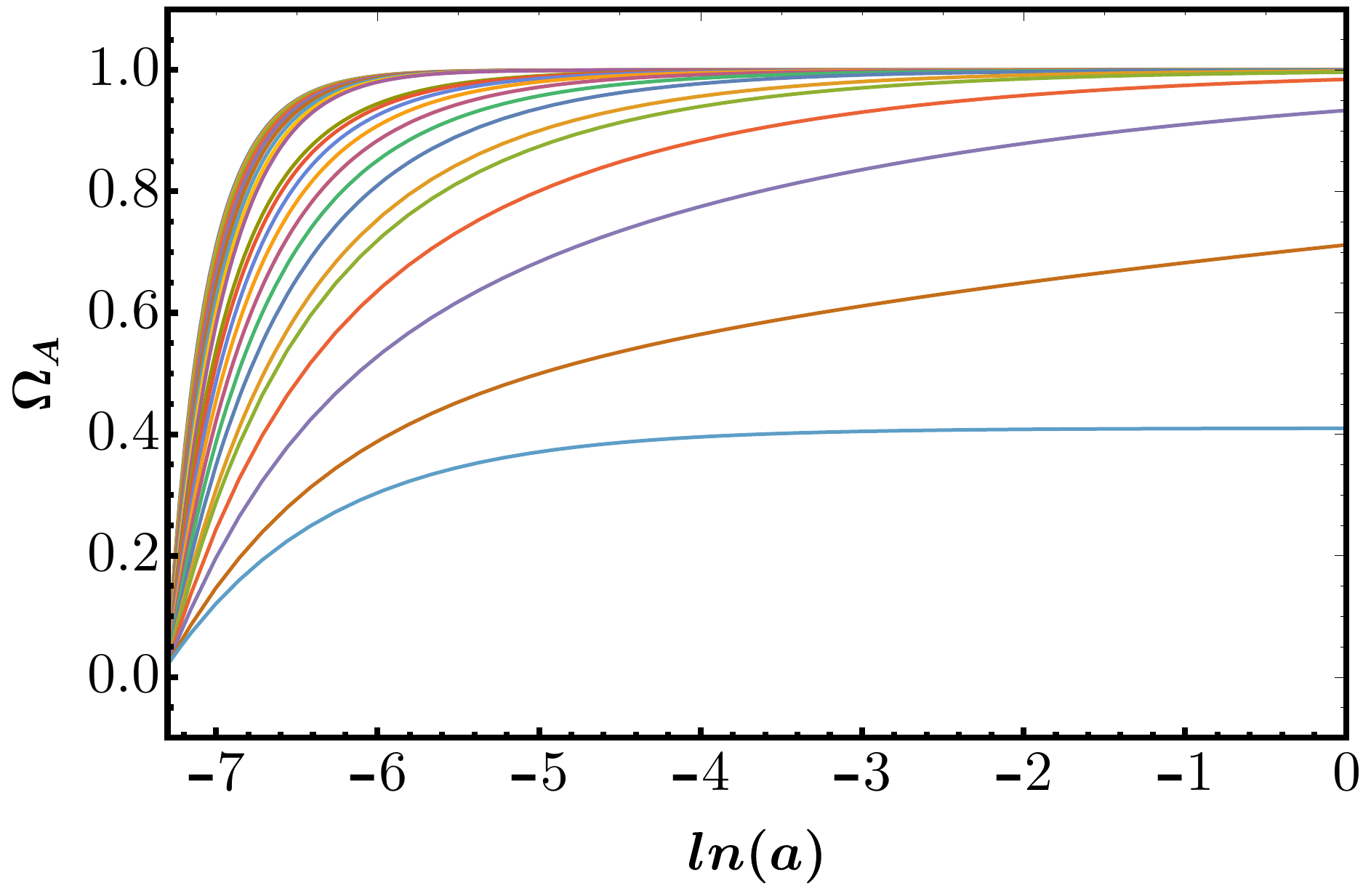}  
  \caption{$\Omega_{A}(a)$ vs ln(a),$\alpha = 4250$}
\end{subfigure}
\caption{Plot of $\xi(a), \Omega_m(a), \Omega_A(a)$ as a function of $\ln(a)$ for $f(R, A) = Re^{\alpha (RA)^{-2}}$. Plots on the left are for $\alpha =  0.9$ and the plots are the right are for $\alpha = 4250$.}
\label{fig:(ra)-2,2}
\end{figure}
Fig. \ref{fig:(ra)-2,2} contains the plots for two different values of $\alpha$. From the plots, it is clear that the features are very similar. From these plots, we notice a 
universal feature of class Ib models: There exists \emph{at least one} singularity point in the $\xi$ range $[-1.2, -1]$. Thus, a Universe from a matter-dominated epoch ($\xi = - 1.5$) to an accelerating Universe  ($\xi = - 0.5$) will not be possible as one will encounter at least one singularity point in the range $\xi$ range $[-1.2, -1]$. Thus, our analysis shows that class Ib models \emph{can not bypass} the no-go theorem.

\subsection{Class Ic models}\label{subsec:Class Ic models}

In this class, $f(R, A)$ is a combination of Class Ia and Class Ib models. In Ref. \cite{2020-Amendola.etal-Phys.Lett.B}, the authors conjectured that Lagrangian with some function of $A$ multiplied with $e^{\alpha(RA)^{n}}$ might smoothly evolve from a matter-dominated Universe to a late-time accelerated universe. 

Before we proceed with the detailed analysis of such models, let us understand the existence of singularity points in the $\xi$ range $[-1.2,-1]$. From Eq. \eqref{A}, we see that $\xi = -1.2, -1$ and $-3$ are special points of the anticurvature scalar. If the function $f(R, A)$ is a proportional to $A$, then 
the action will diverge at $\xi = -1$ and $-3$. If the function $f(R, A)$ is inversely proportional to $A$, then the action will diverge at $\xi = -1.2$.
This implies that as the Universe evolves from the matter-dominated epoch ($\xi = -1.5$) to the current accelerated epoch ($\xi = -0.5$), we will have to encounter a singularity either at $\xi = -1.2$ or at $\xi = -1$. As we show, this feature will be present irrespective of the form of $f(R,A)$.

While the models in this class are extensive, we only consider a class of models that will reduce to Einstein-Hilbert action $(f(R, A) \sim R)$. For example, to confirm/infirm the no-go theorem, we consider the following form of $f(R, A)$:
\begin{equation}
\label{action:Ic}
f(R,A) = R + \alpha (R^{m+1}A^{m})e^{\beta(RA)^{n}}
\end{equation}
Note that in the limit of $\alpha \to 0$ and $\beta \to 0$, the above function reduces to GR. 

Applying the chain rule of differentiation on the exponential term, we have:
\begin{equation}
\frac{d}{d g_{\mu\nu}}( e^{\alpha(RA)^{n}}) = \alpha e^{\alpha(RA)^{n}} n (RA)^{n-1}\left(\frac{dR}{dg_{\mu\nu}} + \frac{dA}{dg_{\mu\nu}}\right) \, .
\end{equation} 
The singularities in $A$ (as can be seen in Eq. \eqref{A}) will propagate to the 
terms including  $dA/dg_{\mu\nu}$.
Depending on the value of $n$ (i. e., $n > 0$ or $n < 0$),  the modified Friedmann equations in the flat-space FLRW metric will always have $(\xi+1)$ or $(5\xi + 6)$ as a factor in the denominator leading to a singularity either at $-1$ or at $-1.2$.

Including $R^{m+1}A^{m}$, we see that the singularities in this can not be canceled by $A^{n - 1}$ and $dA/dg_{\mu\nu}$. In other words, like in Class Ia and Ib models, here again, we can remove only one singularity. To demonstrate this, in 
Fig. \ref{fig:class Ic}, we have plotted $\xi, \Omega_{r}, \Omega_{m}$ and $\Omega_{A}$ as a function of $\ln (a)$ for $\alpha = \beta = 0.04, m= n = 1$ in the action \eqref{action:Ic}.
\begin{figure}[H]
\begin{subfigure}{.55\textwidth}
  %
  \includegraphics[width=.82\linewidth]{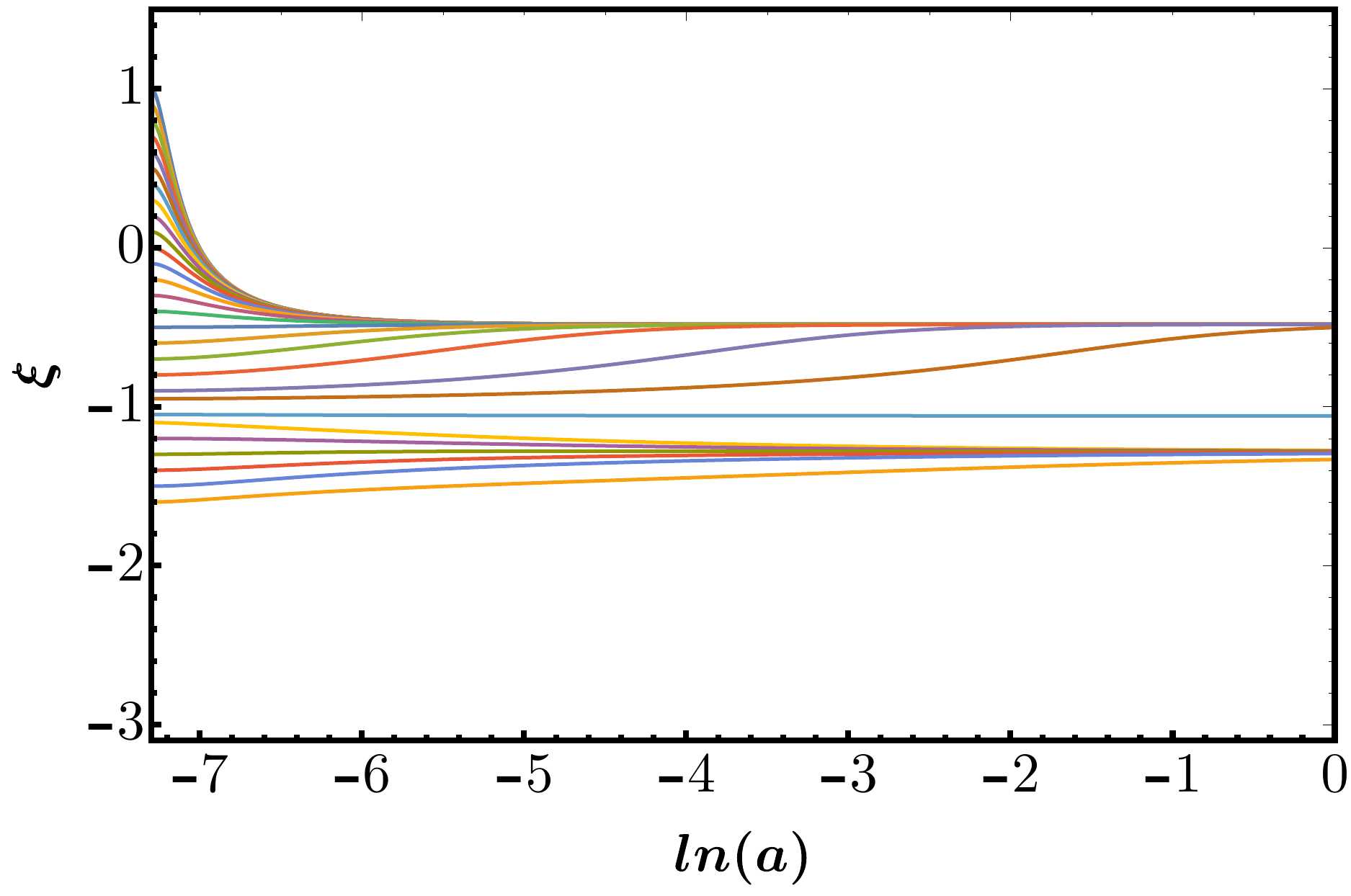}  
  \caption{$\xi(a)$ vs ln(a)}
\end{subfigure}
\begin{subfigure}{.55\textwidth}
  %
  \includegraphics[width=.82\linewidth]{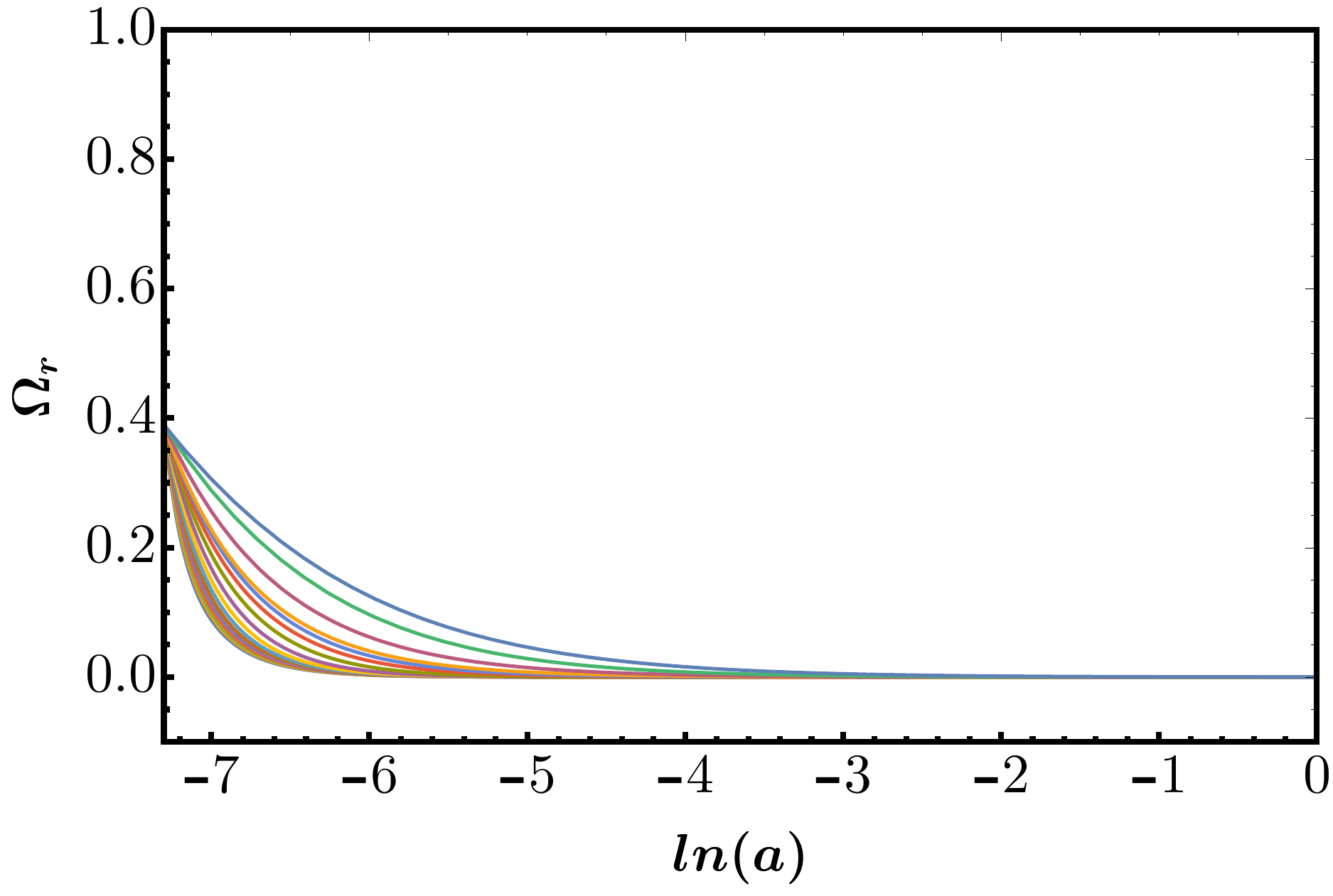}  
  \caption{$\Omega_{r}(a)$ vs ln(a)}
\end{subfigure}
\end{figure}
\begin{figure}[H]\ContinuedFloat
\begin{subfigure}{.55\textwidth}
  %
  \includegraphics[width=.82\linewidth]{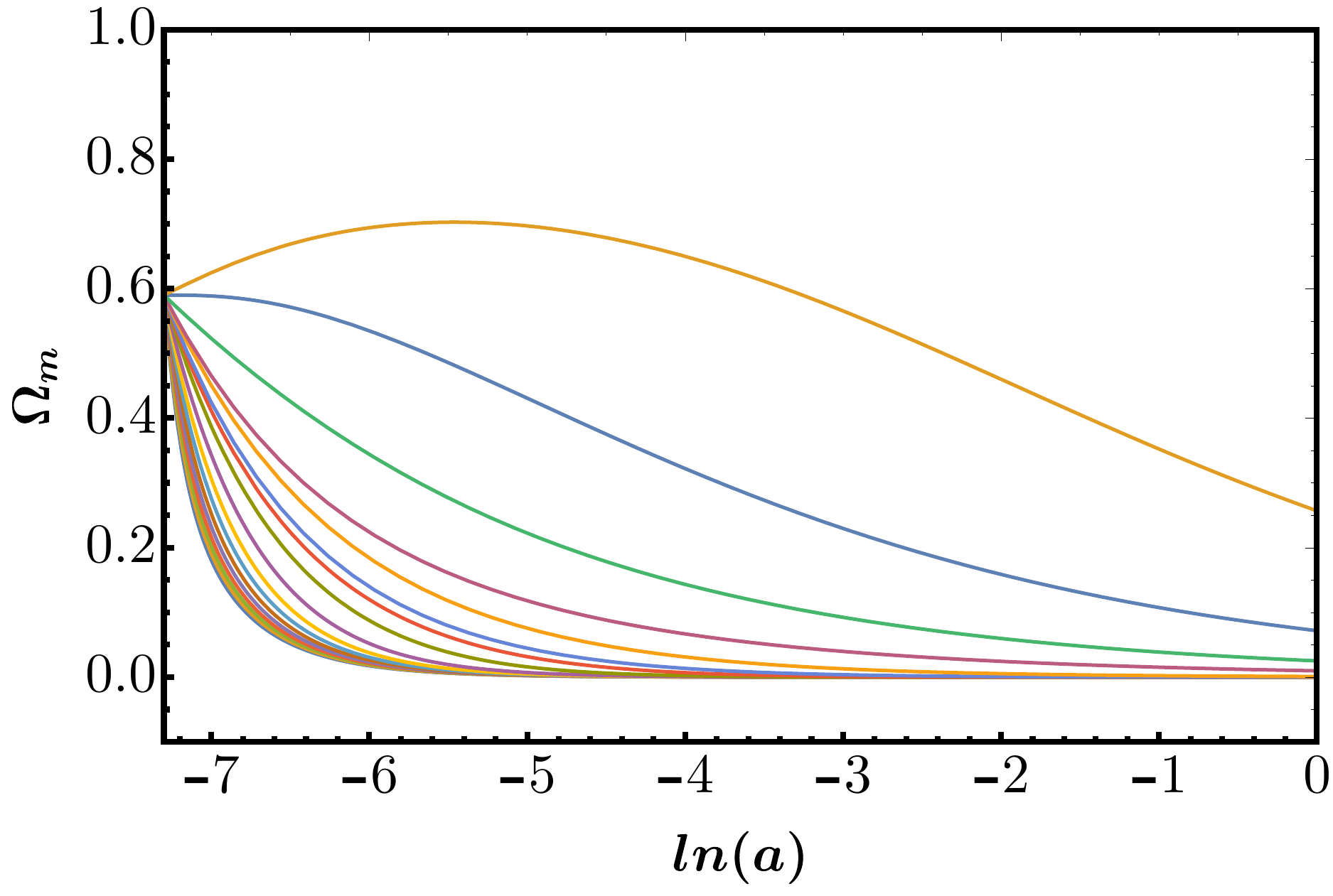}  
  \caption{$\Omega_{m}(a)$ vs ln(a)}
\end{subfigure}
\begin{subfigure}{.55\textwidth}
  %
  \includegraphics[width=.82\linewidth]{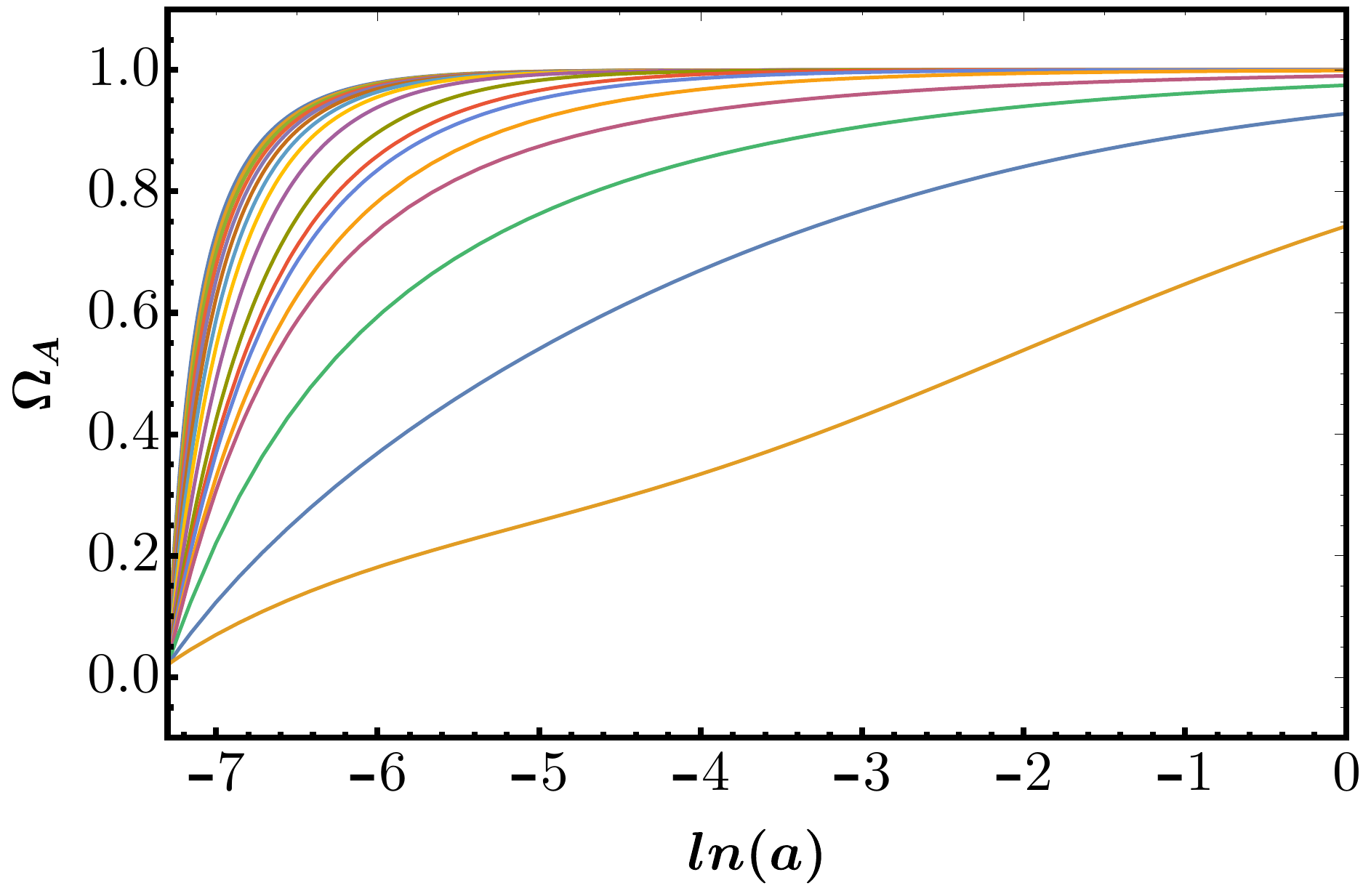}  
  \caption{$\Omega_{A}(a)$ vs ln(a)}
\end{subfigure}
\caption{Plot of $\xi(a), \Omega_r(a), \Omega_m(a), \Omega_A(a)$ as a function of $\ln(a)$ for $f(R, A) = R + \alpha (R^{2}A) \exp[\alpha (RA)]$ and $\alpha = 0.04$.}
\label{fig:class Ic}
\end{figure}
Extending this model by considering a polynomial of $R$ instead of $(R^{m+1}A^{m})$, multiplied with $e^{\alpha(RA)^{n}}$, the above problem will still exist for some terms. Hence, we will always get either one or both singularities  ($\xi = -1, -1.2$) in these kinds of models. 

Note that extending the above model by including a polynomial of $R$ and $A$ in the exponential will lead to singularities in the resulting modified Friedmann equations. This is because of the presence of $A$ in the polynomial. The same analysis and argument apply to models with polynomials of $R$ and $A$ in both the numerator and denominator. 

\subsection{Can Class I models bypass the no-go theorem?}

For certain initial conditions, Class I models lead to  $\xi = -1.5 (w_{\rm eff} = 0)$ (matter-dominated universe) or $\xi = 0 (w_{\rm eff} = -1)$ (de Sitter universe) or $\xi = -0.5 (w_{\rm eff} = -0.67)$ (current Universe) as attractors. However, we can not circumvent the no-go theorem as it is impossible to smoothly evolve 
from $\xi = -1.5$ to $\xi = -0.5$. Hence Class I models can not explain the late-time acceleration of the universe.

From Eq. \eqref{A}, we see that $\xi = -1.2, -1$ and $-3$ are special points 
of the anticurvature scalar. If the function $f(R, A)$ has a term which is proportional to $A$, then 
the action will diverge at $\xi = -1$ and $-3$. If the function $f(R, A)$ has a term which is inversely proportional to $A$, then the action will diverge at $\xi = -1.2$.
This implies that as the Universe evolves from the matter-dominated epoch ($\xi = -1.5$) to the current accelerated epoch ($\xi = -0.5$), we will have to encounter a singularity either at $\xi = -1.2$ or at $\xi = -1$. Choosing the form of $f(R, A)$ can remove one singularity, but not both the singularities. Since, we need to cross both the points to evolve from from the matter-dominated epoch ($\xi = -1.5$) to the current accelerated epoch ($\xi = -0.5$), class I models \emph{can not} bypass the no-go theorem.

\section{Class II models}\label{sec:Class II models}

In this section, we consider the following general action:
\begin{equation}
S = \int d^{4}x \sqrt{-g} \, {\cal F}(R,A^{\mu\nu}A_{\mu\nu})
\end{equation}
where ${\cal F}(R,A^{\mu\nu}A_{\mu\nu})$ is an arbitrary, smooth 
function of Ricci scalar and 
anticurvature tensor square. We further classify this class into two subclasses (Class IIa, and Class IIb). Class IIa models are polynomials in $R$ and $A^{\mu\nu}A_{\mu\nu}$. Class IIb models are of the form $\exp[\alpha (R^2 A^{\mu\nu}A_{\mu\nu})^n]$.

\subsection{Class IIa models}\label{subsec:Class IIa models}

As mentioned above, in this class, we consider the cases where ${\cal F}(R, A^{\mu\nu}A_{\mu\nu})$ is polynomial. Moreover, to keep calculations tractable, we consider the following two forms:
\begin{eqnarray}
{\cal F}_1(R,A^{\mu\nu}A_{\mu\nu}) &=& R  \left( 1 + \frac{\alpha}{R^2 A^{\mu\nu}A_{\mu\nu}} \right) \, ; \nonumber \\
{\cal F}_2(R,A^{\mu\nu}A_{\mu\nu}) &=& R \left(1 + 
\frac{\alpha}{R^2 A^{\mu\nu}A_{\mu\nu}} + 
\frac{\alpha}{R^{4}(A^{\mu\nu}A_{\mu\nu})^{2}} \right)
\end{eqnarray}
where $\alpha$ is a constant. For simplicity, we have fixed the same coefficients for ${\cal F}_2$. In the limit of $\alpha \to 0$, the above models reduce to GR.

\begin{figure}[H]
\begin{subfigure}{.55\textwidth}
  %
  \includegraphics[width=.82\linewidth]{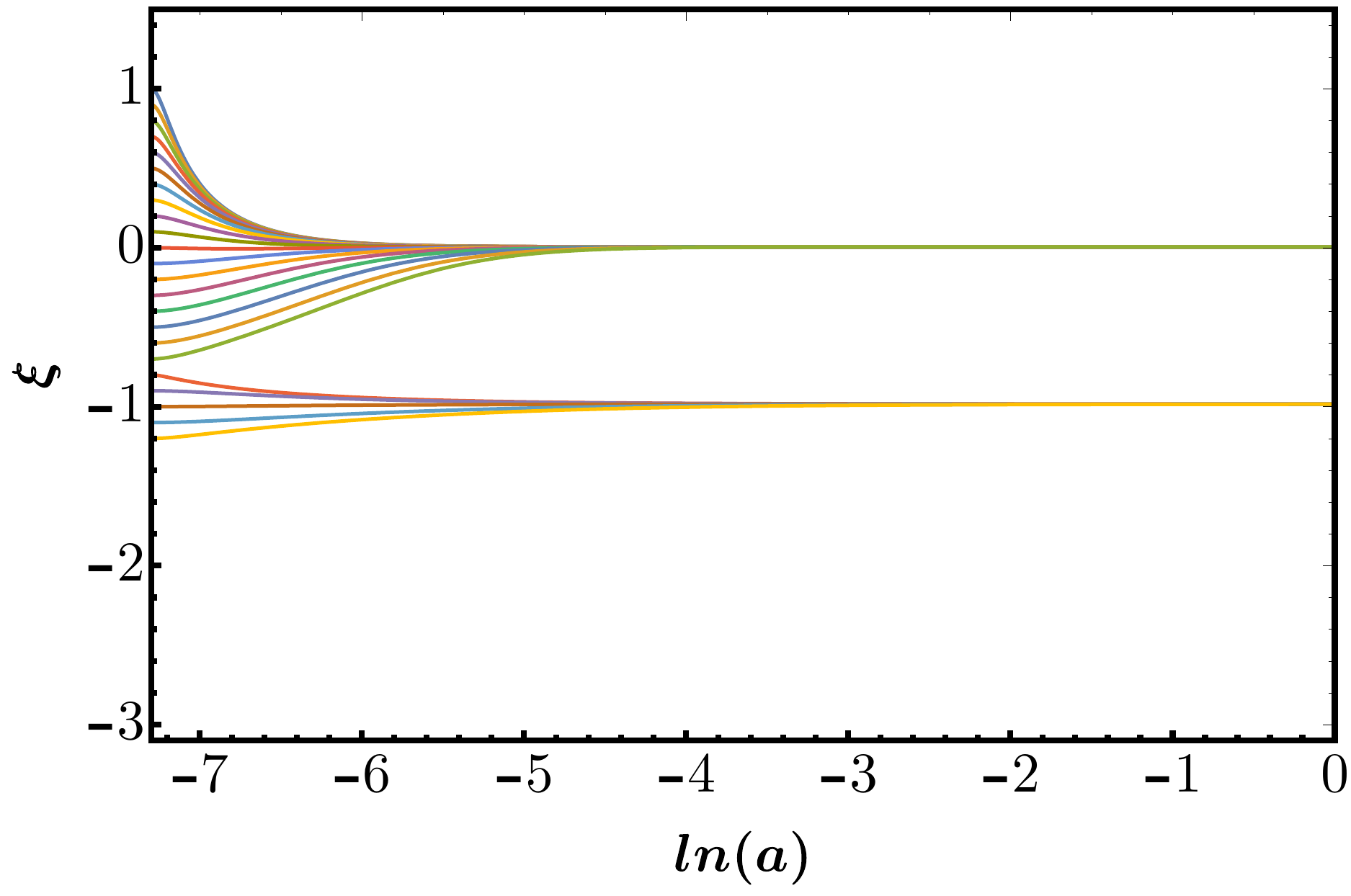}  
  \caption{$\xi(a)$ vs ln(a),$\alpha = -65$}
\end{subfigure}
\begin{subfigure}{.55\textwidth}
  %
  \includegraphics[width=.82\linewidth]{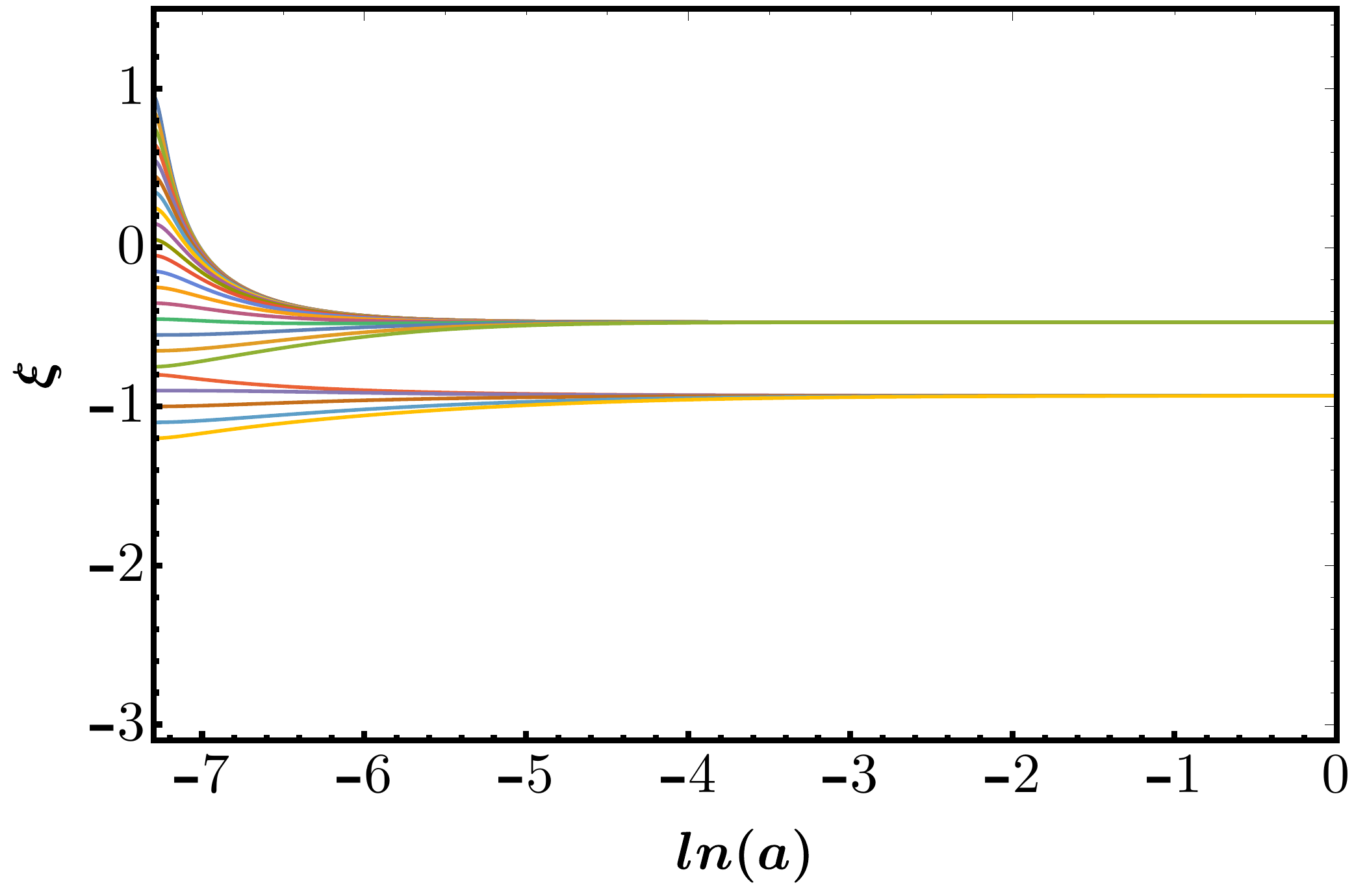}  
  \caption{$\xi(a)$ vs ln(a),$\alpha = -17$}
\end{subfigure}
\begin{subfigure}{.55\textwidth}
  %
  \includegraphics[width=.82\linewidth]{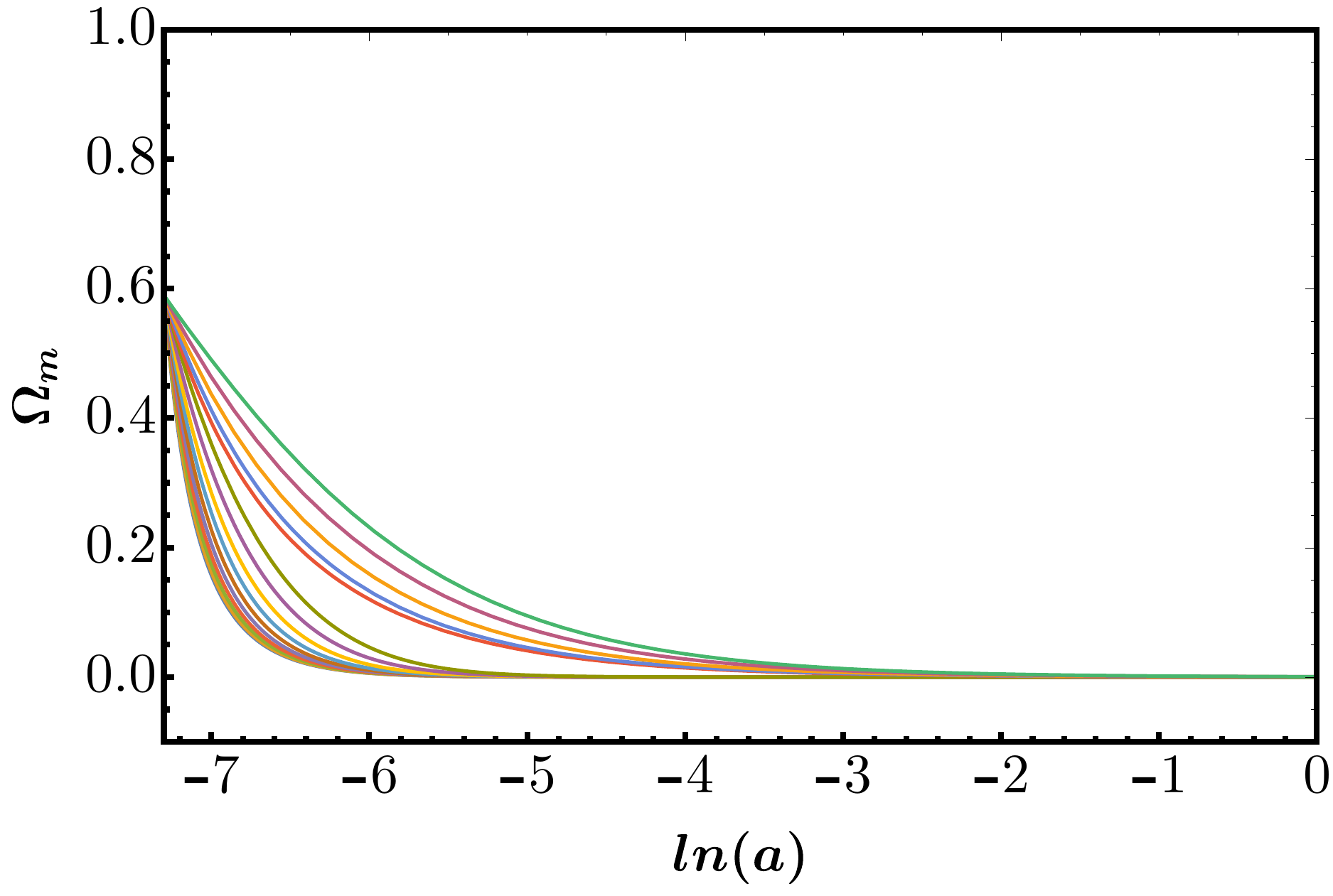}  
  \caption{$\Omega_{m}(a)$ vs ln(a),$\alpha = -65$}
\end{subfigure}
\begin{subfigure}{.55\textwidth}
  %
  \includegraphics[width=.82\linewidth]{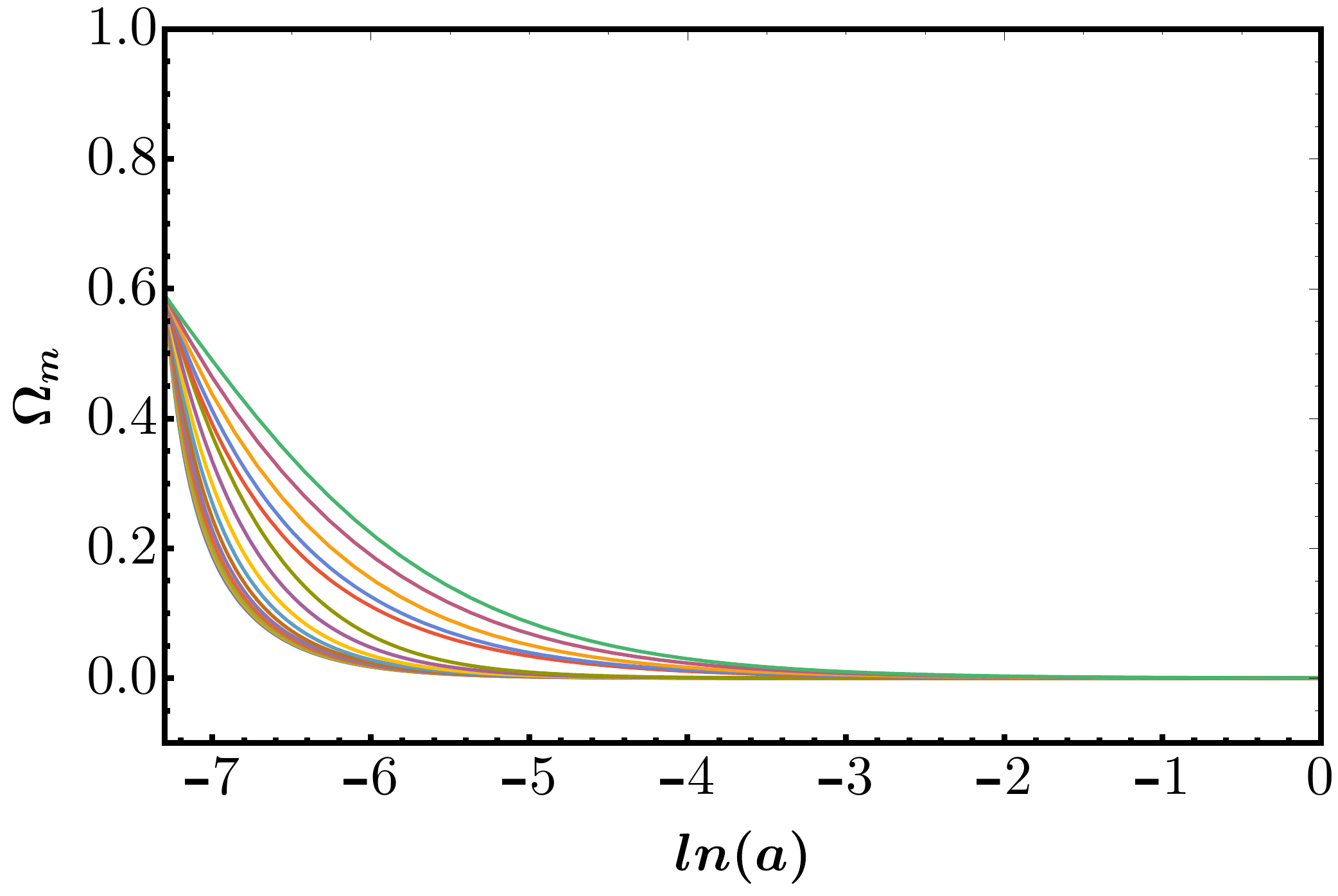}  
  \caption{$\Omega_{m}(a)$ vs ln(a),$\alpha = -17$}
\end{subfigure}
\newline
\begin{subfigure}{.55\textwidth}
  %
  \includegraphics[width=.82\linewidth]{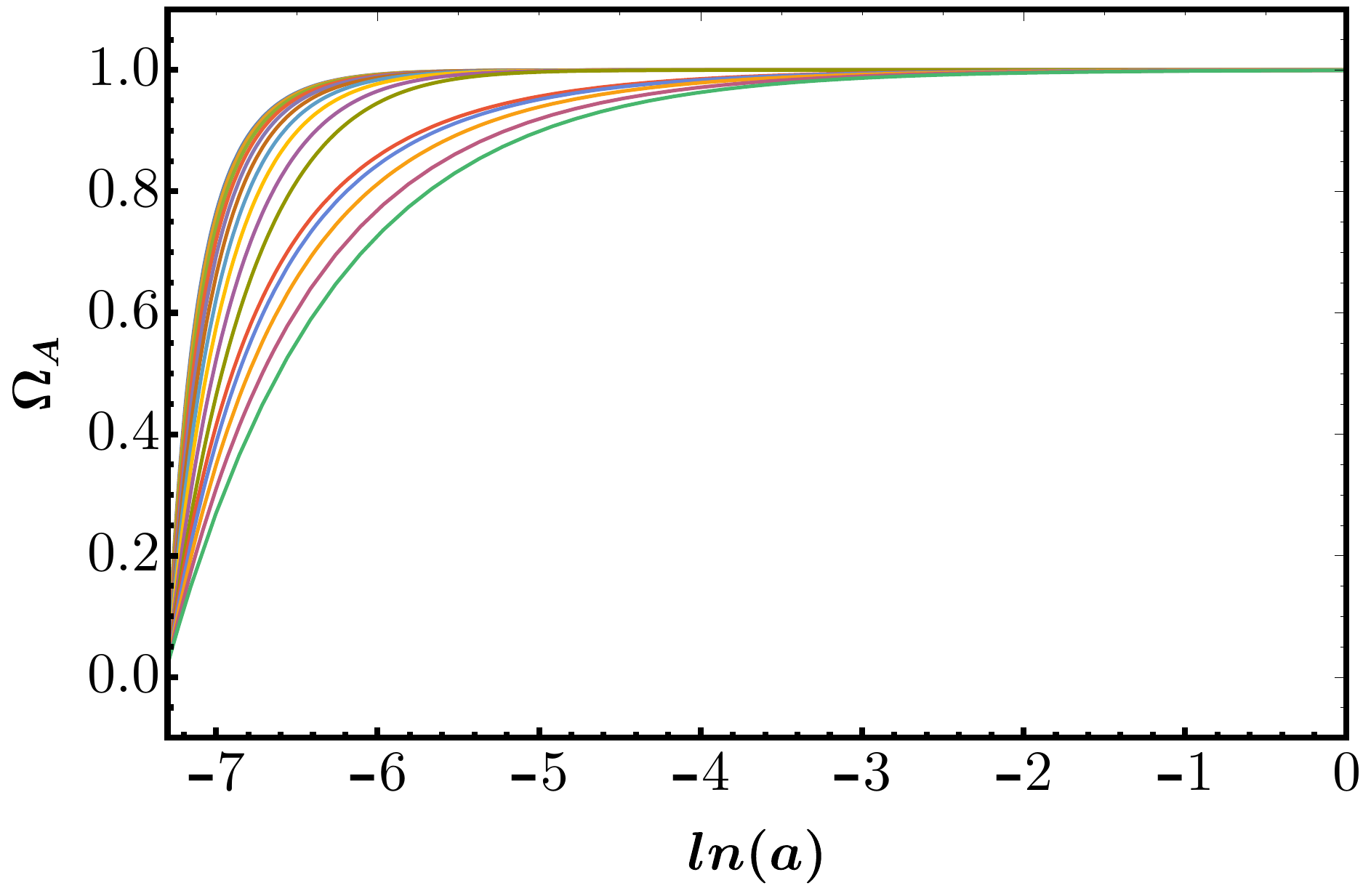}  
  \caption{$\Omega_{A}(a)$ vs ln(a),$\alpha = -65$}
\end{subfigure}
\begin{subfigure}{.55\textwidth}
  %
  \includegraphics[width=.82\linewidth]{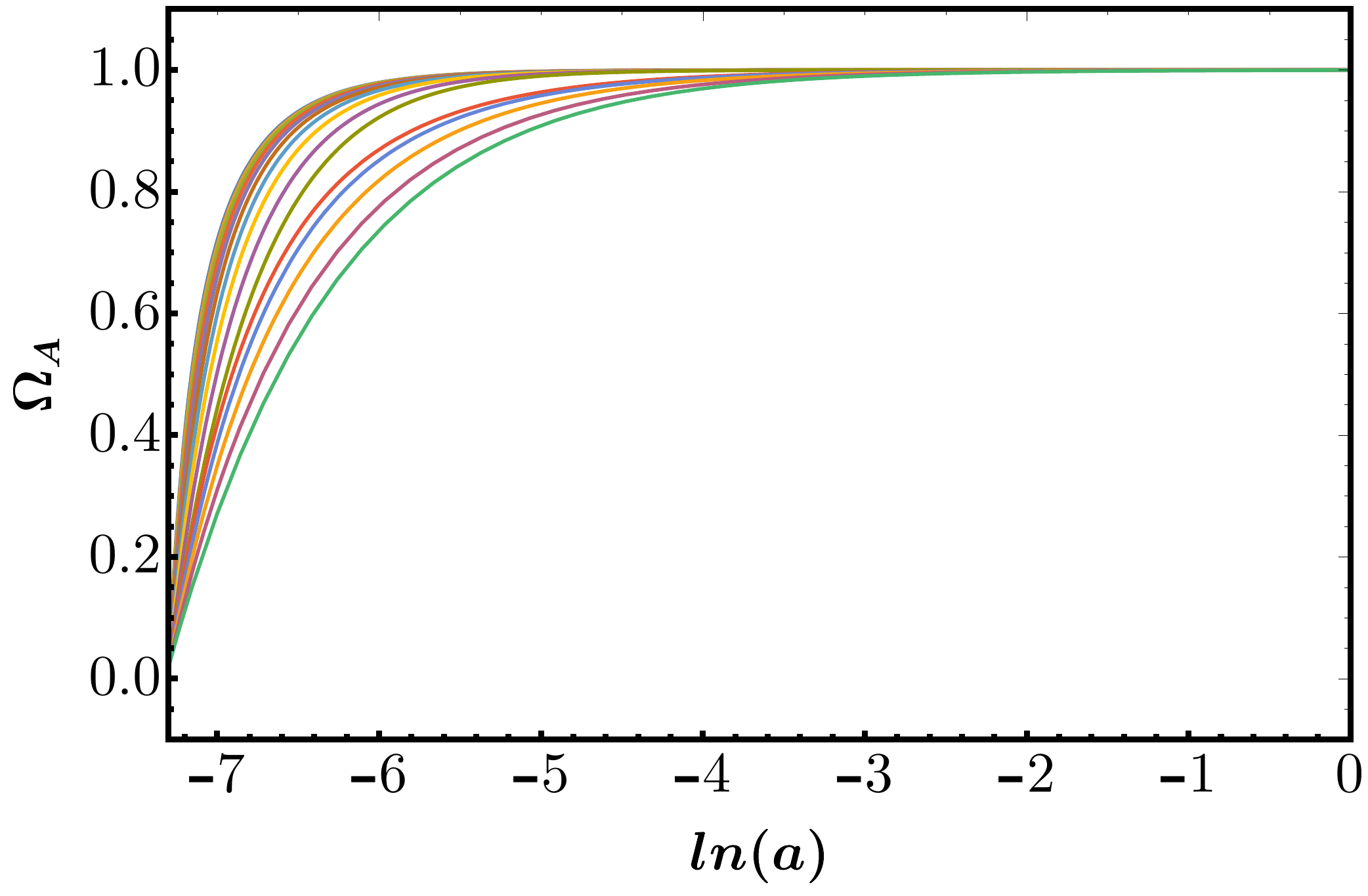}  
  \caption{$\Omega_{A}(a)$ vs ln(a),$\alpha = -17$}
\end{subfigure}
\caption{Plot of $\xi(a), \Omega_m(a), \Omega_A(a)$ as a function of $\ln(a)$ for ${\cal F}_1$. Plots on the left are for $\alpha =  -65$ and the plots are the right are for $\alpha = -17$.}
\label{fig:newmodel1alpha}
\end{figure}

Fig. \ref{fig:newmodel1alpha} contains the plots for ${\cal F}_1$ for two values of $\alpha$. Plots on the left are for $\alpha = -65$ and the plots are the right are for $\alpha = -17$.  Different colors in the plots refer to different initial values 
of $\xi$ in the range $[-5, 0.6]$. From these plots, we infer the following:
\begin{enumerate}
\item For $\alpha = -65$ and initial values of $\xi > -0.7$, we get a de Sitter attractor ($\xi = 0$). For initial values of $\xi$ in the range $[-1.3, -0.8]$, we get another attractor at $\xi = -1$. Corresponding to these initial values, $\Omega_{r}$ and $\Omega_{m}$ starts at $0.39$ and $0.59$ respectively, and quickly converge to $0$; while $\Omega_{A}$ starts at 0.02 and quickly converges to $1$.

\item For $\alpha = -65$ and for the initial values of $\xi$ in the range $[-1.6, -1.3]$, the values ($\xi, \Omega_m, \Omega_r)$ diverge. For initial values of $\xi < -1.6$, we get attractor at $\xi = -2$ (corresponding to radiation-dominated epoch). Note that this is different from that of Class I models. However, for these models, $\Omega_r$ settles to a non-zero value. Hence, this is not consistent with cosmological observations. 

\item For $\alpha = -17$, for different initial values of $\xi$ we observe the same feature. The only difference is that for initial values of $\xi >  -0.7$ we get an attractor at $\xi = -0.5$.
\end{enumerate}

\newpage

\begin{figure}[H]
\begin{subfigure}{.55\textwidth}
  %
  \includegraphics[width=.82\linewidth]{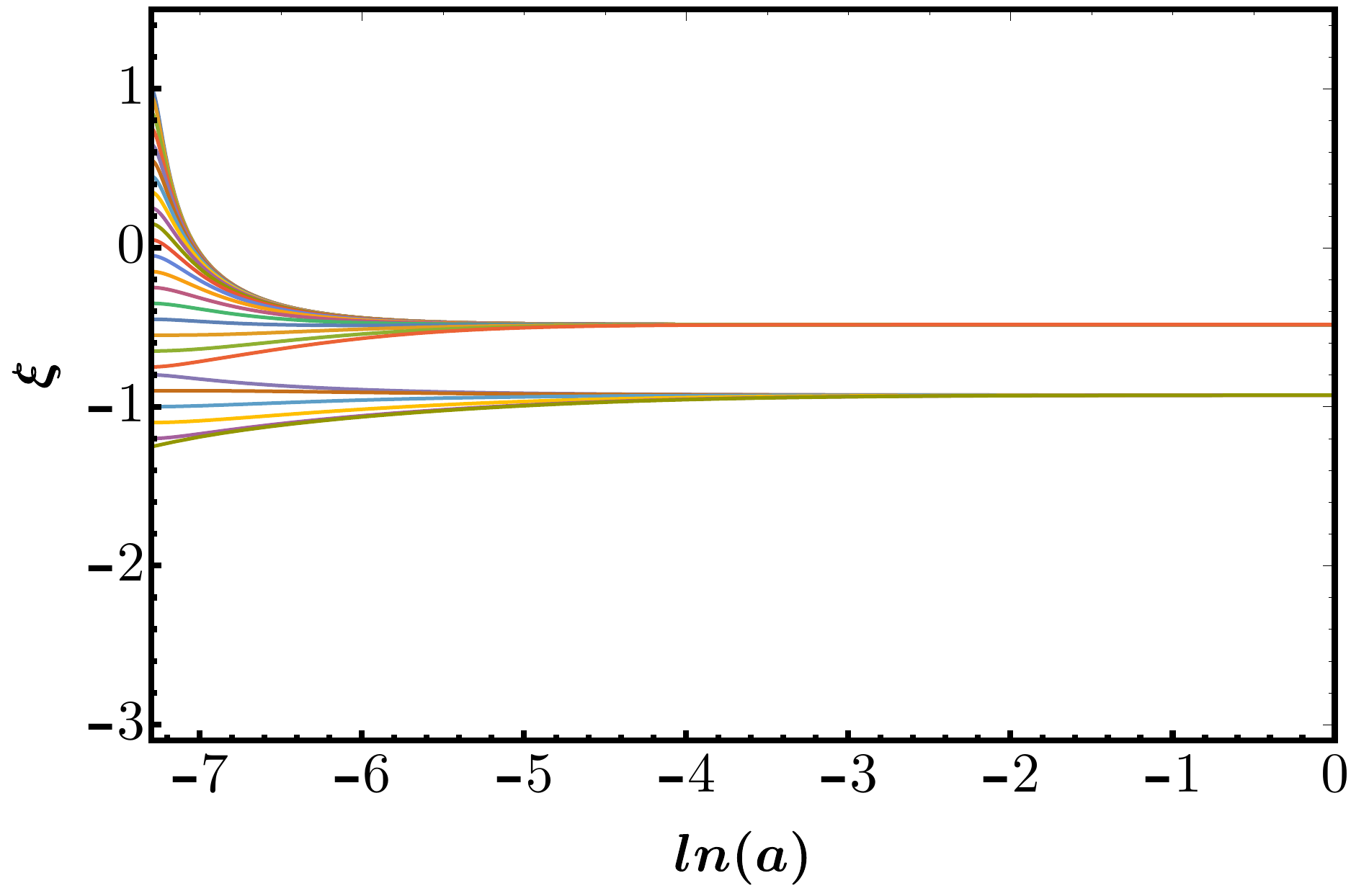}  
  \caption{$\xi(a)$ vs ln(a),$\alpha = -16$}
\end{subfigure}
\begin{subfigure}{.55\textwidth}
  %
  \includegraphics[width=.82\linewidth]{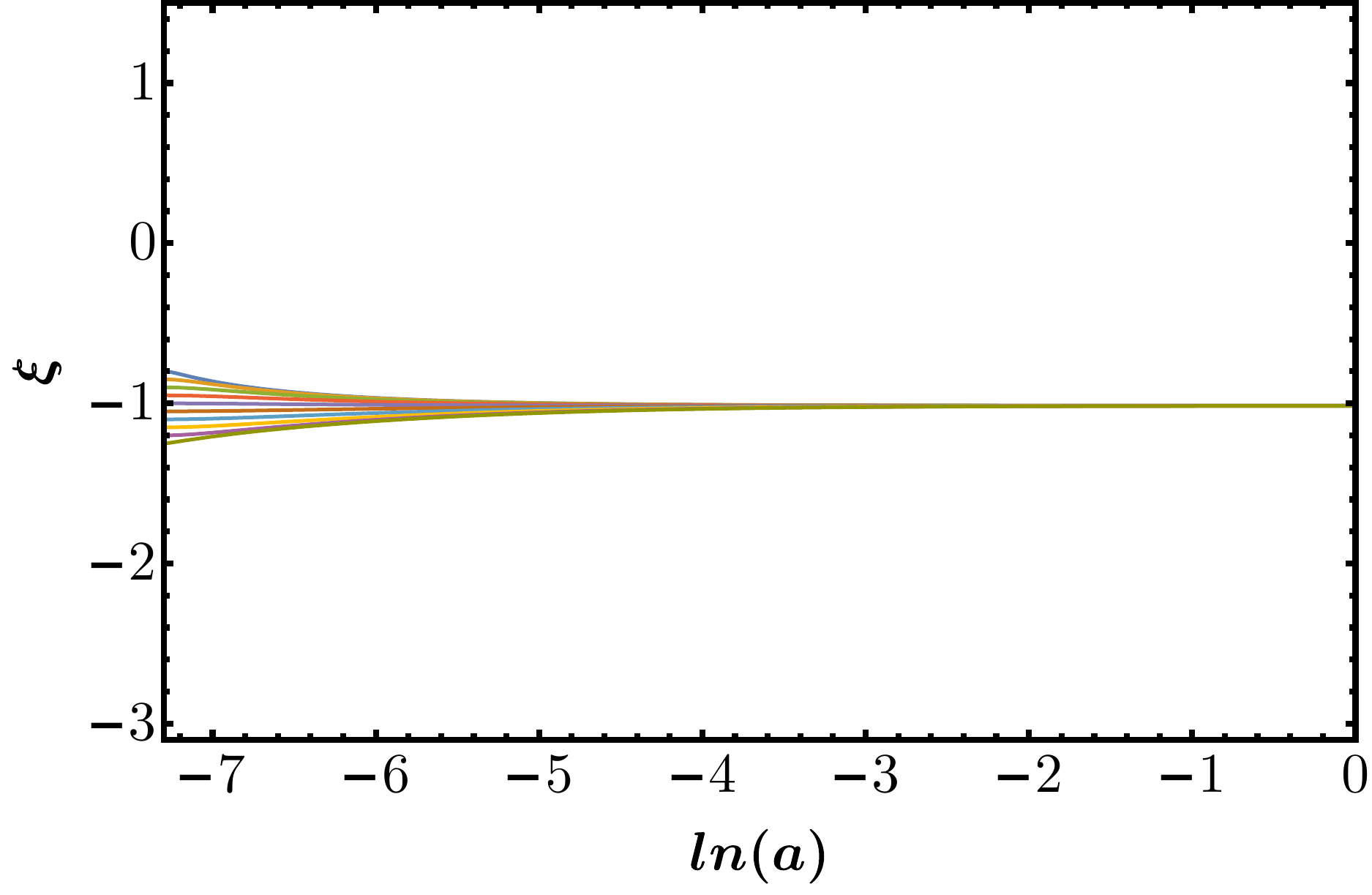}  
  \caption{$\xi(a)$ vs ln(a),$\alpha = 65$}
\end{subfigure}
\newline
\begin{subfigure}{.55\textwidth}
  %
  \includegraphics[width=.82\linewidth]{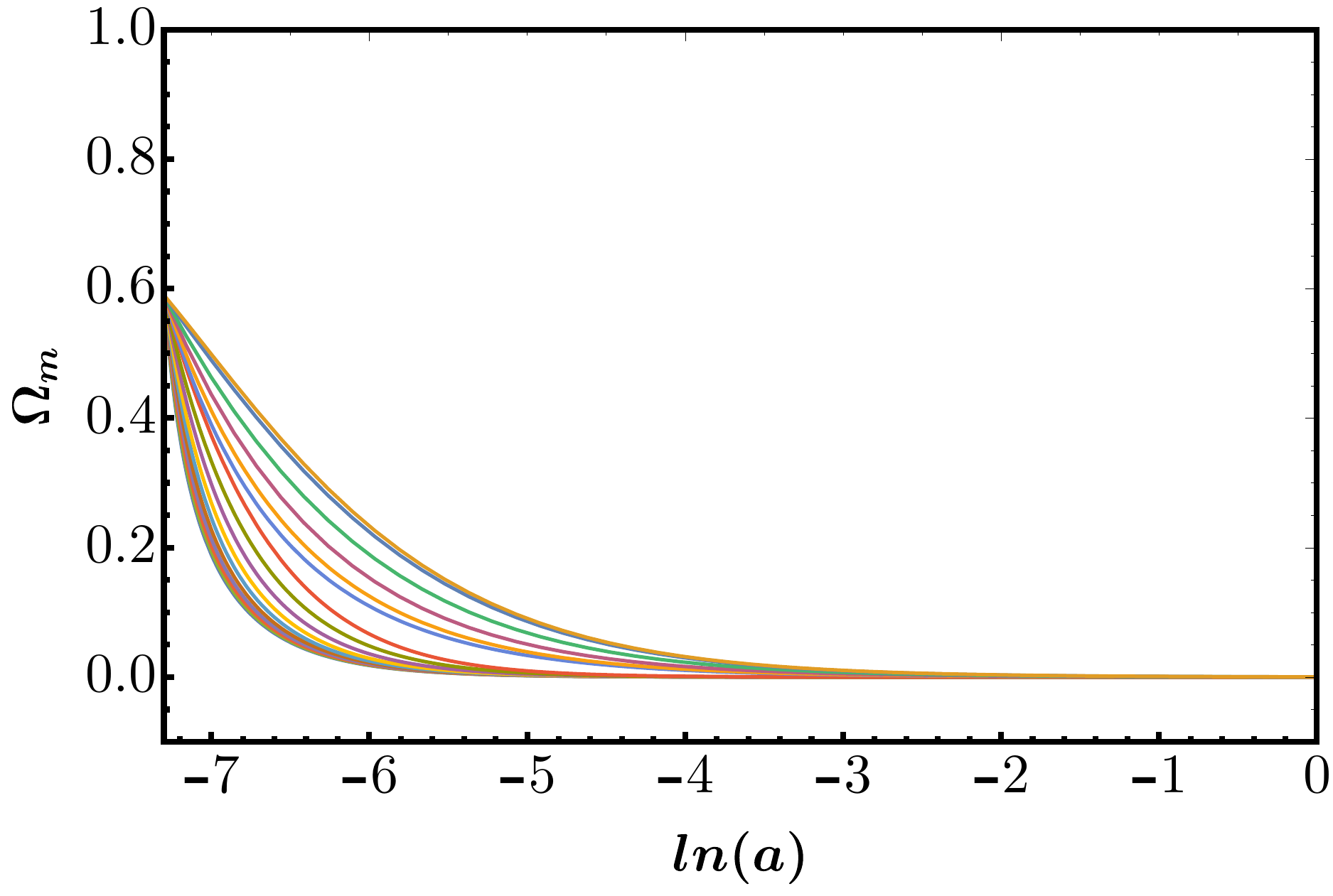}  
  \caption{$\Omega_{m}(a)$ vs ln(a),$\alpha = -16$}
\end{subfigure}
\begin{subfigure}{.55\textwidth}
  %
  \includegraphics[width=.82\linewidth]{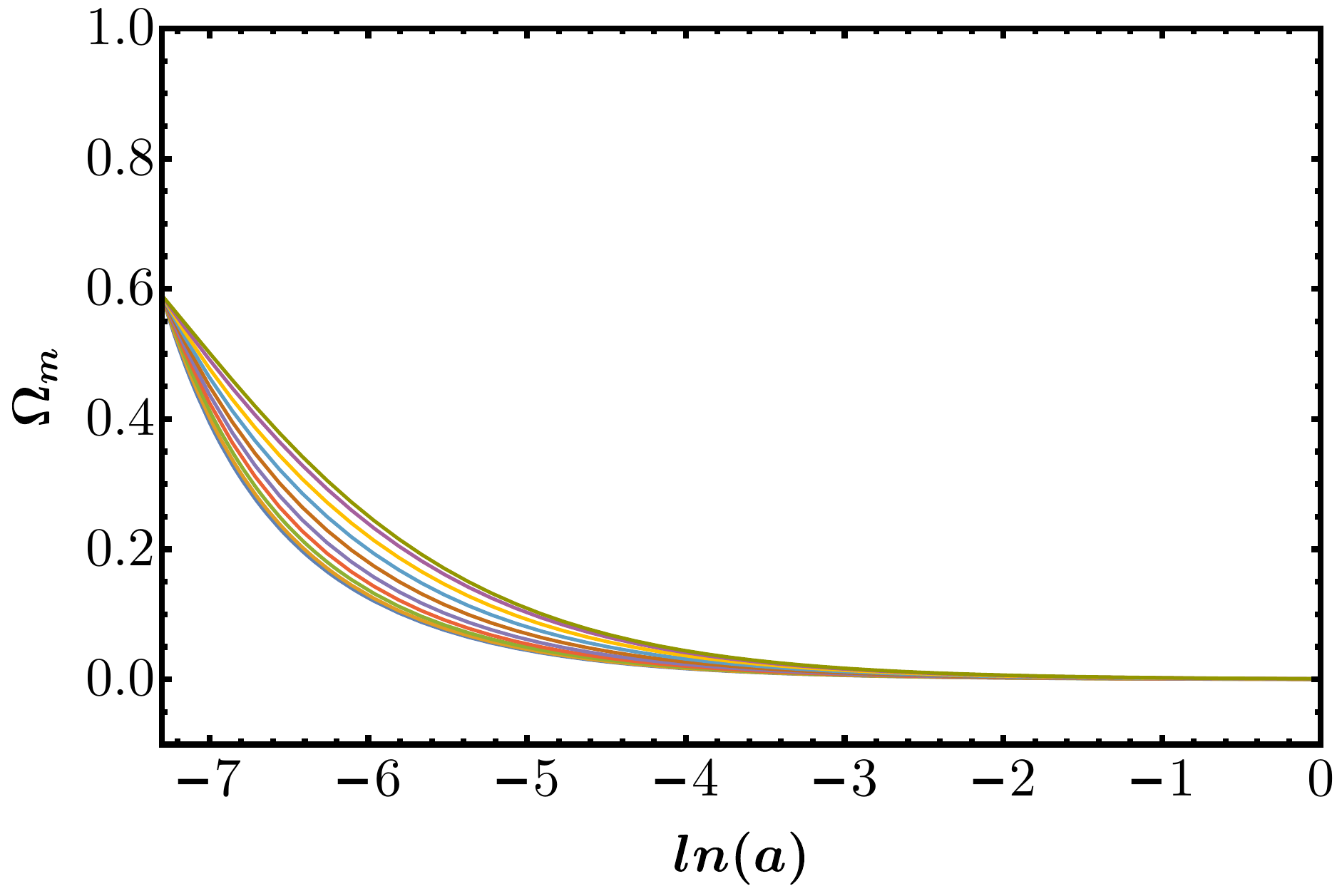}  
  \caption{$\Omega_{m}(a)$ vs ln(a),$\alpha = 65$}
\end{subfigure}
\newline
\begin{subfigure}{.55\textwidth}
  %
  \includegraphics[width=.82\linewidth]{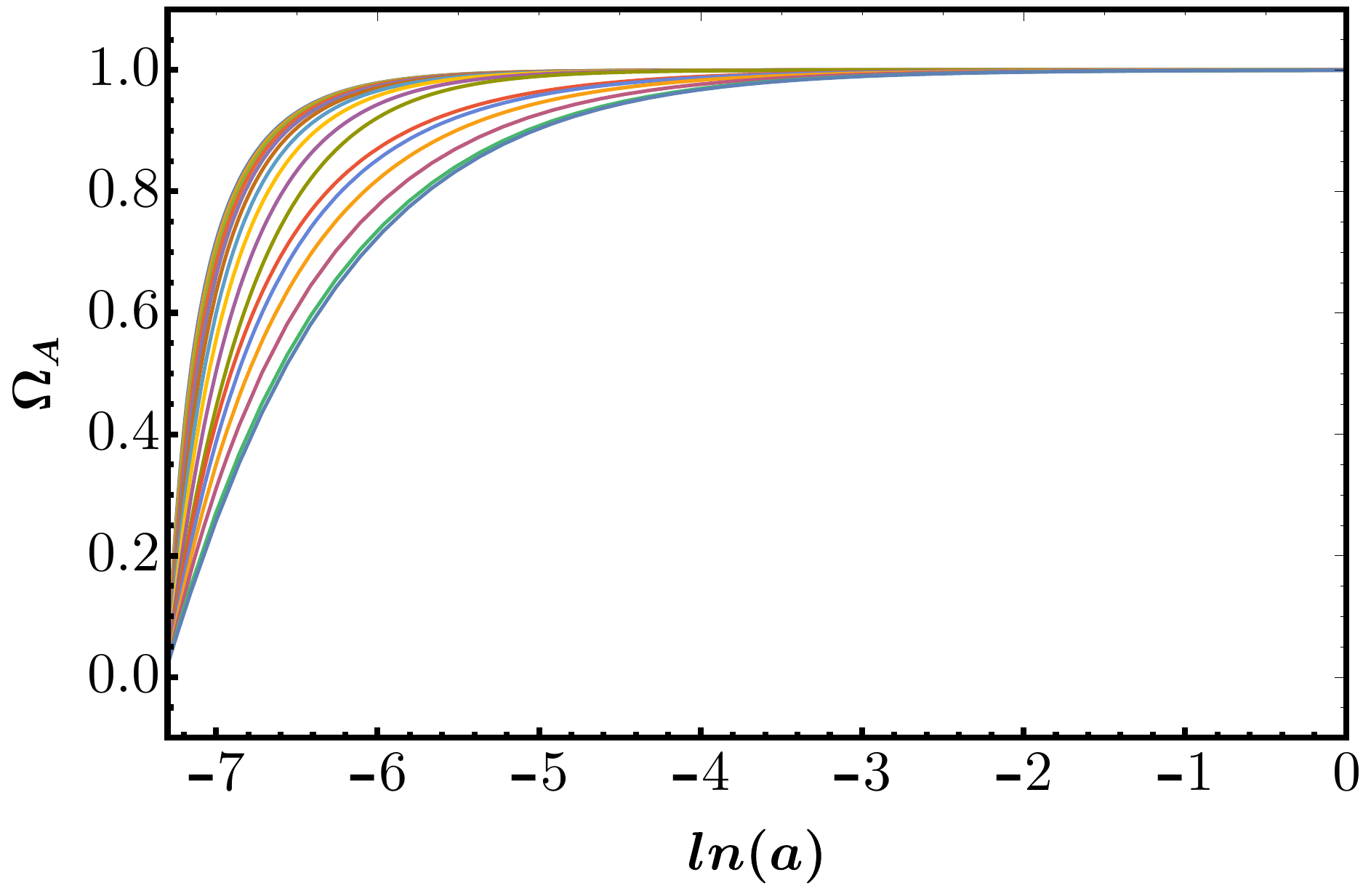}  
  \caption{$\Omega_{A}(a)$ vs ln(a),$\alpha = -16$}
\end{subfigure}
\begin{subfigure}{.55\textwidth}
  %
  \includegraphics[width=.82\linewidth]{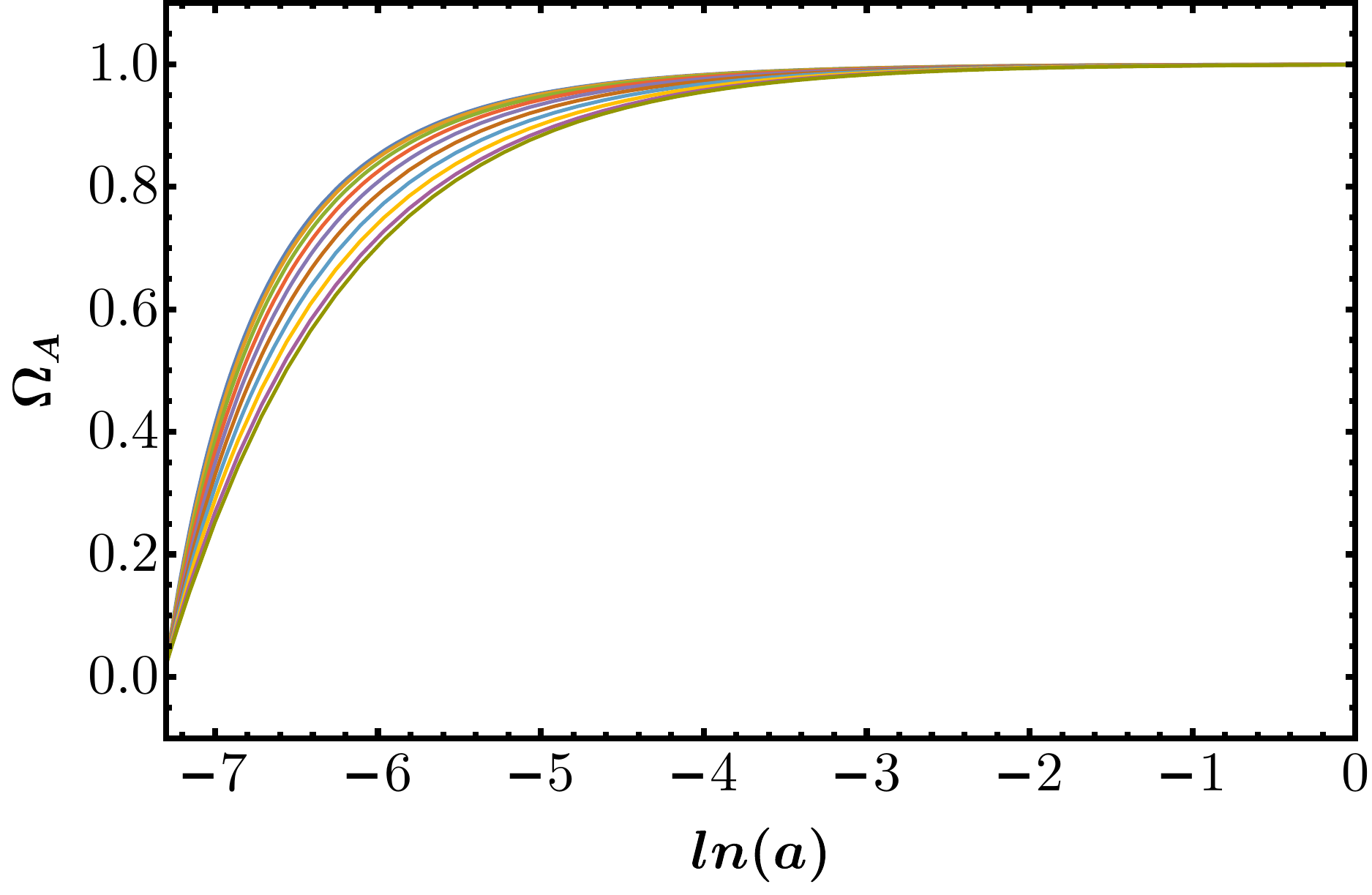}  
  \caption{$\Omega_{A}(a)$ vs ln(a),$\alpha = 65$}
\end{subfigure}
\caption{Plot of $\xi(a), \Omega_m(a), \Omega_A(a)$ as a function of $\ln(a)$ for ${\cal F}_2$. Plots on the left are for $\alpha =  -16$ and the plots are the right are for $\alpha = 65$.}
\label{fig:newmodel3}
\end{figure}

Fig. \ref{fig:newmodel3} contains the plots for ${\cal F}_2$ for two values of $\alpha$. Plots on the left are for $\alpha = -16$ and the plots are the right are for $\alpha = 65$.  For $\alpha = -16$, we observe the same results as in the case of ${\cal F}_1$
for $\alpha = -17$. For $\alpha = 65$, we get the same trend, with the only difference being there are no attractors for initial values greater than $\xi = -0.8$.

\subsection{Class IIb models}\label{subsec:Class IIb models}

In this class, we consider the following exponential forms of ${\cal F}$:
\begin{equation}
\label{action:classIIb}
{\cal F}_3(R,A^{\mu\nu}A_{\mu\nu}) =  R e^{\alpha/(R^{2}A^{\mu\nu}A_{\mu\nu})}
\qquad {\cal F}_4(R,A^{\mu\nu}A_{\mu\nu})  = R e^{\alpha (R^{2}A^{\mu\nu}A_{\mu\nu})}
\end{equation}
where $\alpha$ (a dimensionless constant) decides the deviation from GR.

\begin{figure}[H]
\begin{subfigure}{.55\textwidth}
  %
  \includegraphics[width=.82\linewidth]{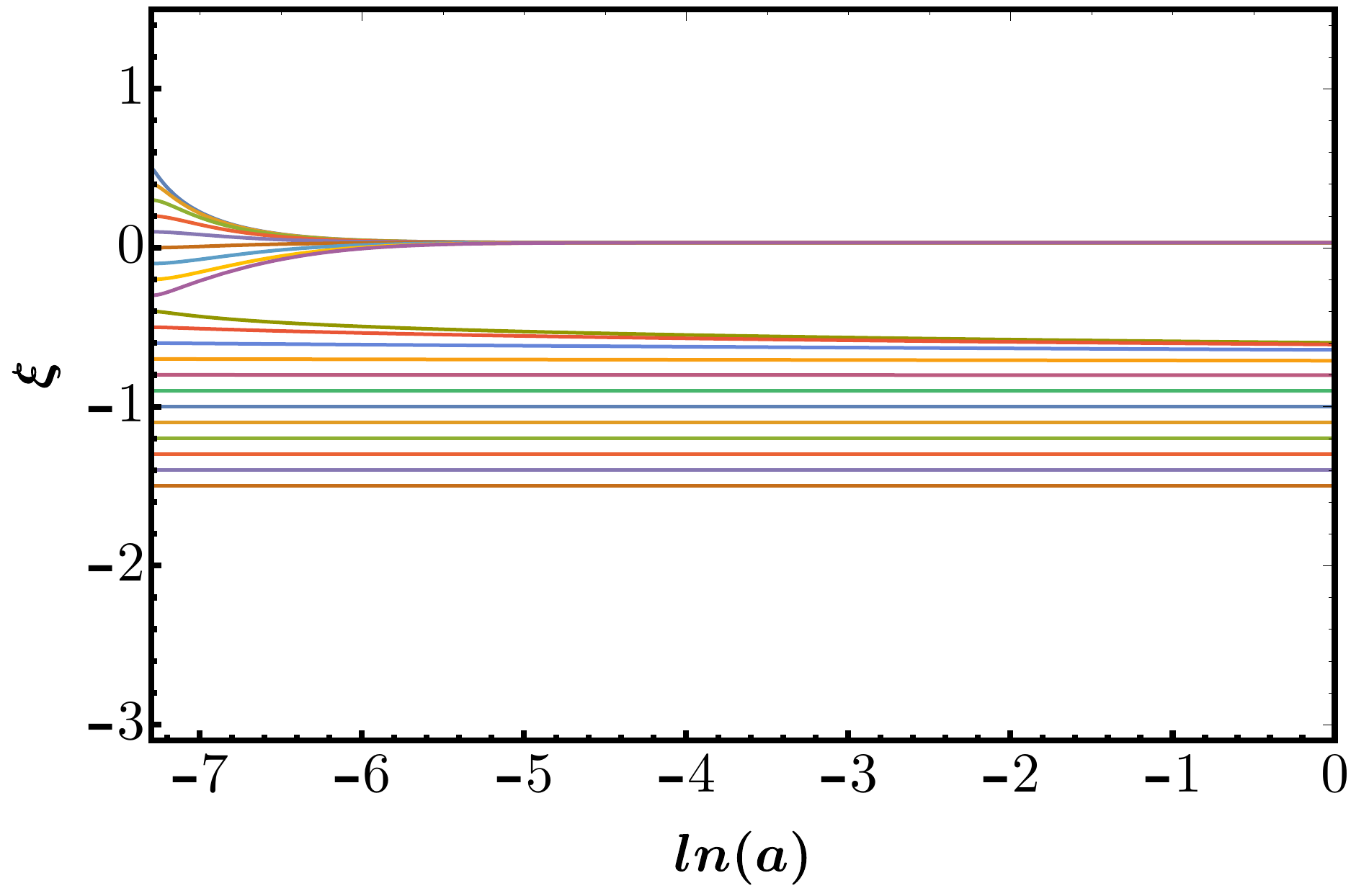}  
  \caption{$\xi(a)$ vs ln(a)}
\end{subfigure}
\begin{subfigure}{.55\textwidth}
  %
  \includegraphics[width=.82\linewidth]{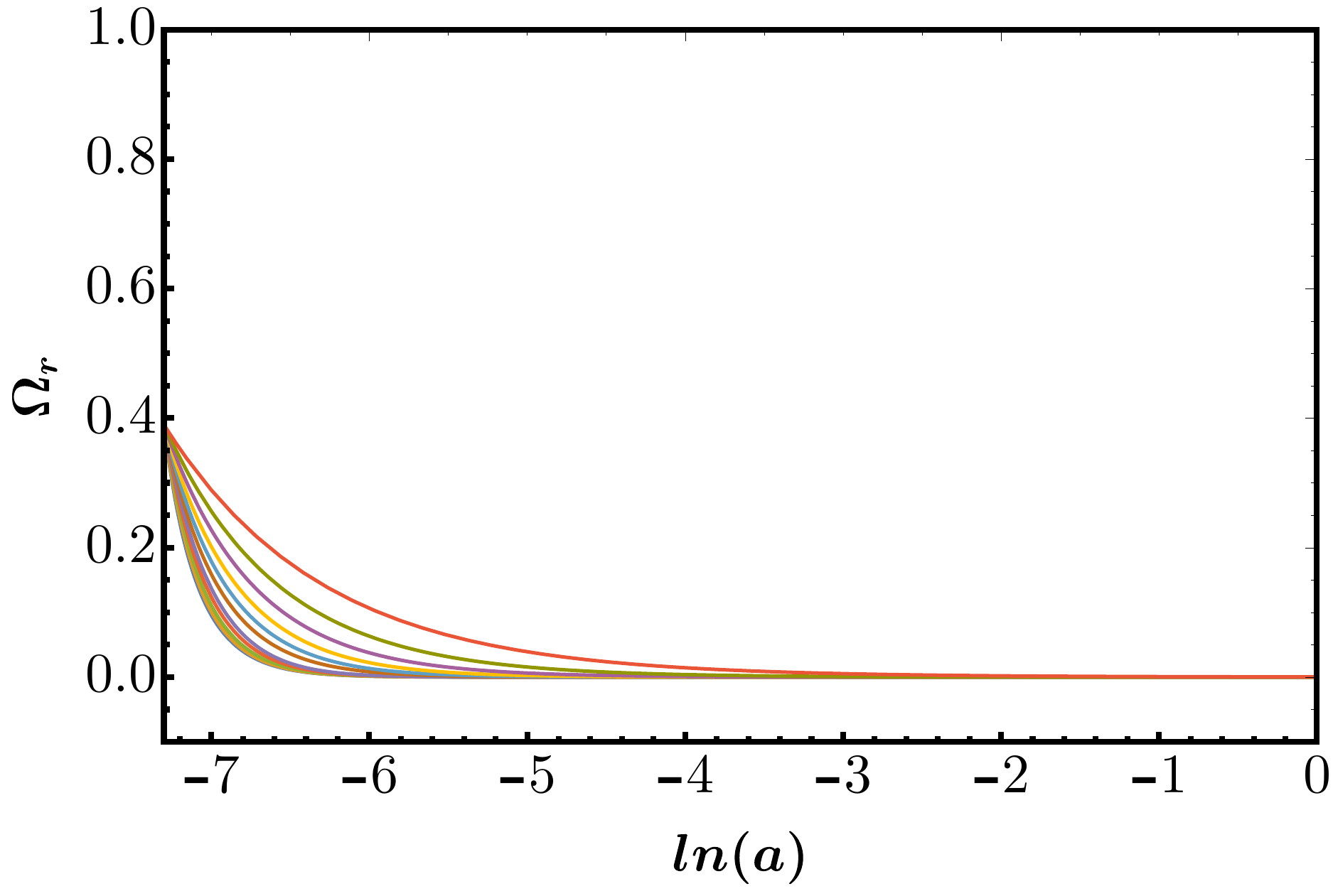}  
  \caption{$\Omega_{r}(a)$ vs ln(a)}
\end{subfigure}
\newline
\begin{subfigure}{.55\textwidth}
  %
  \includegraphics[width=.82\linewidth]{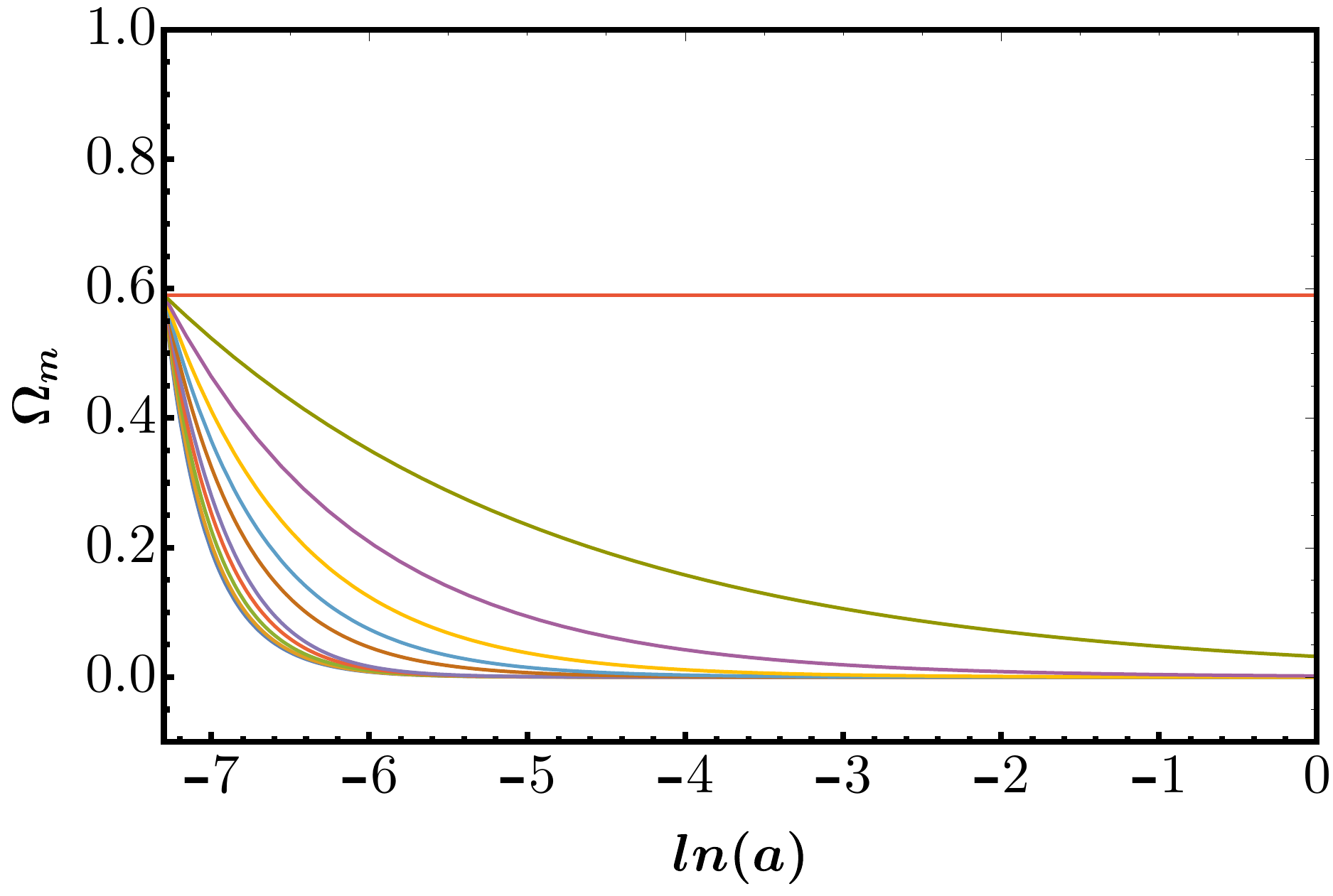}  
  \caption{$\Omega_{m}(a)$ vs ln(a)}
\end{subfigure}
\begin{subfigure}{.55\textwidth}
  %
  \includegraphics[width=.82\linewidth]{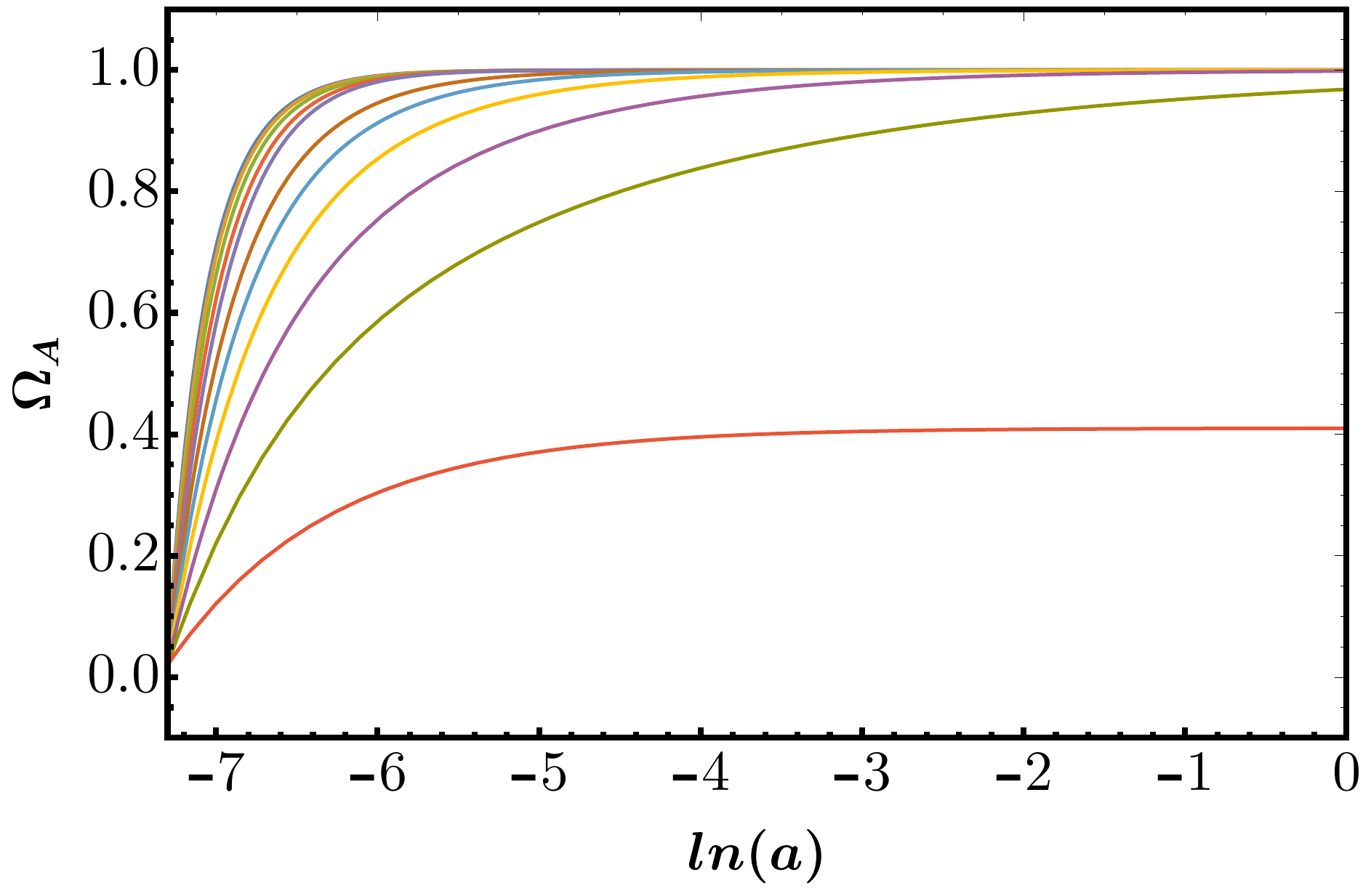}  
  \caption{$\Omega_{A}(a)$ vs ln(a)}
\end{subfigure}
\caption{Plot of $\xi(a), \Omega_r(a), \Omega_m(a), \Omega_A(a)$ as a function of $\ln(a)$ for ${\cal F}_3$ and $\alpha = 700$.}
\label{fig:newmodel5alpha700}
\end{figure}

Fig. \ref{fig:newmodel5alpha700} contains the plots for ${\cal F}_3$ and $\alpha = 700$. Different colors in the plots refer to different initial values 
of $\xi$ in the range $[-5, 0.6]$. From these plots we infer the following:
\begin{enumerate}
\item For initial values of $\xi > -0.4$, we get a de Sitter attractor ($\xi = 0$). 

\item For initial values of $\xi$ in the range $[-2,-0.4]$, $\xi$ is almost a constant.  
Corresponding to these initial values, $\Omega_{r}$ and $\Omega_{m}$ starts at $0.39$ and $0.59$ respectively, and quickly converge to $0$; while, $\Omega_{A}$ starts at 0.02 and quickly converges to $1$.

\item For the initial value of $\xi = -2$, the values ($\xi, \Omega_m, \Omega_r)$ diverge. For initial values of $\xi < -2$, we get attractor at $\xi = -2$ (corresponding to radiation-dominated epoch).  However, for these models, $\Omega_r$ and $\Omega_{m}$ evolve to very large values. Hence, this is not consistent with cosmological observations. 
\end{enumerate}

\begin{figure}[H]
\begin{subfigure}{.55\textwidth}
  %
  \includegraphics[width=.82\linewidth]{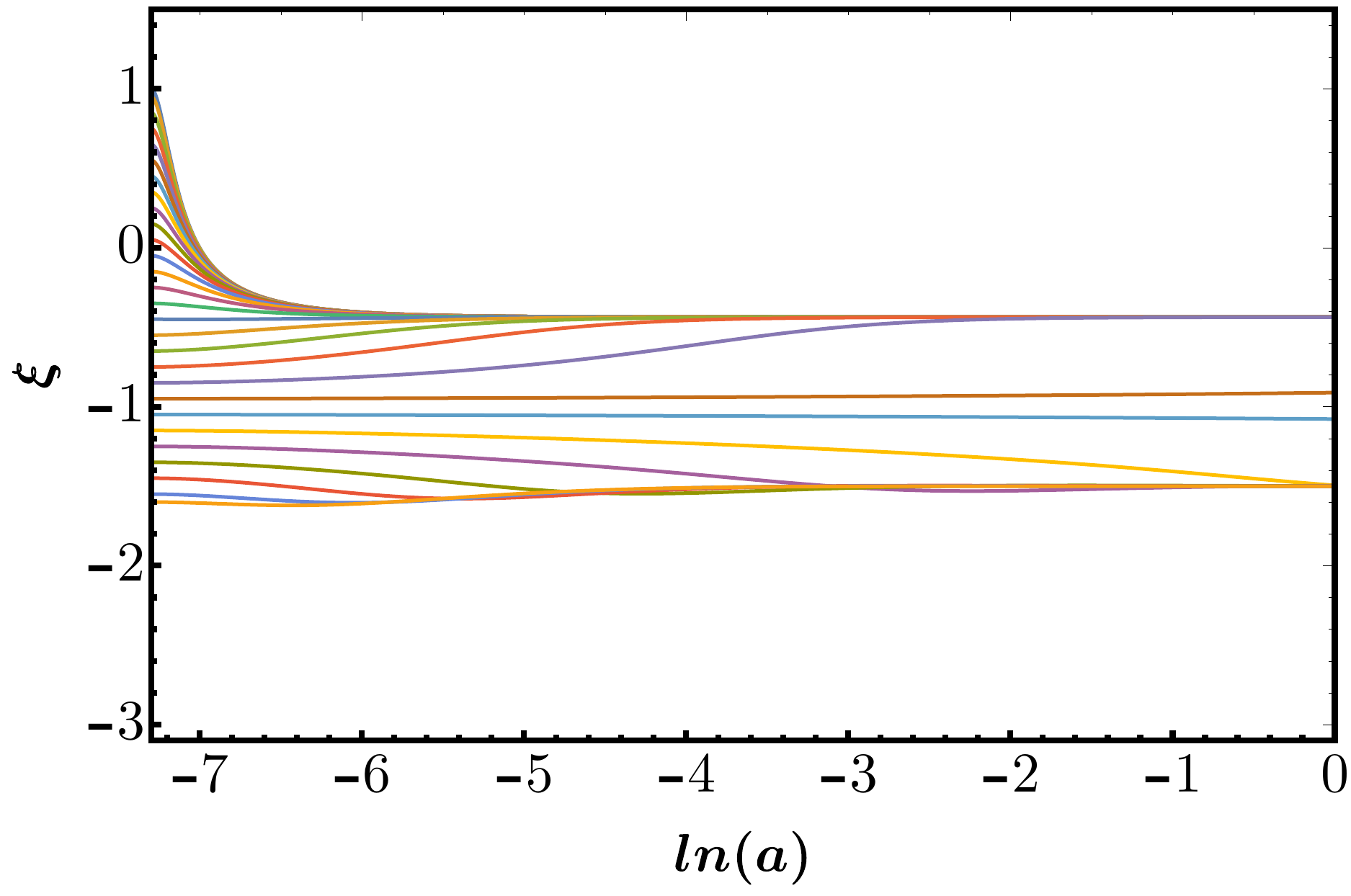}  
  \caption{$\xi(a)$ vs ln(a)}
\end{subfigure}
\begin{subfigure}{.55\textwidth}
  %
  \includegraphics[width=.82\linewidth]{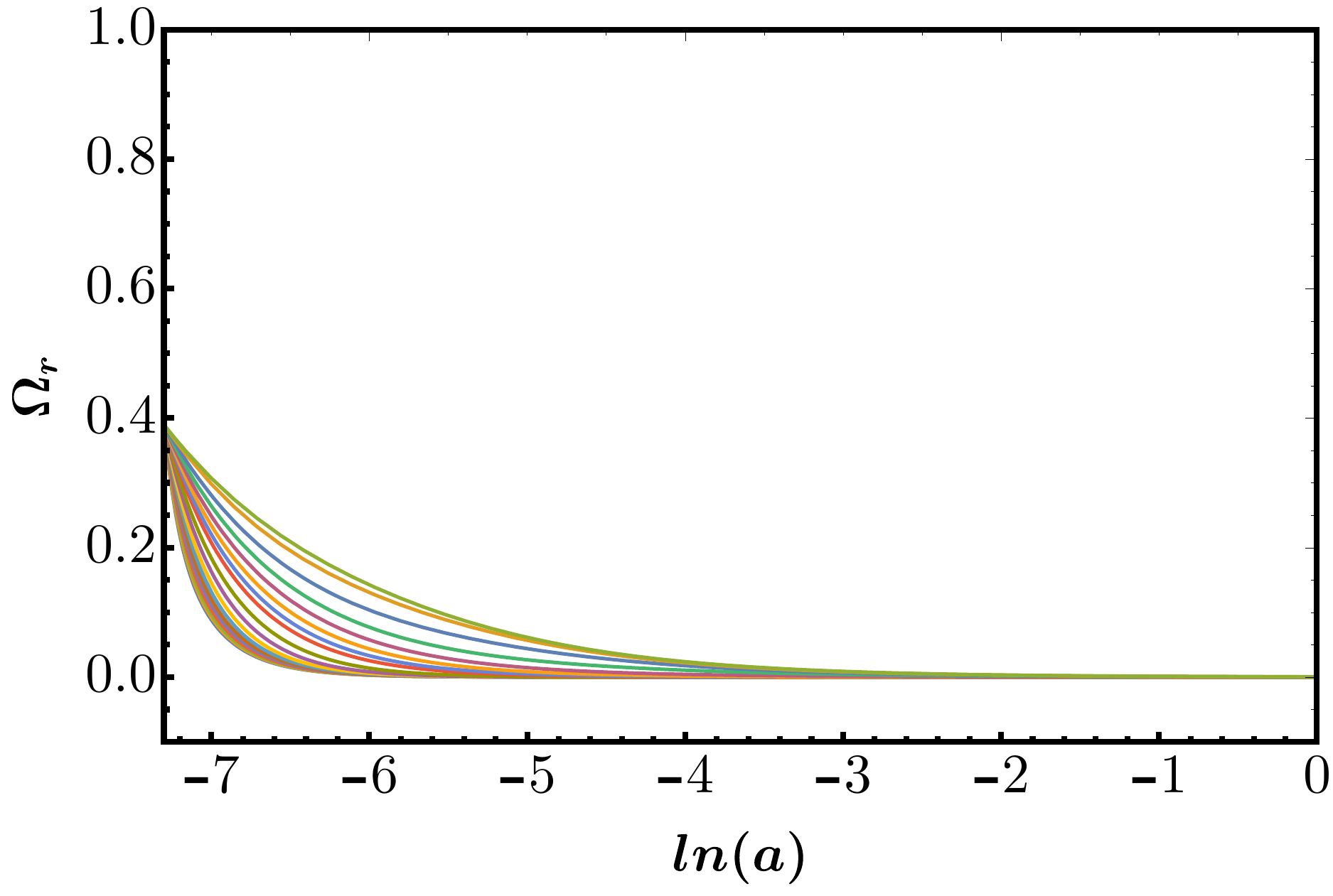}  
  \caption{$\Omega_{r}(a)$ vs ln(a)}
\end{subfigure}
\newline
\begin{subfigure}{.55\textwidth}
  %
  \includegraphics[width=.82\linewidth]{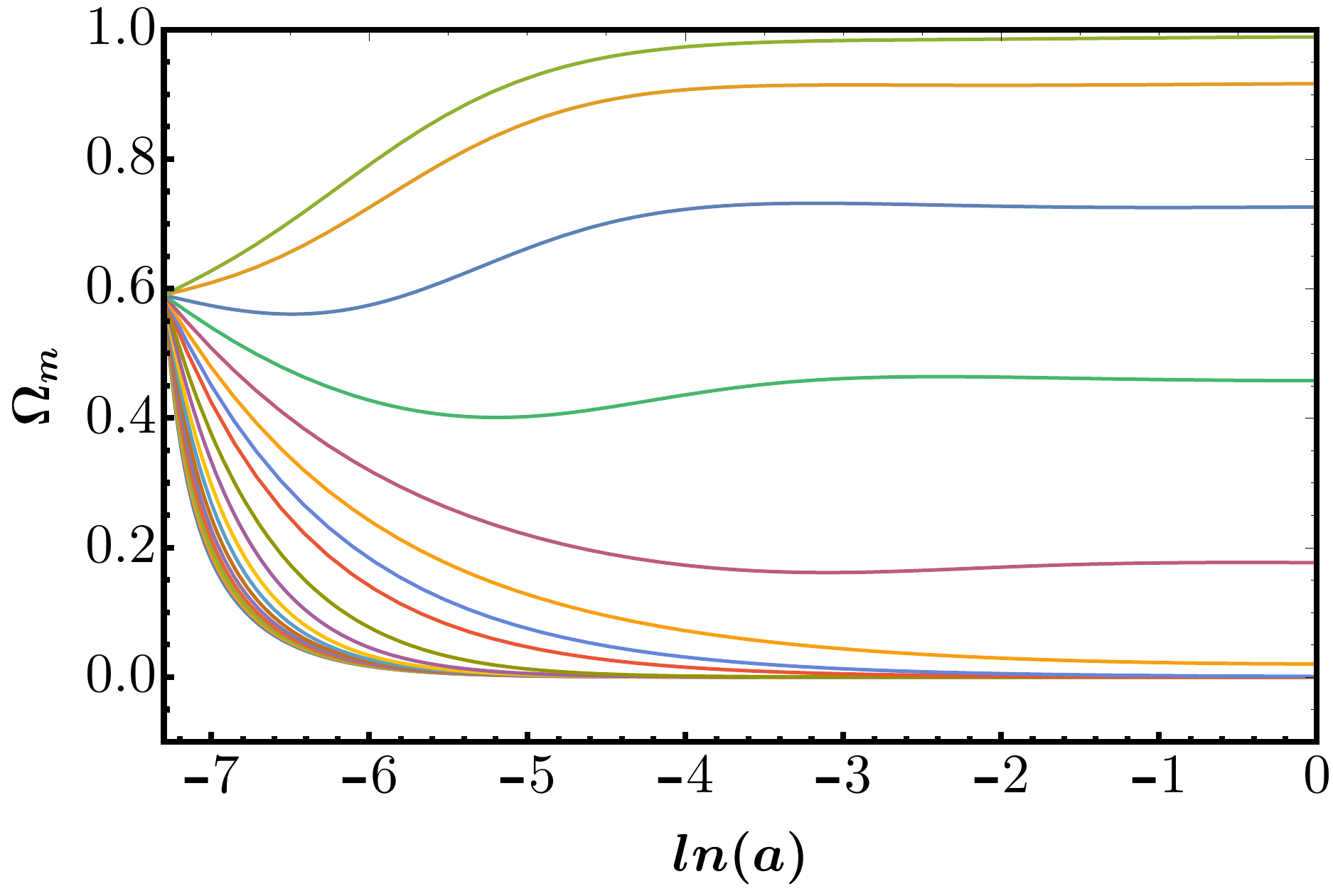}  
  \caption{$\Omega_{m}(a)$ vs ln(a)}
\end{subfigure}
\begin{subfigure}{.55\textwidth}
  %
  \includegraphics[width=.82\linewidth]{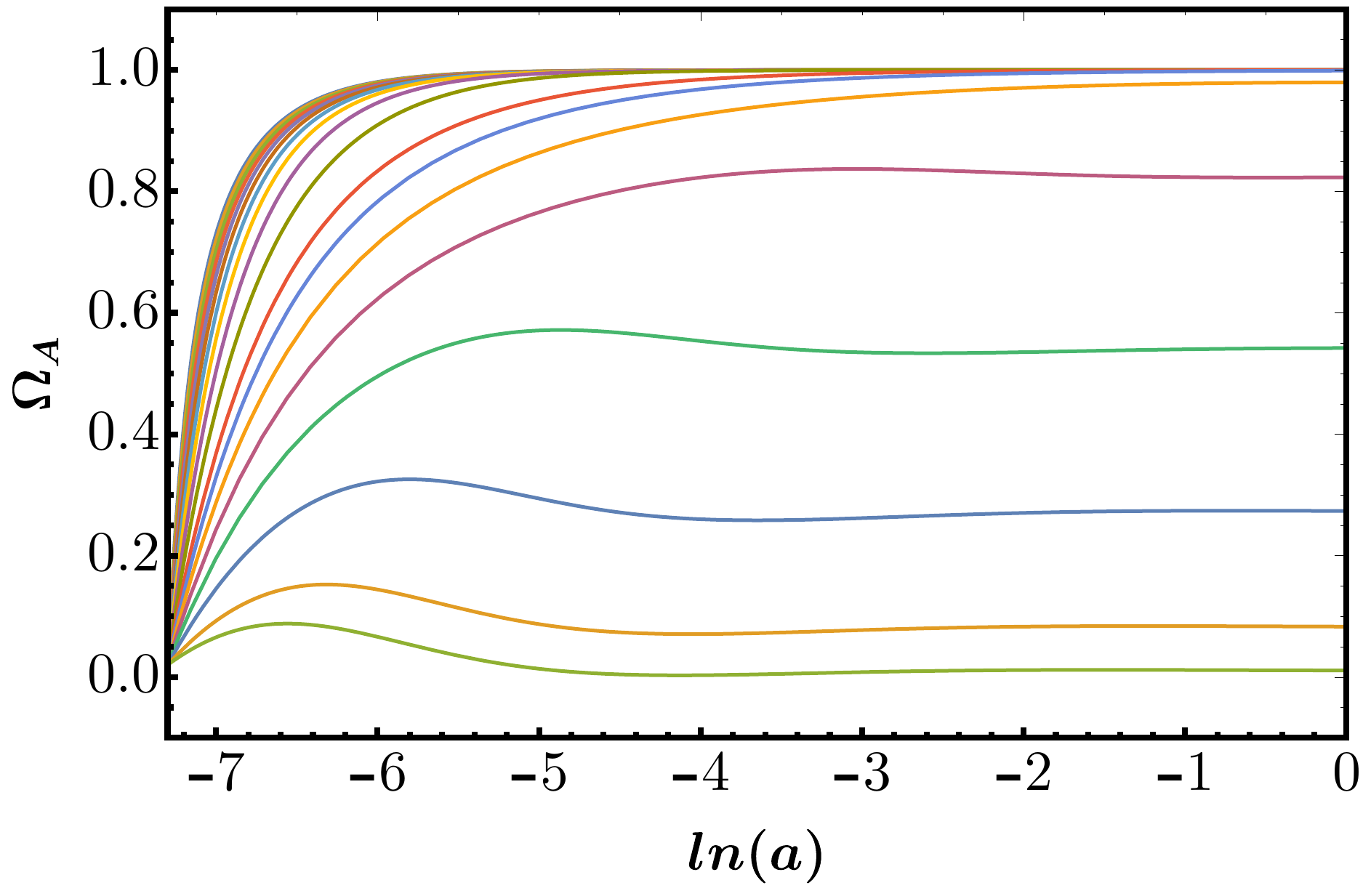}  
  \caption{$\Omega_{A}(a)$ vs ln(a)}
\end{subfigure}
\caption{Plot of $\xi(a), \Omega_r(a), \Omega_m(a), \Omega_A(a)$ as a function of $\ln(a)$ for ${\cal F}_4$ and $\alpha = 0.005$.}
\label{fig:newmodel5alpha0.005}
\end{figure}

Fig. \ref{fig:newmodel5alpha0.005} contains the plots for ${\cal F}_4$ 
and $\alpha = 0.005$. Different colors in the plots refer to different initial values 
of $\xi$ in the range $[-5, 0.6]$. From these plots, we infer the following:
\begin{enumerate}
\item For initial values of $\xi  = -1$, the values ($\xi, \Omega_m, \Omega_r)$ diverge.

\item The attractors are at $\xi \approx -0.5$, $\xi = -1.5$ and $\xi \approx -2.7$. For the attractor point $\xi \approx -2.7$, $\Omega_{r}$ and $\Omega_{m}$ diverge or evolve to unphysical values.

\item For all initial values that lead to an attractor  $\xi \approx -0.5$, 
$\Omega_{r}$ and $\Omega_{m}$ start at $0.39$ and $0.59$ respectively, and quickly converge to $0$; while $\Omega_{A}$ starts at 0.02 and quickly converges to $1$. 

\item For all initial values that lead to an attractor  $\xi \approx -1.5$, 
$\Omega_{r}$ and $\Omega_{m}$ starts at $0.39$ and $0.59$ respectively, and do not converge to $0$; while, $\Omega_{A}$ starts at 0.02 and do not converge to $1$. 
\end{enumerate}

\subsection{Can Class II models bypass the no-go theorem?}

From the analysis of this section, we see that the singularity at $\xi = -1$ is removed in the majority of the cases, and the singularity at $\xi = -1.2$ is removed for all cases. However, we can not bypass the no-go theorem. In other words, points close to $\xi = -1.5$ can not evolve to $\xi = -0.5$. Unlike the Class I models, this is not because of the singularities at $\xi = -1.2$ and $\xi = -1$. In these models, $\xi$ evolves to the attractor $\xi = -2$.
In other words,  $\xi = -2$ ($w_{\rm eff} = 1/3$, corresponding to radiation dominated) is the late-time attractor (in most cases).  More specifically, this model also fails to explain the late-time evolution of the universe. Thus, Class II models \emph{can not bypass} the no-go theorem.

\section{Understanding the No-Go theorem from the reduced action}\label{sec:alternate}

The action principle for the gravity sector reproduces the field equations when all metric components are varied independently. If space-time has a particular symmetry, it is possible to impose this symmetry in the full action and integrate the redundant coordinates leading to the reduced action. Varying the reduced action with respect to the remaining dependent variables may not be equivalent
to the reduced gravity equations, i.e., the equation resulting
from the imposition of the symmetry to the full gravity
field equations. When they are equivalent, the corresponding
reduced Lagrangian is said to be valid. A well-known example
of a non-valid case is the class B family of Bianchi models~\cite{1972-Maccallum.Taub-CMP}. The class of spatially homogeneous  Bianchi cosmologies, including the FLRW space-time and Bianchi IX, are valid cases~\cite{1972-Maccallum.Taub-CMP,2010-Appignani.etal-JCAP,2015-Nandi.Shankaranarayanan-JCAP}. Also, 
in the case of FLRW space-time, this is valid up to second-order in perturbations~\cite{2015-Nandi.Shankaranarayanan-JCAP}. This section shows that the reduced action approach of Ricci-inverse models provides an understanding of the no-go theorem.

Numerical analysis of the modified Friedmann equations of Class I and Class II models shows that these models can lead to a late-time accelerated Universe. However, neither of these models can have a smooth transition from a decelerated universe to a late-time accelerated universe. One possible reason is the appearance of the singularities in the anticurvature scalar either at $\xi = -1.2$ or at $\xi = -1$. With this as a cue, we use the reduced action approach to understand the no-go theorem.

%
\begin{table}[H]
\centering
\begin{tabular}{ccc}
\hline
\textbf{Class} & \bm{$\mathcal{L}$} & \textbf{Reduced form}  \\
\hline
Class I & $\frac{1}{A}$ & $\frac{3(1+\xi)(3+\xi)}{2(6+5\xi)}$ \\ 
\hline  
Class I & $\exp(RA)$ & $\exp\left(\frac{4(2+\xi)(6+5\xi)}{(1+\xi)(3+\xi)}\right)$\\
\hline
Class II & $\frac{1}{R \, A^{\mu\nu}A_{\mu\nu}}$ & 
$\frac{3(1+\xi)^{2}(3+\xi)^{2}}{8(2+\xi)(7\xi^{2}+15\xi+9)} $\\
\hline
\end{tabular}
\caption{Lagrangian corresponding to reduced action of Class I and Class II models \label{tab:2}}
\end{table}

\begin{figure}[H]
\begin{subfigure}{.5\textwidth}
  %
  \includegraphics[width=.82\linewidth]{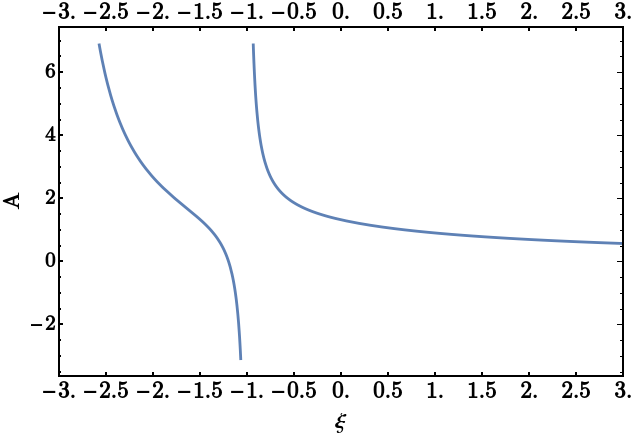}  
  \label{fig:l1}
\end{subfigure}
\begin{subfigure}{.5\textwidth}
  %
  \includegraphics[width=.82\linewidth]{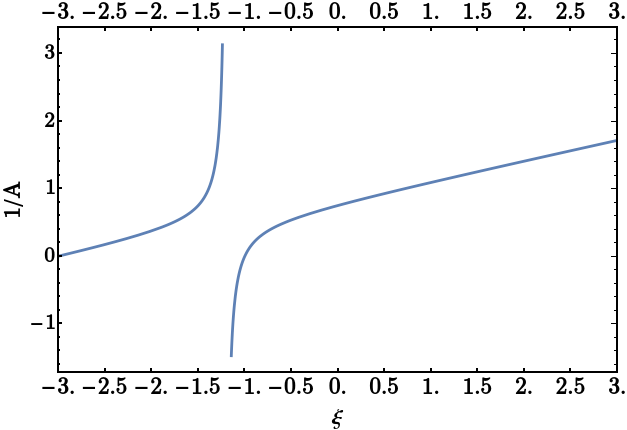}  
  \label{fig:l2}
\end{subfigure}
\newline
\begin{subfigure}{.5\textwidth}
  %
  \includegraphics[width=.82\linewidth]{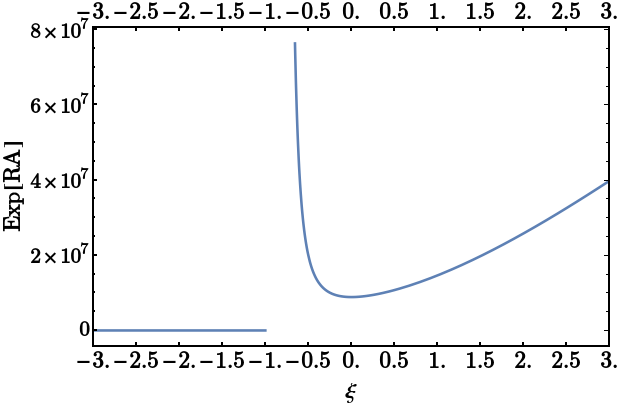}  
  \label{fig:l3}
\end{subfigure}
\begin{subfigure}{.5\textwidth}
  %
  \includegraphics[width=.82\linewidth]{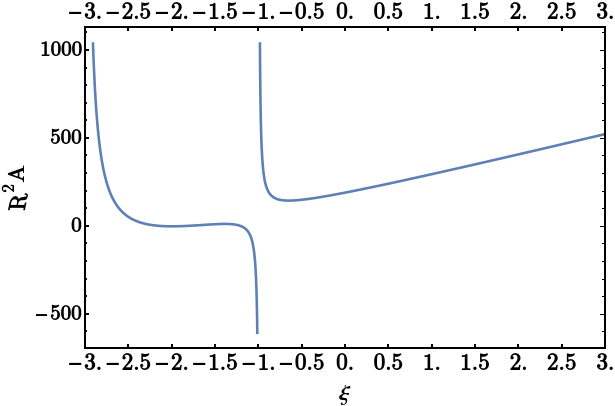}  
  \label{fig:l4}
\end{subfigure}
\newline
\begin{subfigure}{.5\textwidth}
  %
  \includegraphics[width=.82\linewidth]{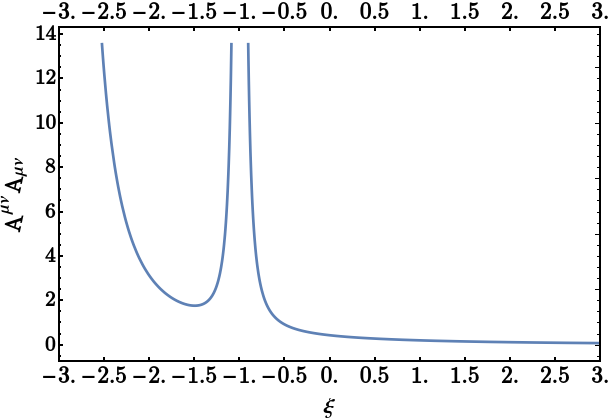}  
  \label{fig:l5}
\end{subfigure}
\begin{subfigure}{.5\textwidth}
  %
  \includegraphics[width=.82\linewidth]{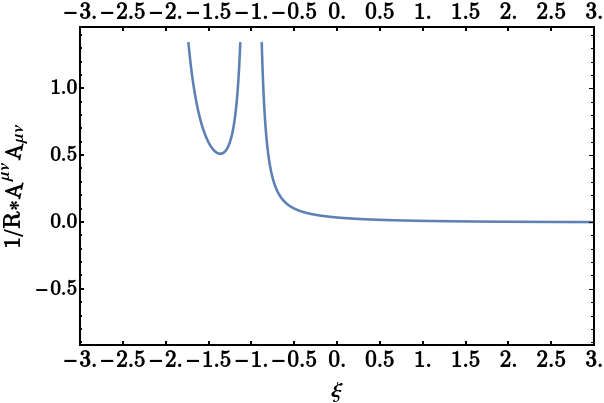}  
  \label{fig:l6}
\end{subfigure}
\newline
\begin{subfigure}{.5\textwidth}
  %
  \includegraphics[width=.82\linewidth]{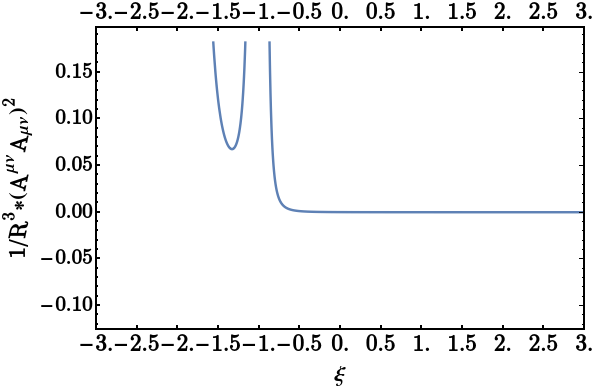}  
  \label{fig:l7}
\end{subfigure}
\begin{subfigure}{.5\textwidth}
  %
  \includegraphics[width=.82\linewidth]{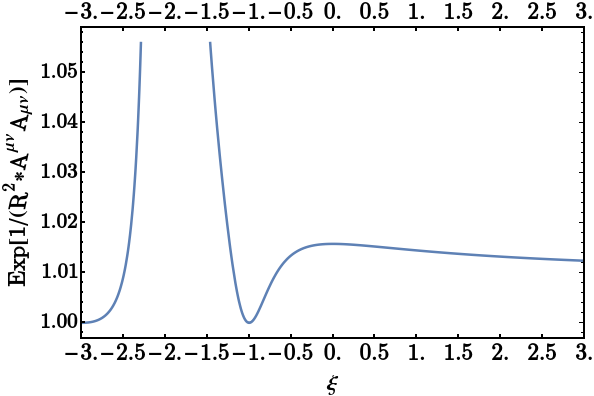}  
  \label{fig:l8}
\end{subfigure}
\caption{Plot of the reduced Lagrangian in FLRW background vs. $\xi$}
\label{fig:lagrangian}
\end{figure}

For both Class I (Eq. \ref{Class I action}) and Class II (Eq. \ref{Class II action}) 
models action, we obtain the corresponding reduced action by applying the symmetry of the FLRW space-time. The reduced action is a function of $\xi$ (excluding the $H^{2}$ factor). Table \ref{tab:2} contains the Lagrangian corresponding to the reduced action for some Class I and Class II models. We do not include the Lagrangian corresponding to the Einstein-Hilbert action as it does not contribute to any singularities during the evolution of the Universe.

Fig. \ref{fig:lagrangian} contains the plots for the different Class I and Class II models discussed in Secs. (\ref{sec:Class I models}, \ref{sec:Class II models}) 
as a function of $\xi$. Since these are plotted as functions of $\xi$, the plots are \emph{independent} of the epoch of the evolution of the Universe. In other words, we can apply the analysis to early-time or late-time. From the plots, we infer the following:

\begin{enumerate}
\item From the plot of $\exp[RA]$ in Fig. \ref{fig:lagrangian} (second row, first column), we notice two features: First, minima occurs at $\xi = -0.5$. Second, 
there's a discontinuity at $\xi = -1$. For $\alpha = 0.07$, these features are seen in Fig. \ref{fig:(ra)}. Specifically, $\xi = -0.5$ acts as an attractor. For the initial value of $\xi = -1$, the values $(\xi, \Omega_m, \Omega_r)$ diverge.  

\item From the plot of $1/A$ in Fig. \ref{fig:lagrangian} (first row, second column), we see that the anticurvature term has a discontinuity at $\xi \approx -1.2$ and two turning points at $\xi \approx -0.7$ and $\xi \approx -1.5$. For $\alpha = -4$, these 
features can be seen in  Fig~\ref{fig:1a1}. We notice from the plot that 
 $\xi \simeq -0.7$ and $\xi \simeq -1.5$ are attractor. For the initial value of $\xi = -1.2$, the values $(\xi, \Omega_m, \Omega_r)$ diverge. 
 
\item Thus, we can conclude that we can map the effective action approach to the features obtained by solving the Eqs. (\ref{density eom}, \ref{pressure eom}, \ref{energy density equation}) numerically in the redshift range $1500 < z < 0$.  For different values of $\alpha$, the attractor point or initial values where the physical quantities diverge might change slightly; however, the overall nature remains the same. 

\item The above conclusion is also true for the Class II models. From the plot of $\exp[1/(R^{2}A^{\mu\nu}A_{\mu\nu})]$ in Fig. \ref{fig:lagrangian} (fourth row, second column), we see that the anticurvature tensor term has three turning points $\xi \approx -0.5, -1, -2.7$ and has two discontinuities at $\xi = -1.5, -2.5$. 
For $\alpha = 0.005$, these features are seen in Fig \ref{fig:newmodel5alpha0.005}. 

\item From the plots in Fig. \ref{fig:lagrangian} of the reduced action, we notice 
that there exists discontinuity while going from $\xi = -1.5~\text{to}~\xi = -0.5$. This confirms from the analysis (in Secs. \eqref{sec:Class I models} and \eqref{sec:Class II models}) that it is not possible to smoothly evolve from the matter-dominated epoch ($\xi = -1.5$) to the current accelerated epoch ($\xi = -0.5$). Thus, the reduced action approach provides an alternative understanding of the no-go theorem.
\end{enumerate}

\section{Mapping Modified gravity theories and Ricci-inverse models }
\label{sec:modgrav}

The above results lead us to the following question: Can we map Ricci-inverse gravity to modified theories of gravity? In other words, can modified gravity theories mimic the features of Ricci-Inverse gravity? While the analysis is not possible for a generic space-time, it is possible for FRW space-time. Using the reduced action approach, we identify which modified theories of gravity mimic Ricci-inverse models.

We consider six modified gravity models that can be written as a function of the Ricci scalar, Ricci tensor, and Riemann tensor. In that sense, these models can be the closest representations of Ricci-inverse models. 

\subsection*{$\mathcal{L} = R + \frac{\alpha}{R^{2}}$, $[\alpha] = [L^{-6}]$}
\begin{figure}[H]
  \begin{center}
  \includegraphics[width=.62\linewidth]{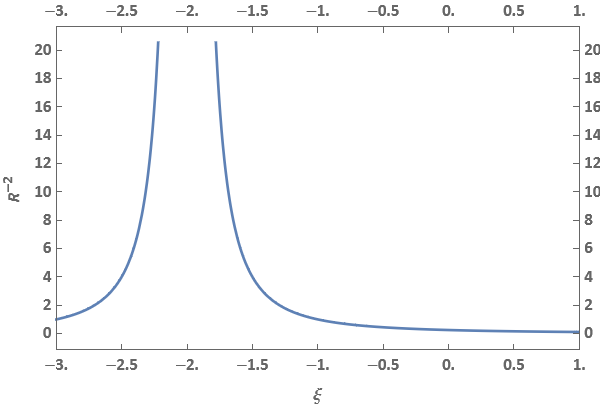} 
  \caption{Reduced action for $R^{-2}$ vs $\xi(a)$}\label{fig:Rsqinverse}  
  \end{center}
\end{figure}

Fig. \eqref{fig:Rsqinverse} is the plot of reduced action for the above Lagrangian versus $\xi$. This model leads to a discontinuity at $\xi = -2$ and no minima around $\xi = -1.5$ or $\xi = -0.5$. Thus, this will neither have a radiation-dominated attractor nor explain the evolution from radiation to matter-dominated epoch or matter-dominated epoch to the current accelerated epoch.

\subsection*{$\mathcal{L} = R + \alpha R^{2}$, $[\alpha] = [L^{6}]$}
\begin{figure}[H]
  \begin{center}
  \includegraphics[width=.62\linewidth]{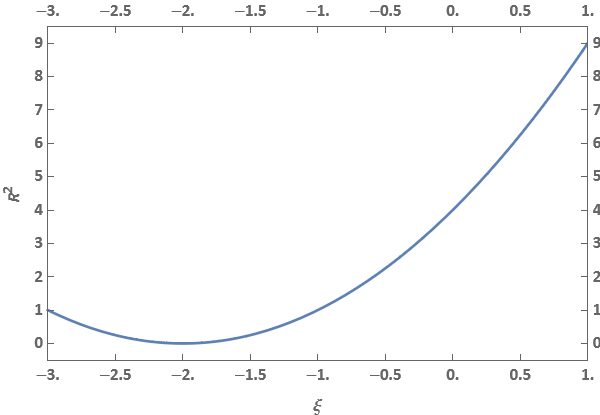}  
  \caption{Reduced action for $R^{2}$ vs $\xi(a)$}\label{Rsquare}
  \end{center}
\end{figure}

Fig. \ref{Rsquare} is the plot of reduced action for the Starobinsky model versus $\xi$. From the plot, we see that this modified gravity model leads to a minimum at $\xi = -2$ corresponding to radiation dominated phase. This modified gravity model can possibly explain the evolution from the early Universe to radiation dominated. Note that $\xi = -\epsilon$, where $\epsilon$ is the slow roll parameter. 

\subsection*{Inverse Gauss-Bonnet models} 

We consider  the following two such models that include the inverse of the Gauss-Bonnet term:
\begin{eqnarray}
\mathcal{L}_{\rm GB}^{(1)} &=& R + \frac{\alpha R}{(R^{2} - 4R^{\rho\sigma}R_{\rho\sigma} + R^{\mu\nu\alpha\beta}R_{\mu\nu\alpha\beta})}\qquad [\alpha] = \text{dimensionless} \\
\mathcal{L}_{\rm GB}^{(2)} & = & R + \frac{\alpha}{(R^{2} - 4R^{\rho\sigma}R_{\rho\sigma} + R^{\mu\nu\alpha\beta}R_{\mu\nu\alpha\beta})}\qquad [\alpha] = [L^{-2}]
\end{eqnarray}
\begin{figure}[H]
  \begin{center}
  \includegraphics[width=.62\linewidth]{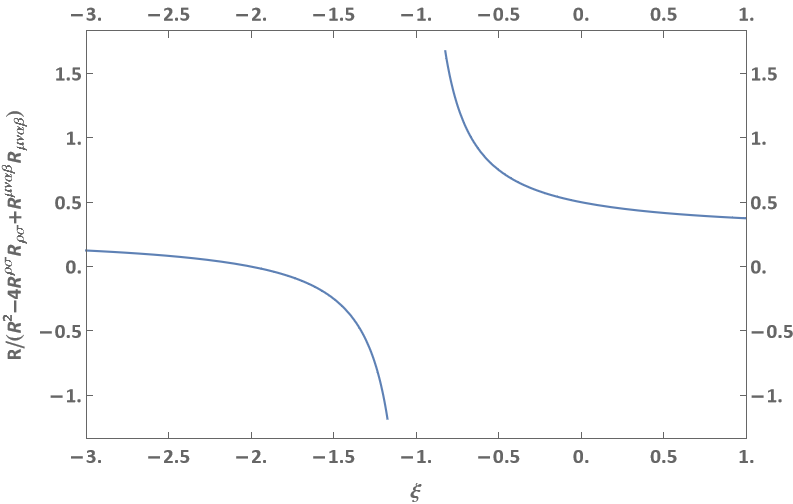}  
  \caption{Reduced action for $\mathcal{L}_{\rm GB}^{(1)}$ vs $\xi(a)$}\label{R_by_gauss_bonnet}
  \end{center}
\end{figure}
\begin{figure}[H]
  \begin{center}
  \includegraphics[width=.62\linewidth]{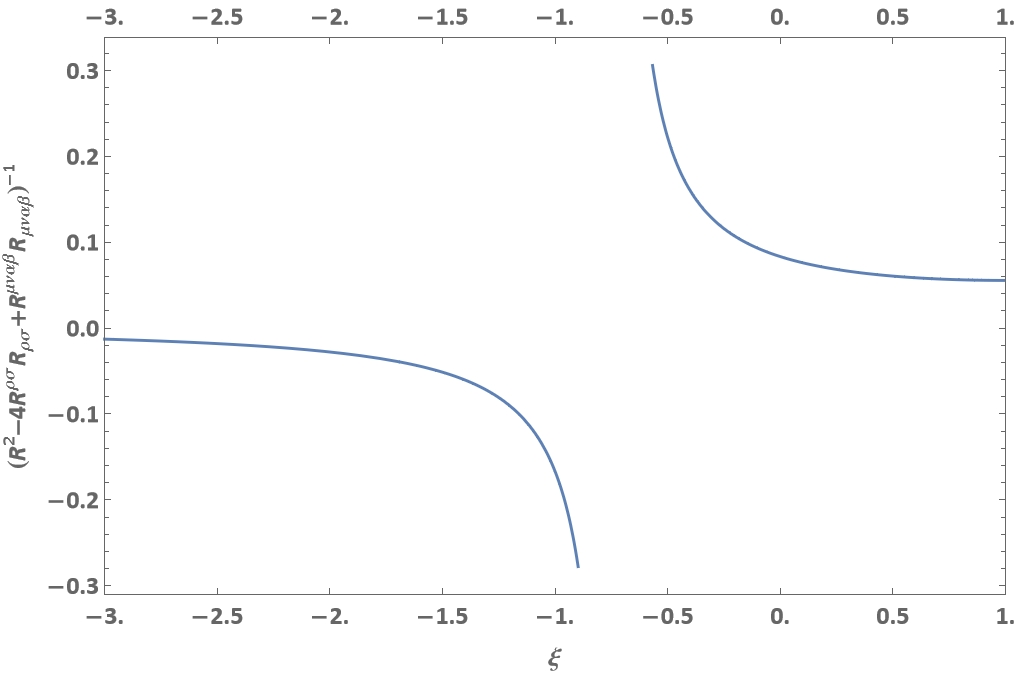}  
  \caption{Reduced action for $\mathcal{L}_{\rm GB}^{(2)}$ vs $\xi(a)$}\label{1_by_gauss_bonnet}
  \end{center}
\end{figure}

Fig. \ref{R_by_gauss_bonnet} and Fig. \ref{1_by_gauss_bonnet} are the plots of reduced action for the inverse Gauss-Bonnet models versus $\xi$.
From the figures, we see that the inverse of Gauss-Bonnet terms leads to 
a discontinuity at approximately $\xi \simeq -1$. Thus, it is not possible to evolve smoothly from the matter-dominated epoch ($\xi = -1.5$) to the current accelerated epoch ($\xi = -0.5$) through this model.

\subsection*{Generalized curvature models} 

Let us now consider the case where the coefficients of  the Riemann, Ricci tensor and Ricci scalar  are different from that of Gauss-Bonnet ---  $(aR^{2} + bR^{\rho\sigma}R_{\rho\sigma} + cR^{\mu\nu\alpha\beta}R_{\mu\nu\alpha\beta})$  where $a, b$ and $c$ are not equal to $1, -4$ and $1$, respectively. Specifically, we consider the following two models: 
\begin{eqnarray}
\mathcal{L}_{\rm Gen}^{(1)} & =&  R + \frac{\alpha R}{(R^{2} - 3 \, R^{\rho\sigma}R_{\rho\sigma} - 10 \, R^{\mu\nu\alpha\beta}R_{\mu\nu\alpha\beta})} \qquad [\alpha] = \text{dimensionless} \\
\mathcal{L}_{\rm Gen}^{(2)} & =& R + \frac{\alpha}{(R^{2} - 2.35R^{\rho\sigma}R_{\rho\sigma} - 7.55R^{\mu\nu\alpha\beta}R_{\mu\nu\alpha\beta})} \qquad [\alpha] = [L^{-2}]
\end{eqnarray}
\begin{figure}[H]
  \begin{center}
  \includegraphics[width=.62\linewidth]{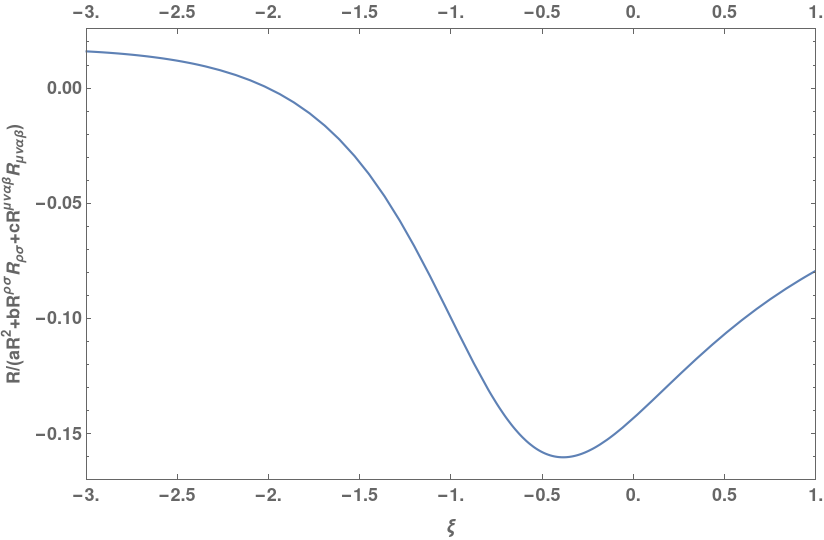}  
  \caption{Reduced action for $\mathcal{L}_{\rm Gen}^{(1)}$  with $a = 1$, $b = -3$, $c = -10$}\label{R_by_general}
  \end{center}
\end{figure}
\begin{figure}[H]
\begin{center}
  \includegraphics[width=.62\linewidth]{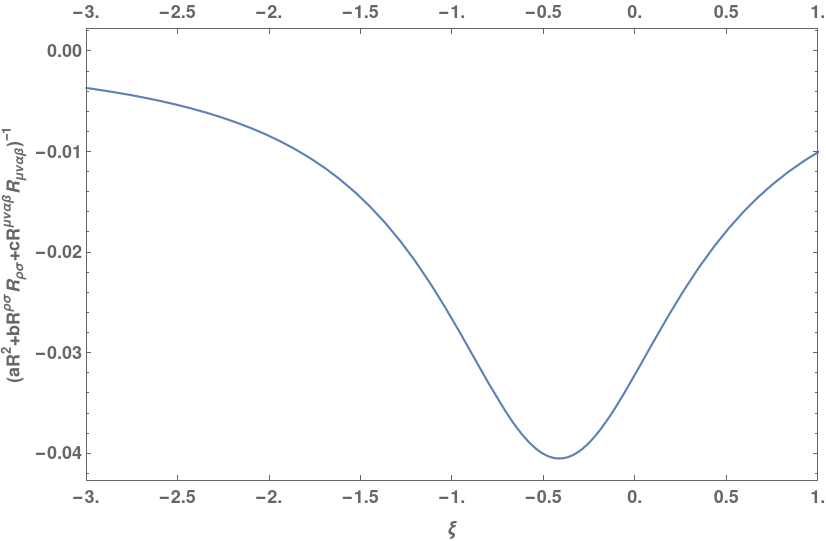}  
  \caption{Reduced action for $\mathcal{L}_{\rm Gen}^{(2)}$  vs $\xi(a)$ with $a = 1$, $b = -2.35$, $c = -7.55$}\label{1_by_general}
  \end{center}
\end{figure}

Fig. \ref{R_by_general} and Fig. \ref{1_by_general} are the plots of reduced action for the generalized curvature models versus $\xi$.
From the figures, we see that for some specific values of $a, b$, and $c$, we get a minimum at $\xi = -0.5$. Hence, the evolution from the matter-dominated epoch to the current accelerated epoch may likely be explained through this model. We infer that the generalized curvature models can provide similar features as that of Ricci-inverse gravity models. The analysis needs to be for a generic space-time to do the complete mapping. We plan to do this in future work. 

However, this does \emph{not} imply that the inverse Ricci tensor can be expressed as a combination of the Polynomials of Ricci tensor or Riemann tensor. Similarly, it is not possible to generally express inverse Ricci scalar as a function of Ricci scalar. For instance, as we have seen in section (\ref{sec:framework}), for the FLRW metric, the inverse Ricci scalar and Ricci scalar are given by
\begin{equation}
A= \dfrac{2 a(t)(\dot{a}^2(t) + 5 a(t)\ddot{a}(t))}{3\ddot{a}(t)(2\dot{a}^2(t)+a(t)\ddot{a}(t))} ,\quad R= \dfrac{6}{a^2(t)} ( \dot{a}(t)^2 + \ddot{a}(t)a(t))
\end{equation}
As one can see, it is not possible to express $A$ as a function of $R$.

For the FLRW metric, the $00$ component of the anticurvature tensor is obtained by calculating the matrix inverse of the Ricci tensor:
\begin{equation}
A^{00} = \frac{R_{11}R_{22}R_{33}}{R_{00}R_{11}R_{22}R_{33}}
\end{equation}
We get this simple expression since the Ricci tensor is diagonal for the FLRW metric. However, we see that the expression is a combination of the individual components of the Ricci tensor. Even in this case, we cannot express the anticurvature tensor as a function of the Ricci tensor. Hence an action that is a function of anticurvature scalar $A$ cannot be expressed as a function of Riemann and Ricci tensors and Ricci scalar. Thus, Ricci-inverse gravity is a novel modified theory of gravity.

\section{Conclusions}\label{sec:conclusion}

In this work, we considered two classes of Ricci-inverse models --- Class I and Class II models. Class I models ($f(R, A)$) are an arbitrary, smooth function of Ricci and anticurvature scalars. Class II models ${\cal F}(R, A^{\mu\nu}A_{\mu\nu})$ are arbitrary, smooth functions of Ricci scalar and square of anticurvature tensor.
Class I models are classified into three subclasses --- Class Ia, Class Ib, and Class Ic. Class Ia models are polynomials in $R$, and $A$~\cite{2020-Amendola.etal-Phys.Lett.B,2021-Do-Eur.Phys.J.C,2022-Do-Eur.Phys.J.C}. Class Ib models are of the form $\exp[(R A)^n]$. Class Ic models are non-polynomial functions of $R$ and $A$. Class II models are further
classified into two subclasses (Class IIa, and Class IIb). 
Class II models are further classified into two subclasses --- Class IIa, and  Class IIb. Class IIa models are polynomials in $R$ and $A^{\mu\nu}A_{\mu\nu}$. Class IIb models are of the form $\exp[\alpha (R^2 A^{\mu\nu}A_{\mu\nu})^n]$. 

We evolved the modified Friedmann equations (\ref{density eom}, \ref{pressure eom}, \ref{energy density equation}) numerically in the redshift range $1500 < z < 0$ and  obtained $\xi(a), \Omega_r(a), \Omega_m(a), \Omega_A(a)$ as a function of $\ln(a)$ (corresponding to the redshift range $1500 < z < 0$) for different values of $\alpha$. We showed that although these models can lead to the late-time accelerated Universe, it can not evolve from a decelerated expansion to an accelerated expansion. Hence, these two classes of Ricci-inverse models are \emph{not} consistent with the observed background evolution of the Universe.

In the case of Class I models we found that as the Universe evolves from the matter-dominated epoch ($\xi = -1.5$) to the current accelerated epoch ($\xi = -0.5$), we encounter a singularity either at $\xi = -1.2$ or at $\xi = -1$. Choosing the form of $f(R, A)$ can remove one singularity but not both singularities. Since, we need to cross both the points to evolve from from the matter-dominated epoch ($\xi = -1.5$) to the current accelerated epoch ($\xi = -0.5$), class I models \emph{can not} bypass the no-go theorem. Unlike the Class I models, Class II models do not have singularities at $\xi = -1.2$ and $\xi = -1$. However, in these models, $\xi$ evolves to the attractor $\xi = -2$ and fails to explain the late-time acceleration of the Universe.

Using the reduced action approach, we obtained an alternative way of understanding the no-go theorem. Specifically, we showed that the features that the effective action approach exhibits can be mapped to the features obtained by solving the Eqs. (\ref{density eom}, \ref{pressure eom}, \ref{energy density equation}) numerically in the redshift range $1500 < z < 0$.  For different values of $\alpha$, the attractor point or initial values where the physical quantities diverge might change slightly; however, the overall nature remains the same. Hence, we can not bypass the no-go theorem for Ricci-inverse gravity models.

We considered several other modified theories of gravity and applied the reduced action approach to see if they could mimic  Ricci-Inverse Gravity. We found models of the form $(aR^{2} + bR^{\rho\sigma}R_{\rho\sigma} + cR^{\mu\nu\alpha\beta}R_{\mu\nu\alpha\beta})^{-1}$  with values of $a, b$ and $c$ different than that of Gauss-Bonnet can potentially explain the evolution of the Universe from the matter-dominated phase to the accelerated expansion phase for some specific values of $a, b$ and $c$~\cite{2005-Carroll.etal-Phys.Rev.D}.

Many modified gravity models lead to extra degrees of freedom which cause instabilities. From our analysis for FLRW space-time, the extra degree of freedom is not apparent. To understand the properties of the plausible extra degree of freedom, one must do perturbation theory. The perturbation theory is essential if the background analysis provides a possible explanation for the accelerated expansion. In this work, we have explicitly shown that the inverse Ricci model cannot explain the cosmic expansion history starting from the radiation-dominated epoch to the matter-dominated epoch to the dark energy-dominated epoch. Hence we have not done the stability analysis for Ricci-inverse gravity.

We have not done a detailed analysis for the early Universe --- Universe evolving from an inflationary epoch to a radiation-dominated epoch. However, the reduced action approach in section (\ref{sec:alternate}) is independent of the epoch. Hence, our analysis can be extended and leads to the \emph{same conclusion} --- Ricci-inverse gravity cannot smoothly evolve from an inflationary epoch to a radiation-dominated epoch in the flat FLRW space-time. In other words, Ricci-inverse gravity can not provide a satisfactory explanation for a smooth exit from inflation.

Thus, from these results, we can conclude that Ricci-inverse gravity can not explain the late-time acceleration of the Universe in the FLRW metric. However, if one can construct a scalar quantity with no singularities or no singularities in the range  
$\xi \in [-2, 0]$, then it is possible to construct a Lagrangian such that it has minima around $\xi = -0.5$, which can act as an attractor leading to the late-time acceleration of the Universe. One possibility is to consider Riemann-inverse gravity. In this gravity, we can define an ``Anti-Riemann" tensor as:
\begin{equation}
A_{\nu}^{\alpha\rho\gamma}R^{\mu}_{\alpha\rho\beta} = \delta^{\mu}_{\nu}\delta^{\gamma}_{\beta}
\end{equation}
This is beyond the scope of this work but is currently being investigated.

\section{Acknowledgements}
ID acknowledges the financial support provided by DST, Govt. of India through INSPIRE scheme. JPJ is supported by CSIR Senior Research Fellowship, India. The work is supported by  ISRO-Respond Grant, India. 

\section*{Data availability}
Data sharing is not applicable to this article as no datasets were generated or analysed during the current study.

\appendix
\section{Derivation of Eq. \eqref{eomII} for Class II models}
\label{sec:Appendix A}
Consider the action:
\begin{equation}
S = \int\sqrt{-g} \hspace{2pt}d^{4}x\left({\cal F}(R,A^{\mu\nu}A_{\mu\nu})\right)
\end{equation}
Varying the action w.r.t the metric leads to:
\begin{equation}
\begin{split}
\delta S = \int d^{4}x\left[ {\cal F} \delta\sqrt{-g}+\sqrt{-g}\hspace{2pt}\delta {\cal F} \right]\hspace{140pt}\\
= \int \sqrt{-g} \hspace{2pt}d^{4}x\left[\frac{-1}{2} {\cal F} g_{\mu\nu}\delta g^{\mu\nu}+{\cal F}_{R}R_{\mu\nu}\delta g^{\mu\nu} + {\cal F}_{R}g^{\mu\nu}\delta R_{\mu\nu} + {\cal F}_{A^{2}} A^{\mu\nu}\delta A_{\mu\nu}
+ {\cal F}_{A^{2}}A_{\mu\nu}\delta A^{\mu\nu}
\right]
\end{split}
\end{equation}
where ${\cal F}_{R} = \partial {\cal F}/\partial R$ and ${\cal F}_{A^{2}} = \partial {\cal F}/\partial (A^{\mu\nu}A_{\mu\nu})$.
We will calculate each term in the RHS separately:  

$$ \hspace*{-7.5cm}\textbf{I term :} \qquad \qquad \frac{-1}{2} {\cal F} g_{\mu\nu}\delta g^{\mu\nu} = \frac{1}{2} {\cal F} 
g^{\mu\nu}\delta g_{\mu\nu}$$ 

\vspace*{5pt}

$$\hspace*{-7.5cm}\textbf{II term :} \qquad \qquad
{\cal F}_{R}R_{\mu\nu}\delta g^{\mu\nu} = -{\cal F}_{R}R^{\mu\nu}\delta g_{\mu\nu}$$
\vspace*{5pt}

\textbf{III term :} $\qquad \qquad {\cal F}_{R}g^{\mu\nu}\delta R_{\mu\nu}$
\begin{gather*}
\delta R^{a}_{bcd} = \nabla_{c}(\delta \Gamma^{a}_{db}) - \nabla_{d}(\delta \Gamma^{a}_{cb})\\
= \frac{1}{2}\nabla_{c}[g^{ai}(-\nabla_{i}\delta g_{db} + \nabla_{d}\delta g_{bi} + \nabla_{b}\delta g_{di}) + \delta g^{ai}\Gamma_{idb}]
-\frac{1}{2}\nabla_{d}[g^{ai}(-\nabla_{i}\delta g_{cb} + \nabla_{c}\delta g_{bi} + \nabla_{b}\delta g_{ci}) + \delta g^{ai}\Gamma_{icb}]\\
= \frac{g^{ai}}{2}\nabla_{c}[-\nabla_{i}\delta g_{db} + \nabla_{d}\delta g_{bi} + \nabla_{b}\delta g_{di}]
 - \frac{g^{ai}}{2}\nabla_{d}[-\nabla_{i}\delta g_{cb} + \nabla_{c}\delta g_{bi} + \nabla_{b}\delta g_{ci}]
\end{gather*}
We have:
\begin{gather*}
\delta R_{ab} = \frac{1}{2}[-\nabla^{d}\nabla_{d}\delta g_{ab} + \nabla^{d}\nabla_{a}\delta g_{bd} + \nabla^{d}\nabla_{b}\delta g_{ad} - \nabla_{a}\nabla_{b}(g^{cd}\delta g_{cd})]
\end{gather*}
After contracting with metric, we get:
\begin{gather*}
g^{ab}\delta R_{ab} = \nabla^{b}\nabla^{a}\delta g_{ab} - \nabla^{b}\nabla_{b}(g^{cd}\delta g_{cd})\\
{\cal F}_{R}g^{\mu\nu}\delta R_{\mu\nu} = {\cal F}_{R}\nabla^{\nu}\nabla^{\mu}\delta g_{\mu\nu} - {\cal F}_{R}\nabla^{\mu}\nabla_{\mu}(g^{\beta\alpha}\delta g_{\beta\alpha})
\end{gather*}
Rewriting, we get:
\begin{gather*}
{\cal F}_{R}\nabla^{\nu}\nabla^{\mu}\delta g_{\mu\nu} = \nabla^{\nu}[{\cal F}_{R}\nabla^{\mu}\delta g_{\mu\nu}] - \nabla^{\mu}[\nabla^{\nu}{\cal F}_{R}\hspace{2pt}\delta g_{\mu\nu}] + \delta g_{\mu\nu} \nabla^{\mu} \nabla^{\nu}{\cal F}_{R}\\
-{\cal F}_{R}\nabla^{\mu}\nabla_{\mu}(g^{\beta\alpha}\delta g_{\beta\alpha}) = -\nabla^{\mu}[{\cal F}_{R}\nabla_{\mu}(g^{\beta\alpha}\delta g_{\beta\alpha})] + \nabla_{\mu}[\nabla^{\mu}{\cal F}_{R}\hspace{2pt}g^{\beta\alpha}\delta g_{\beta\alpha}]-g^{\beta\alpha}\delta g_{\beta\alpha}\nabla_{\mu}\nabla^{\mu}{\cal F}_{R}
\end{gather*}
The first two terms in both of the above formulas, when multiplied with $\sqrt{-g}$, become a total derivative and hence won't contribute to the resulting equations (boundary terms). We then have:
\begin{gather*}
{\cal F}_{R}g^{\mu\nu}\delta R_{\mu\nu} = [\nabla^{\mu} \nabla^{\nu}{\cal F}_{R} - g^{\mu\nu} \nabla^{2}{\cal F}_{R}]\delta g_{\mu\nu}
\end{gather*}

\textbf{IV term :} ${\cal F}_{A^{2}} A^{\mu\nu}\delta A_{\mu\nu}$
\begin{gather*}
A_{\mu\nu} = g_{\nu\beta}g_{\mu\alpha}A^{\alpha\beta}\,;~~~~
\delta A_{\mu\nu} = g_{\nu\beta}g_{\mu\alpha}\delta A^{\alpha\beta} + g_{\nu\beta}\delta g_{\mu\alpha}A^{\alpha\beta} + \delta g_{\nu\beta}g_{\mu\alpha}A^{\alpha\beta}\\
{\cal F}_{A^{2}} A^{\mu\nu}\delta A_{\mu\nu} = {\cal F}_{A^{2}}[A^{\mu\nu}g_{\nu\beta}g_{\mu\alpha}\delta A^{\alpha\beta} + A^{\mu\nu}g_{\mu\alpha}A^{\alpha\beta}\delta g_{\nu\beta} + A^{\mu\nu}g_{\nu\beta}A^{\alpha\beta}\delta g_{\mu\alpha}]\\
= {\cal F}_{A^{2}}A_{\mu\nu}\delta A^{\mu\nu} + {\cal F}_{A^{2}}(A^{\rho\nu}A^{\mu}_{\rho}\delta g_{\mu\nu} + A^{\mu\rho}A^{\nu}_{\rho}\delta g_{\mu\nu})
\end{gather*}

Interchanging $\mu$, $\nu$ in the last term and using the properties that $\delta g_{\mu\nu} = \delta g_{\nu\mu}$ and $A^{\alpha\beta} = A^{\beta\alpha}$. Then the last and second terms are identical. We then finally have:
\begin{gather*}
{\cal F}_{A^{2}} A^{\mu\nu}\delta A_{\mu\nu} = {\cal F}_{A^{2}}A_{\mu\nu}\delta A^{\mu\nu} + 2{\cal F}_{A^{2}} A^{\rho\nu}A^{\mu}_{\rho}\delta g_{\mu\nu}
\end{gather*}

\textbf{V term :} ${\cal F}_{A^{2}}A_{\mu\nu}\delta A^{\mu\nu}$
\begin{gather*}
\delta A^{\mu\nu} = \frac{-1}{2}A^{\mu\alpha}g^{\rho\lambda}(\nabla_{\rho}\nabla_{\alpha}\delta g_{\beta\lambda} - \nabla_{\rho}\nabla_{\lambda}\delta g_{\alpha\beta} - \nabla_{\beta}\nabla_{\alpha}\delta g_{\rho\lambda} + \nabla_{\beta}\nabla_{\lambda}\delta g_{\alpha\rho} + [\nabla_{\beta},\nabla_{\rho}]\delta g_{\lambda\alpha})A^{\beta\nu}
\end{gather*}
Let's look at the first term here: $-\frac{1}{2}{\cal F}_{A^{2}}A_{\mu\nu}A^{\mu\alpha}g^{\rho\lambda}\nabla_{\rho}\nabla_{\alpha}\delta g_{\beta\lambda}A^{\beta\nu}$. Just like we did for the third term (write it as the sum of three terms, two of them become total derivatives, and only the third term contributes), we can do the same here. The final result is:
\begin{gather*}
-\frac{1}{2}{\cal F}_{A^{2}}A_{\mu\nu}A^{\mu\alpha}g^{\rho\lambda}\nabla_{\rho}\nabla_{\alpha}\delta g_{\beta\lambda} = -\frac{1}{2}\delta g_{\beta\lambda}g^{\rho\lambda}\nabla_{\rho}\nabla_{\alpha}({\cal F}_{A^{2}}A_{\mu\nu}A^{\mu\alpha}A^{\beta\nu})
\end{gather*}
In arriving, we have used $\nabla_{c}g^{ab} = 0$. We repeat the same for the other terms:
\begin{gather*}
{\cal F}_{A^{2}}A_{\mu\nu}\delta A^{\mu\nu} = \frac{-1}{2}\delta g_{\mu\nu}[g^{\rho\nu}\nabla_{\alpha}\nabla_{\rho}({\cal F}_{A^{2}}A_{\sigma\kappa}A^{\sigma\alpha}A^{\mu\kappa})
-\nabla^{2}({\cal F}_{A^{2}}A_{\sigma\kappa}A^{\sigma\mu}A^{\nu\kappa})\\
-g^{\mu\nu}\nabla_{\alpha}\nabla_{\rho}({\cal F}_{A^{2}}A_{\sigma\kappa}A^{\sigma\alpha}A^{\rho\kappa})
+g^{\rho\nu}\nabla_{\rho}\nabla_{\alpha}({\cal F}_{A^{2}}A_{\sigma\kappa}A^{\sigma\mu}A^{\alpha\kappa})\\
+g^{\rho\nu}\nabla_{\rho}\nabla_{\alpha}({\cal F}_{A^{2}}A_{\sigma\kappa}A^{\sigma\mu}A^{\alpha\kappa})
-g^{\rho\nu}\nabla_{\alpha}\nabla_{\rho}({\cal F}_{A^{2}}A_{\sigma\kappa}A^{\sigma\mu}A^{\alpha\kappa})]
\end{gather*}
We get:
\begin{gather*}
{\cal F}_{A^{2}}A_{\mu\nu}\delta A^{\mu\nu} = \frac{-1}{2}\delta g_{\mu\nu}[g^{\rho\nu}\nabla_{\alpha}\nabla_{\rho}({\cal F}_{A^{2}}A_{\sigma\kappa}A^{\sigma\alpha}A^{\mu\kappa})
-\nabla^{2}({\cal F}_{A^{2}}A_{\sigma\kappa}A^{\sigma\mu}A^{\nu\kappa})\\
-g^{\mu\nu}\nabla_{\alpha}\nabla_{\rho}({\cal F}_{A^{2}}A_{\sigma\kappa}A^{\sigma\alpha}A^{\rho\kappa})
+2g^{\rho\nu}\nabla_{\rho}\nabla_{\alpha}({\cal F}_{A^{2}}A_{\sigma\kappa}A^{\sigma\mu}A^{\alpha\kappa})
-g^{\rho\nu}\nabla_{\alpha}\nabla_{\rho}({\cal F}_{A^{2}}A_{\sigma\kappa}A^{\sigma\mu}A^{\alpha\kappa})]
\end{gather*}

Combining all the terms, we have:
{\small
\begin{equation}
\begin{split}
{\cal F}_{R}R^{\mu\nu} - 2{\cal F}_{A^{2}}A^{\mu\rho}A^{\nu}_{\rho} -\frac{1}{2}fg^{\mu\nu} - \nabla^{\mu}\nabla^{\nu}{\cal F}_{R} + g^{\mu\nu}\nabla^{2}{\cal F}_{R} + g^{\rho\nu}\nabla_{\alpha}\nabla_{\rho}({\cal F}_{A^{2}}A_{\sigma\kappa}A^{\sigma\alpha}A^{\mu\kappa}) \\
-\nabla^{2}({\cal F}_{A^{2}}A_{\sigma\kappa}A^{\sigma\mu}A^{\nu\kappa}) 
-g^{\mu\nu}\nabla_{\alpha}\nabla_{\rho}({\cal F}_{A^{2}}A_{\sigma\kappa}A^{\sigma\alpha}A^{\rho\kappa})
+2g^{\rho\nu}\nabla_{\rho}\nabla_{\alpha}({\cal F}_{A^{2}}A_{\sigma\kappa}A^{\sigma\mu}A^{\alpha\kappa}) \\
-g^{\rho\nu}\nabla_{\alpha}\nabla_{\rho}({\cal F}_{A^{2}}A_{\sigma\kappa}A^{\sigma\mu}A^{\alpha\kappa}) = 0
\end{split}
\end{equation}
}

\section{Class Ia models}\label{sec:Appendix B}

Substituting the following form~\cite{2020-Amendola.etal-Phys.Lett.B}:
\begin{equation}
f(R,A) = R + \frac{\alpha}{A}
\end{equation}
in Eq. \eqref{eomI}, we obtain the following modified Friedmann equations:
\begin{eqnarray}
\rho_{t} &=& 3\alpha H^{2}\frac{(\xi + 3)^{2}(5\xi +6)-18\xi '}{4(5\xi+6)^{3}} + 3H^{2} \\
p_{t} &=& -\frac{\alpha H^{2}}{4(5\xi+6)^{4}}
\left[(5\xi+6)\left((\xi+3)^{2}(2\xi+3)(5\xi+6)-18\xi''\right)+270(\xi')^{2}-54(\xi+2)(5\xi+6)\xi'\right] \nonumber \\
& & - 2H^{2}\xi - 3H^{2}
\end{eqnarray}

\begin{figure}[H]
\begin{subfigure}{.55\textwidth}
  %
  \includegraphics[width=.82\linewidth]{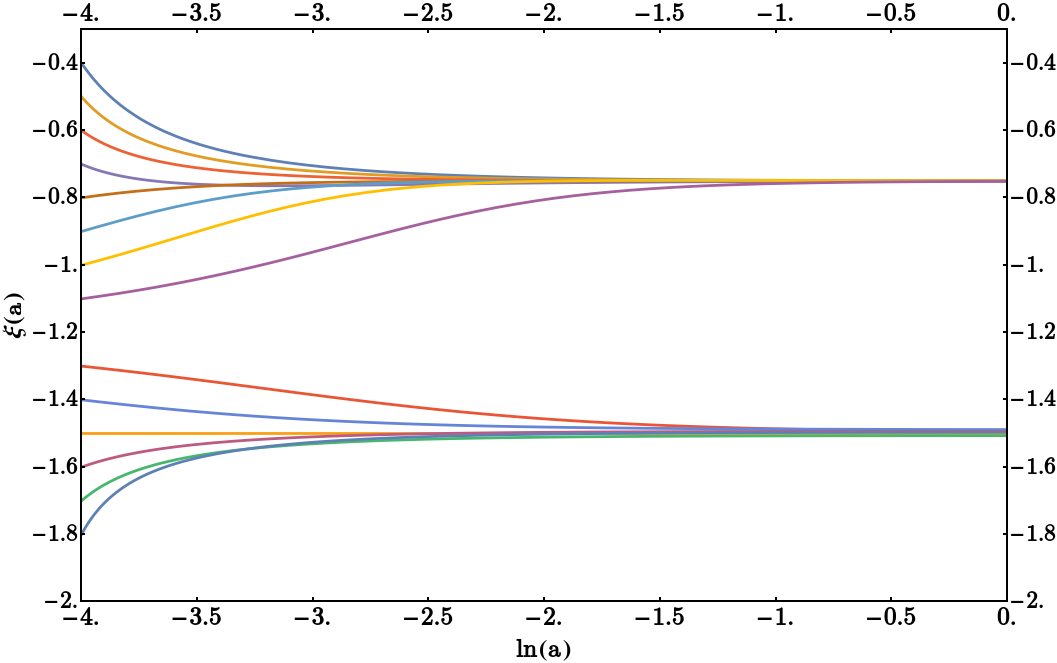}  
  \caption{$\Omega_{m}+\Omega_{A} = 1 $}
  \label{fig:1a1}
\end{subfigure}
\begin{subfigure}{.55\textwidth}
 %
  \includegraphics[width=.82\linewidth]{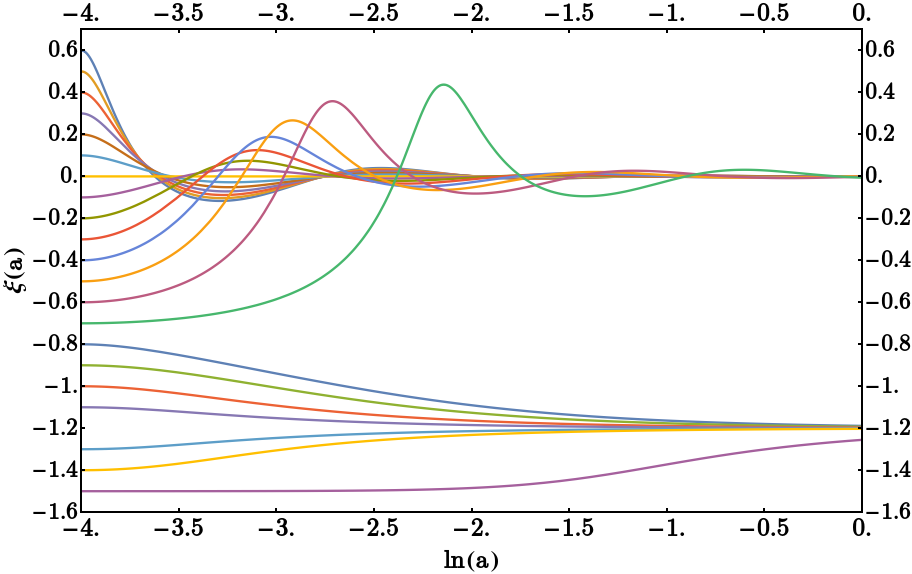}  
  \caption{$\Omega_{m}+\Omega_{A}+\Omega_{\Lambda} = 1$}
  \label{fig:1a3}
\end{subfigure}
\caption{Plot of $\xi~{\rm vs}~\ln(a)$ for $f(R,A) = R + \alpha/A$ and $\alpha = -4$. 
The left plot is for $\Omega_{m}+\Omega_{A} = 1$ and the right plot is for $\Omega_{m}+\Omega_{A}+\Omega_{\Lambda} = 1$.}
\label{fig:1a}
\end{figure}

In Fig. \eqref{fig:1a}, we have plotted $\xi(a)$ as a function of 
$\ln(a)$. The behaviour is identical to that derived in Ref. \cite{2020-Amendola.etal-Phys.Lett.B}. We see from the plots that the initial values of $\xi$ close to $\xi = -1.5$ will never smoothly evolve to  $\xi = -0.5$. This leads to the no-go theorem~\cite{2020-Amendola.etal-Phys.Lett.B}. 

\end{document}